\pdfoutput=1
\documentclass[11pt,twoside,a4paper,pdftex,cmspaper,final,collab]{cms-tdr}

\begin{document}\cmsNoteHeader{CFT-09-014}
%
%
%

%
%
\hyphenation{env-iron-men-tal}
\hyphenation{had-ron-i-za-tion}
\hyphenation{cal-or-i-me-ter}
\hyphenation{de-vices}
%
%
\RCS$Revision: 1.81 $
\RCS$Date: 2010/01/28 16:50:55 $
\RCS$Name:  $
%
%
%

\newcommand {\etal}{\mbox{et al.}\xspace} 
\newcommand {\ie}{\mbox{i.e.}\xspace}     
\newcommand {\eg}{\mbox{e.g.}\xspace}     
\newcommand {\etc}{\mbox{etc.}\xspace}     
\newcommand {\vs}{\mbox{\sl vs.}\xspace}      
\newcommand {\mdash}{\ensuremath{\mathrm{-}}} 

\newcommand {\Lone}{Level-1\xspace} 
\newcommand {\Ltwo}{Level-2\xspace}
\newcommand {\Lthree}{Level-3\xspace}

\providecommand{\ACERMC} {\textsc{AcerMC}\xspace}
\providecommand{\ALPGEN} {{\textsc{alpgen}}\xspace}
\providecommand{\CHARYBDIS} {{\textsc{charybdis}}\xspace}
\providecommand{\CMKIN} {\textsc{cmkin}\xspace}
\providecommand{\CMSIM} {{\textsc{cmsim}}\xspace}
\providecommand{\CMSSW} {{\textsc{cmssw}}\xspace}
\providecommand{\COBRA} {{\textsc{cobra}}\xspace}
\providecommand{\COCOA} {{\textsc{cocoa}}\xspace}
\providecommand{\COMPHEP} {\textsc{CompHEP}\xspace}
\providecommand{\EVTGEN} {{\textsc{evtgen}}\xspace}
\providecommand{\FAMOS} {{\textsc{famos}}\xspace}
\providecommand{\GARCON} {\textsc{garcon}\xspace}
\providecommand{\GARFIELD} {{\textsc{garfield}}\xspace}
\providecommand{\GEANE} {{\textsc{geane}}\xspace}
\providecommand{\GEANTfour} {{\textsc{geant4}}\xspace}
\providecommand{\GEANTthree} {{\textsc{geant3}}\xspace}
\providecommand{\GEANT} {{\textsc{geant}}\xspace}
\providecommand{\HDECAY} {\textsc{hdecay}\xspace}
\providecommand{\HERWIG} {{\textsc{herwig}}\xspace}
\providecommand{\HIGLU} {{\textsc{higlu}}\xspace}
\providecommand{\HIJING} {{\textsc{hijing}}\xspace}
\providecommand{\IGUANA} {\textsc{iguana}\xspace}
\providecommand{\ISAJET} {{\textsc{isajet}}\xspace}
\providecommand{\ISAPYTHIA} {{\textsc{isapythia}}\xspace}
\providecommand{\ISASUGRA} {{\textsc{isasugra}}\xspace}
\providecommand{\ISASUSY} {{\textsc{isasusy}}\xspace}
\providecommand{\ISAWIG} {{\textsc{isawig}}\xspace}
\providecommand{\MADGRAPH} {\textsc{MadGraph}\xspace}
\providecommand{\MCATNLO} {\textsc{mc@nlo}\xspace}
\providecommand{\MCFM} {\textsc{mcfm}\xspace}
\providecommand{\MILLEPEDE} {{\textsc{millepede}}\xspace}
\providecommand{\ORCA} {{\textsc{orca}}\xspace}
\providecommand{\OSCAR} {{\textsc{oscar}}\xspace}
\providecommand{\PHOTOS} {\textsc{photos}\xspace}
\providecommand{\PROSPINO} {\textsc{prospino}\xspace}
\providecommand{\PYTHIA} {{\textsc{pythia}}\xspace}
\providecommand{\SHERPA} {{\textsc{sherpa}}\xspace}
\providecommand{\TAUOLA} {\textsc{tauola}\xspace}
\providecommand{\TOPREX} {\textsc{TopReX}\xspace}
\providecommand{\XDAQ} {{\textsc{xdaq}}\xspace}

\newcommand {\DZERO}{D\O\xspace}     


\newcommand{\de}{\ensuremath{^\circ}}
\newcommand{\ten}[1]{\ensuremath{\times \text{10}^\text{#1}}}
\newcommand{\unit}[1]{\ensuremath{\text{\,#1}}\xspace}
\newcommand{\mum}{\ensuremath{\,\mu\text{m}}\xspace}
\newcommand{\micron}{\ensuremath{\,\mu\text{m}}\xspace}
\newcommand{\cm}{\ensuremath{\,\text{cm}}\xspace}
\newcommand{\mm}{\ensuremath{\,\text{mm}}\xspace}
\newcommand{\mus}{\ensuremath{\,\mu\text{s}}\xspace}
\newcommand{\keV}{\ensuremath{\,\text{ke\hspace{-.08em}V}}\xspace}
\newcommand{\MeV}{\ensuremath{\,\text{Me\hspace{-.08em}V}}\xspace}
\newcommand{\GeV}{\ensuremath{\,\text{Ge\hspace{-.08em}V}}\xspace}
\newcommand{\TeV}{\ensuremath{\,\text{Te\hspace{-.08em}V}}\xspace}
\newcommand{\PeV}{\ensuremath{\,\text{Pe\hspace{-.08em}V}}\xspace}
\newcommand{\keVc}{\ensuremath{{\,\text{ke\hspace{-.08em}V\hspace{-0.16em}/\hspace{-0.08em}c}}}\xspace}
\newcommand{\MeVc}{\ensuremath{{\,\text{Me\hspace{-.08em}V\hspace{-0.16em}/\hspace{-0.08em}c}}}\xspace}
\newcommand{\GeVc}{\ensuremath{{\,\text{Ge\hspace{-.08em}V\hspace{-0.16em}/\hspace{-0.08em}c}}}\xspace}
\newcommand{\TeVc}{\ensuremath{{\,\text{Te\hspace{-.08em}V\hspace{-0.16em}/\hspace{-0.08em}c}}}\xspace}
\newcommand{\keVcc}{\ensuremath{{\,\text{ke\hspace{-.08em}V\hspace{-0.16em}/\hspace{-0.08em}c}^\text{2}}}\xspace}
\newcommand{\MeVcc}{\ensuremath{{\,\text{Me\hspace{-.08em}V\hspace{-0.16em}/\hspace{-0.08em}c}^\text{2}}}\xspace}
\newcommand{\GeVcc}{\ensuremath{{\,\text{Ge\hspace{-.08em}V\hspace{-0.16em}/\hspace{-0.08em}c}^\text{2}}}\xspace}
\newcommand{\TeVcc}{\ensuremath{{\,\text{Te\hspace{-.08em}V\hspace{-0.16em}/\hspace{-0.08em}c}^\text{2}}}\xspace}

\newcommand{\pbinv} {\mbox{\ensuremath{\,\text{pb}^\text{$-$1}}}\xspace}
\newcommand{\fbinv} {\mbox{\ensuremath{\,\text{fb}^\text{$-$1}}}\xspace}
\newcommand{\nbinv} {\mbox{\ensuremath{\,\text{nb}^\text{$-$1}}}\xspace}
\newcommand{\percms}{\ensuremath{\,\text{cm}^\text{$-$2}\,\text{s}^\text{$-$1}}\xspace}
\newcommand{\lumi}{\ensuremath{\mathcal{L}}\xspace}
\newcommand{\Lumi}{\ensuremath{\mathcal{L}}\xspace}
%
\newcommand{\LvLow}  {\ensuremath{\mathcal{L}=\text{10}^\text{32}\,\text{cm}^\text{$-$2}\,\text{s}^\text{$-$1}}\xspace}
\newcommand{\LLow}   {\ensuremath{\mathcal{L}=\text{10}^\text{33}\,\text{cm}^\text{$-$2}\,\text{s}^\text{$-$1}}\xspace}
\newcommand{\lowlumi}{\ensuremath{\mathcal{L}=\text{2}\times \text{10}^\text{33}\,\text{cm}^\text{$-$2}\,\text{s}^\text{$-$1}}\xspace}
\newcommand{\LMed}   {\ensuremath{\mathcal{L}=\text{2}\times \text{10}^\text{33}\,\text{cm}^\text{$-$2}\,\text{s}^\text{$-$1}}\xspace}
\newcommand{\LHigh}  {\ensuremath{\mathcal{L}=\text{10}^\text{34}\,\text{cm}^\text{$-$2}\,\text{s}^\text{$-$1}}\xspace}
\newcommand{\hilumi} {\ensuremath{\mathcal{L}=\text{10}^\text{34}\,\text{cm}^\text{$-$2}\,\text{s}^\text{$-$1}}\xspace}


\newcommand{\zp}{\ensuremath{\mathrm{Z}^\prime}\xspace}


\newcommand{\kt}{\ensuremath{k_{\mathrm{T}}}\xspace}
\newcommand{\BC}{\ensuremath{{B_{\mathrm{c}}}}\xspace}
\newcommand{\bbarc}{\ensuremath{{\overline{b}c}}\xspace}
\newcommand{\bbbar}{\ensuremath{{b\overline{b}}}\xspace}
\newcommand{\ccbar}{\ensuremath{{c\overline{c}}}\xspace}
\newcommand{\JPsi}{\ensuremath{{J}/\psi}\xspace}
\newcommand{\bspsiphi}{\ensuremath{B_s \to \JPsi\, \phi}\xspace}
\newcommand{\AFB}{\ensuremath{A_\mathrm{FB}}\xspace}
\newcommand{\EE}{\ensuremath{e^+e^-}\xspace}
\newcommand{\MM}{\ensuremath{\mu^+\mu^-}\xspace}
\newcommand{\TT}{\ensuremath{\tau^+\tau^-}\xspace}
\newcommand{\wangle}{\ensuremath{\sin^{2}\theta_{\mathrm{eff}}^\mathrm{lept}(M^2_\mathrm{Z})}\xspace}
\newcommand{\ttbar}{\ensuremath{{t\overline{t}}}\xspace}
\newcommand{\stat}{\ensuremath{\,\text{(stat.)}}\xspace}
\newcommand{\syst}{\ensuremath{\,\text{(syst.)}}\xspace}

\newcommand{\HGG}{\ensuremath{\mathrm{H}\to\gamma\gamma}}
\newcommand{\gev}{\GeV}
\newcommand{\GAMJET}{\ensuremath{\gamma + \mathrm{jet}}}
\newcommand{\PPTOJETS}{\ensuremath{\mathrm{pp}\to\mathrm{jets}}}
\newcommand{\PPTOGG}{\ensuremath{\mathrm{pp}\to\gamma\gamma}}
\newcommand{\PPTOGAMJET}{\ensuremath{\mathrm{pp}\to\gamma +
\mathrm{jet}
}}
\newcommand{\MH}{\ensuremath{\mathrm{M_{\mathrm{H}}}}}
\newcommand{\RNINE}{\ensuremath{\mathrm{R}_\mathrm{9}}}
\newcommand{\DR}{\ensuremath{\Delta\mathrm{R}}}


\newcommand{\PT}{\ensuremath{p_{\mathrm{T}}}\xspace}
\newcommand{\pt}{\ensuremath{p_{\mathrm{T}}}\xspace}
\newcommand{\ET}{\ensuremath{E_{\mathrm{T}}}\xspace}
\newcommand{\HT}{\ensuremath{H_{\mathrm{T}}}\xspace}
\newcommand{\et}{\ensuremath{E_{\mathrm{T}}}\xspace}
\newcommand{\Em}{\ensuremath{E\!\!\!/}\xspace}
\newcommand{\Pm}{\ensuremath{p\!\!\!/}\xspace}
\newcommand{\PTm}{\ensuremath{{p\!\!\!/}_{\mathrm{T}}}\xspace}
\newcommand{\ETm}{\ensuremath{E_{\mathrm{T}}^{\mathrm{miss}}}\xspace}
\newcommand{\MET}{\ensuremath{E_{\mathrm{T}}^{\mathrm{miss}}}\xspace}
\newcommand{\ETmiss}{\ensuremath{E_{\mathrm{T}}^{\mathrm{miss}}}\xspace}
\newcommand{\VEtmiss}{\ensuremath{{\vec E}_{\mathrm{T}}^{\mathrm{miss}}}\xspace}

%

\newcommand{\ga}{\ensuremath{\gtrsim}}
\newcommand{\la}{\ensuremath{\lesssim}}
\newcommand{\swsq}{\ensuremath{\sin^2\theta_W}\xspace}
\newcommand{\cwsq}{\ensuremath{\cos^2\theta_W}\xspace}
\newcommand{\tanb}{\ensuremath{\tan\beta}\xspace}
\newcommand{\tanbsq}{\ensuremath{\tan^{2}\beta}\xspace}
\newcommand{\sidb}{\ensuremath{\sin 2\beta}\xspace}
\newcommand{\alpS}{\ensuremath{\alpha_S}\xspace}
\newcommand{\alpt}{\ensuremath{\tilde{\alpha}}\xspace}

\newcommand{\QL}{\ensuremath{Q_L}\xspace}
\newcommand{\sQ}{\ensuremath{\tilde{Q}}\xspace}
\newcommand{\sQL}{\ensuremath{\tilde{Q}_L}\xspace}
\newcommand{\ULC}{\ensuremath{U_L^C}\xspace}
\newcommand{\sUC}{\ensuremath{\tilde{U}^C}\xspace}
\newcommand{\sULC}{\ensuremath{\tilde{U}_L^C}\xspace}
\newcommand{\DLC}{\ensuremath{D_L^C}\xspace}
\newcommand{\sDC}{\ensuremath{\tilde{D}^C}\xspace}
\newcommand{\sDLC}{\ensuremath{\tilde{D}_L^C}\xspace}
\newcommand{\LL}{\ensuremath{L_L}\xspace}
\newcommand{\sL}{\ensuremath{\tilde{L}}\xspace}
\newcommand{\sLL}{\ensuremath{\tilde{L}_L}\xspace}
\newcommand{\ELC}{\ensuremath{E_L^C}\xspace}
\newcommand{\sEC}{\ensuremath{\tilde{E}^C}\xspace}
\newcommand{\sELC}{\ensuremath{\tilde{E}_L^C}\xspace}
\newcommand{\sEL}{\ensuremath{\tilde{E}_L}\xspace}
\newcommand{\sER}{\ensuremath{\tilde{E}_R}\xspace}
\newcommand{\sFer}{\ensuremath{\tilde{f}}\xspace}
\newcommand{\sQua}{\ensuremath{\tilde{q}}\xspace}
\newcommand{\sUp}{\ensuremath{\tilde{u}}\xspace}
\newcommand{\suL}{\ensuremath{\tilde{u}_L}\xspace}
\newcommand{\suR}{\ensuremath{\tilde{u}_R}\xspace}
\newcommand{\sDw}{\ensuremath{\tilde{d}}\xspace}
\newcommand{\sdL}{\ensuremath{\tilde{d}_L}\xspace}
\newcommand{\sdR}{\ensuremath{\tilde{d}_R}\xspace}
\newcommand{\sTop}{\ensuremath{\tilde{t}}\xspace}
\newcommand{\stL}{\ensuremath{\tilde{t}_L}\xspace}
\newcommand{\stR}{\ensuremath{\tilde{t}_R}\xspace}
\newcommand{\stone}{\ensuremath{\tilde{t}_1}\xspace}
\newcommand{\sttwo}{\ensuremath{\tilde{t}_2}\xspace}
\newcommand{\sBot}{\ensuremath{\tilde{b}}\xspace}
\newcommand{\sbL}{\ensuremath{\tilde{b}_L}\xspace}
\newcommand{\sbR}{\ensuremath{\tilde{b}_R}\xspace}
\newcommand{\sbone}{\ensuremath{\tilde{b}_1}\xspace}
\newcommand{\sbtwo}{\ensuremath{\tilde{b}_2}\xspace}
\newcommand{\sLep}{\ensuremath{\tilde{l}}\xspace}
\newcommand{\sLepC}{\ensuremath{\tilde{l}^C}\xspace}
\newcommand{\sEl}{\ensuremath{\tilde{e}}\xspace}
\newcommand{\sElC}{\ensuremath{\tilde{e}^C}\xspace}
\newcommand{\seL}{\ensuremath{\tilde{e}_L}\xspace}
\newcommand{\seR}{\ensuremath{\tilde{e}_R}\xspace}
\newcommand{\snL}{\ensuremath{\tilde{\nu}_L}\xspace}
\newcommand{\sMu}{\ensuremath{\tilde{\mu}}\xspace}
\newcommand{\sNu}{\ensuremath{\tilde{\nu}}\xspace}
\newcommand{\sTau}{\ensuremath{\tilde{\tau}}\xspace}
\newcommand{\Glu}{\ensuremath{g}\xspace}
\newcommand{\sGlu}{\ensuremath{\tilde{g}}\xspace}
\newcommand{\Wpm}{\ensuremath{W^{\pm}}\xspace}
\newcommand{\sWpm}{\ensuremath{\tilde{W}^{\pm}}\xspace}
\newcommand{\Wz}{\ensuremath{W^{0}}\xspace}
\newcommand{\sWz}{\ensuremath{\tilde{W}^{0}}\xspace}
\newcommand{\sWino}{\ensuremath{\tilde{W}}\xspace}
\newcommand{\Bz}{\ensuremath{B^{0}}\xspace}
\newcommand{\sBz}{\ensuremath{\tilde{B}^{0}}\xspace}
\newcommand{\sBino}{\ensuremath{\tilde{B}}\xspace}
\newcommand{\Zz}{\ensuremath{Z^{0}}\xspace}
\newcommand{\sZino}{\ensuremath{\tilde{Z}^{0}}\xspace}
\newcommand{\sGam}{\ensuremath{\tilde{\gamma}}\xspace}
\newcommand{\chiz}{\ensuremath{\tilde{\chi}^{0}}\xspace}
\newcommand{\chip}{\ensuremath{\tilde{\chi}^{+}}\xspace}
\newcommand{\chim}{\ensuremath{\tilde{\chi}^{-}}\xspace}
\newcommand{\chipm}{\ensuremath{\tilde{\chi}^{\pm}}\xspace}
\newcommand{\Hone}{\ensuremath{H_{d}}\xspace}
\newcommand{\sHone}{\ensuremath{\tilde{H}_{d}}\xspace}
\newcommand{\Htwo}{\ensuremath{H_{u}}\xspace}
\newcommand{\sHtwo}{\ensuremath{\tilde{H}_{u}}\xspace}
\newcommand{\sHig}{\ensuremath{\tilde{H}}\xspace}
\newcommand{\sHa}{\ensuremath{\tilde{H}_{a}}\xspace}
\newcommand{\sHb}{\ensuremath{\tilde{H}_{b}}\xspace}
\newcommand{\sHpm}{\ensuremath{\tilde{H}^{\pm}}\xspace}
\newcommand{\hz}{\ensuremath{h^{0}}\xspace}
\newcommand{\Hz}{\ensuremath{H^{0}}\xspace}
\newcommand{\Az}{\ensuremath{A^{0}}\xspace}
\newcommand{\Hpm}{\ensuremath{H^{\pm}}\xspace}
\newcommand{\sGra}{\ensuremath{\tilde{G}}\xspace}
\newcommand{\mtil}{\ensuremath{\tilde{m}}\xspace}
\newcommand{\rpv}{\ensuremath{\rlap{\kern.2em/}R}\xspace}
\newcommand{\LLE}{\ensuremath{LL\bar{E}}\xspace}
\newcommand{\LQD}{\ensuremath{LQ\bar{D}}\xspace}
\newcommand{\UDD}{\ensuremath{\overline{UDD}}\xspace}
\newcommand{\Lam}{\ensuremath{\lambda}\xspace}
\newcommand{\Lamp}{\ensuremath{\lambda'}\xspace}
\newcommand{\Lampp}{\ensuremath{\lambda''}\xspace}
\newcommand{\spinbd}[2]{\ensuremath{\bar{#1}_{\dot{#2}}}\xspace}

\newcommand{\MD}{\ensuremath{{M_\mathrm{D}}}\xspace}
\newcommand{\Mpl}{\ensuremath{{M_\mathrm{Pl}}}\xspace}
\newcommand{\Rinv} {\ensuremath{{R}^{-1}}\xspace}

%
%
\hyphenation{en-viron-men-tal}

\newcommand{\fix}[1]{{\bf <<< #1 !!! }}%
\newcommand{\invpt}{\ensuremath{1/\pt}}
\cmsNoteHeader{09-014}
\title{Performance of CMS Muon Reconstruction\\ in Cosmic-Ray Events}
\address[cern]{CERN}
\author[cern]{A. Cern Person}

\date{\today}

\abstract{
The performance of muon reconstruction in CMS is evaluated using
a large data sample of cosmic-ray muons recorded in 2008.
Efficiencies of various
high-level trigger, identification, and reconstruction algorithms have
been measured for a broad range of muon momenta, and were found to be
in good agreement with expectations from Monte Carlo simulation.
The relative momentum resolution for muons crossing the barrel part of
the detector is better than 1\% at 10~GeV/$c$ and is about 8\% at
500~GeV/$c$, the latter being only a factor of two worse than
expected with ideal alignment conditions.  Muon charge misassignment
ranges from less than 0.01\% at 10~GeV/$c$ to about 1\% at 500~GeV/$c$.}

\hypersetup{%
pdfauthor={R. Bellan, V. Valuev},%
pdftitle={Performance of CMS Muon Reconstruction in Cosmic-Ray Events},%
pdfsubject={CMS},%
pdfkeywords={CMS, CRAFT, muon, muon system, muon reconstruction}}

\maketitle 

\section{Introduction}
The primary goal of the Compact Muon Solenoid (CMS)
experiment~\cite{:2008zzk} is to explore particle physics at the TeV
energy scale, by exploiting the proton-proton collisions delivered by the
Large Hadron Collider (LHC)~\cite{Evans:2008zzb}.
During October-November 2008, the CMS Collaboration conducted a month-long
data-taking exercise known as the Cosmic Run At Four Tesla
(CRAFT).  The main goals of CRAFT were to test \textit{in situ}
the CMS magnet at the nominal current and to commission the
experiment for extended operation~\cite{CRAFTGeneral}.
These goals
were met successfully, and a dataset of 270 million cosmic-ray
events was recorded with the solenoid at an axial field
of 3.8~T and with all installed detector systems participating in
taking data.

This paper describes studies of the muon reconstruction
performance carried out using a data\-set of muons collected during CRAFT.
The muon system performs three main tasks:
triggering on muons, identifying muons, and assisting the CMS tracker
in measuring the momentum and charge of high-\pt muons.
Four stations of muon detectors are
embedded in the steel yoke of the superconducting solenoid, covering
the pseudorapidity region $|\eta| <$ 2.4.  Each station consists of
several layers of drift tubes (DT) in the region $|\eta| <$ 1.2
and cathode strip chambers (CSC) in the $|\eta|$ interval between
0.9 and 2.4.  The region $|\eta| <$ 0.8 is covered by the DT chambers
in all 4 stations and is referred to as the barrel region.
In the region $|\eta| <$ 1.6, the
DT and CSC subsystems are complemented by resistive plate chambers (RPC).
A detailed description of the CMS muon system can be found
elsewhere~\cite{:2008zzk}.

The CMS experiment has a two-level trigger system consisting of the
hardware-based \Lone Trigger and the software-based High-Level Trigger
(HLT).  During CRAFT, the \Lone trigger was configured to yield
maximum efficiency for cosmic muons~\cite{CMS_CFT_09_013}, while the HLT
was primarily in "pass-through" mode, transferring to storage the
events accepted by the \Lone trigger without additional processing or
selection~\cite{CMS_CFT_09_020}.

Studies of the performance of the individual muon subsystems carried
out on CRAFT data, as well as of the \Lone muon trigger, are
described in Refs.~\cite{CMS_CFT_09_010, CMS_CFT_09_011, CMS_CFT_09_012,
CMS_CFT_09_013, CMS_CFT_09_022}.  This paper focuses on the results related
to the tasks of muon identification and
reconstruction for high-level trigger and physics analysis.  Initial
event selection and
Monte Carlo (MC) simulation are summarized in Section~\ref{sec:data_sim}.
Section~\ref{sec:algos} describes muon reconstruction and identification
algorithms.  Data-simulation comparisons of distributions
of the main kinematic variables of cosmic muons and of several other basic
quantities
are shown in Section~\ref{sec:data-sim}.
Section~\ref{sec:eff} summarizes results obtained for
muon reconstruction and identification efficiencies.
Section~\ref{sec:pres} describes studies of muon momentum and
angular resolutions.  Section~\ref{sec:chargeid} is dedicated to the
measurement of the muon charge.  Performance of the muon high-level
trigger is discussed in Section~\ref{sec:hlt}.

\section{Cosmic Muons in CRAFT: Data Selection and Event Simulation}
\label{sec:data_sim}

In order to compare CRAFT data with the predictions obtained from
simulations, a large sample of cosmic-muon events was generated using
the CMSCGEN package~\cite{CMS_Note_2007-024, Biallass:2009ev}.
This generator includes a detailed description of the CMS access
shafts and other surroundings (composition of material above
the cavern, geometry of the service cavern, etc.) and reproduces
correctly angular and momentum distributions of cosmic-ray muons in
the CMS cavern (see Section~\ref{sec:data-sim-kine}).
The detector response to the cosmic muons was simulated with the
standard CMS software based on the GEANT4 package~\cite{Agostinelli:2002hh}.

The reconstruction of both CRAFT data and simulated cosmic-muon events was
performed with the CMS full-reconstruction CMSSW package~\cite{PTDR1}.
Modifications of the standard reconstruction code necessary for
efficient reconstruction of cosmic-muon events are summarized in
Ref.~\cite{CMS_CFT_09_007}.
Tracker and muon reconstruction algorithms used in the processing
of the samples are described in the next section.
An improved 
description of the magnetic field in the
muon system~\cite{CMS_CFT_09_015}, as well as a CRAFT-based
alignment for the silicon tracker~\cite{CMS_CFT_09_003} and for the
muon chambers~\cite{CMS_CFT_09_016} were used as input for the reconstruction.

To facilitate studies performed by various analysis groups, several
subsets of CRAFT data and corresponding Monte Carlo events were produced
according to different selection criteria~\cite{CMS_CFT_09_007}.
Unless explicitly stated, the analyses described in this paper used the
``super-pointing'' dataset designed to contain muons similar to those
expected from collisions at the LHC.  It was formed by requiring that
there be at least one track reconstructed in the tracker or in the
muon system that crosses a cylindrical volume of 10~cm radius and
100~cm length along the beam line centered at the nominal interaction
point, roughly corresponding to the outer boundaries of the pixel
detector.  The total number of events in good-quality runs in
this dataset was about 500\,000.
In a few cases mentioned below, the ``tracker-pointing'' dataset
formed by less stringent selection criteria was used instead: it contained
events with any track reconstructed in the tracker, or at
least one track in the muon system crossing the outer boundaries of
the tracker barrel, which is approximated by a cylinder of 90~cm radius
and 260~cm length.  The total number of these events was about 8 million.
Both the super-pointing and the tracker-pointing datasets
predominantly consist of muons fully contained in the DT region: the
fraction of muons with at least one CSC hit is of the order of 7\%.

\section{Muon Reconstruction and Identification Algorithms in CRAFT}
\label{sec:algos}
Using the data from CRAFT, various muon reconstruction and
identification algorithms were studied.  In addition to the standard
algorithms designed for muons produced at the LHC, dedicated
cosmic-muon reconstruction algorithms were provided, optimized for
muons not pointing at the nominal beam-interaction region.
While the standard reconstruction algorithms typically yield two tracks
(``2 legs'') for a single cosmic muon, one in each of the top and bottom
halves of the detector, the dedicated cosmic-muon algorithms are also
capable of fitting tracks traversing the whole detector (``1 leg'').

This section gives
an overview of the different algorithms.  Depending on the information
used, they can roughly be divided into three groups: standalone muon fits
using only information from the muon system (Section~\ref{sec:sta}),
global muon reconstruction algorithms based on a combined fit to selected
hits in the muon system and the silicon tracker (Sections~\ref{sec:glb} and
\ref{sec:tevmus}), and muon identification algorithms checking whether tracks
reconstructed in the silicon tracker have signatures compatible with that of
a muon in the calorimeters and in the muon system
(Section~\ref{sec:trackermus}).

\subsection{Standalone muon reconstruction} \label{sec:sta}
In the offline reconstruction, track segments are built within individual
muon chambers using a linear fit to the positions of the hits reconstructed in
each of the 8-12 (in case of DT) or 6 (in case of CSC) layers of the chamber.
These segments are used to generate ``seeds'' consisting of position and
direction vectors and an initial estimate of the muon transverse
momentum. The seeds serve as starting points for the track fits in the
muon system, which are performed using segments and hits from DTs,
CSCs, and RPCs, and are based on the Kalman-filter
technique~\cite{Fruhwirth:1987fm}.  To improve the momentum resolution, a
beam-spot constraint is applied in the fit for beam collision data.
The result is a collection of
tracks reconstructed using only information from the muon system,
which are referred to as ``standalone muons''.

The standard reconstruction algorithm for standalone muons is
described in Section~9.1.1 of Ref.~\cite{PTDR1}.  Since it was
developed to reconstruct muons produced at the LHC, it makes
use of the fact that such muons are produced at or close to the
nominal interaction point and travel from the center of the detector
to its periphery.  Therefore, a number of modifications to the
standard algorithm were necessary to reconstruct efficiently muons
coming from outside, in particular those traversing
the detector far from its center.  A detailed description of the modifications
implemented at various stages of the standalone muon reconstruction
(in seeding, navigation, and trajectory building) to build the dedicated
cosmic-muon reconstruction algorithm can be found in Ref.~\cite{Liu:2007zz}.

Since the standard algorithm is still applicable to a subset of cosmic
muons crossing the detector close to the nominal interaction point,
both dedicated cosmic-muon (CosmicSTA) and beam collision (ppSTA) algorithms
were used to reconstruct standalone muons in CRAFT.  The CosmicSTA
algorithm could also be configured to attempt to combine
muon hits in both detector hemispheres into one single standalone-muon
track spanning the whole detector and revert to single-hemisphere
CosmicSTA tracks should such an attempt be unsuccessful; this algorithm
is referred to as 1-leg CosmicSTA.

\subsection{Global muon reconstruction} \label{sec:glb}
At values of the transverse momentum \pt below about 200~GeV/$c$,
the momentum resolution for a muon track is driven by measurements
in the silicon tracker.  As a particle's momentum increases and
the curvature of its corresponding track decreases, however, momentum
resolution in the tracker becomes limited by position measurement
resolution (including misalignment effects).  One can then benefit
from the large lever arm and 3.8~T magnetic field in the region between
the tracker and the muon system by including hits in the muon chambers.

Each standalone-muon track is compared with the tracks
reconstructed in the tracker (referred to as
``tracker tracks''), and the best-matching tracker track is
selected. For each ``tracker track'' -- ``standalone muon'' pair, the
track fit using all hits in both tracks is performed, again based on the
Kalman-filter technique and taking into account the average expected
energy losses, the magnetic field, and multiple scattering in the
detector materials.   The result is a collection of tracks referred
to as ``global muons''.  More details on the standard reconstruction of
global muons can be found in Section~9.1.2 of Ref.~\cite{PTDR1}.  As
in the case of the standalone muon reconstruction, some modifications
to the standard global muon algorithm were implemented for cosmic
muons, notably to enable reconstruction of tracks consisting of two
standalone-muon tracks at opposite sides of the detector and a single
track that traverses the entire tracker sandwiched between them.  These
modifications are described in Ref.~\cite{Liu:2007zz}.

Reconstruction of tracker tracks in CRAFT data was performed using two
algorithms, the standard Combinatorial Kalman Finder (CKF)
and the specialized Cosmic Track Finder (CosmicTF).  Both
algorithms are described in Refs.~\cite{CMS_CFT_09_002, Adam:2009di}.
The CKF algorithm can be used in its standard configuration intended
for proton-proton collisions (ppCKF), or re-configured specifically
for cosmic-muon events in two ways: to reconstruct muons as single
tracks (1-leg CosmicCKF) or as two separate tracks in the two
hemispheres of the detector (2-leg CosmicCKF).
The CosmicTF algorithm was designed to reconstruct cosmic muons
crossing the tracker as single tracks;
furthermore, tracks produced by the CosmicTF can be split at the point
of their closest approach to the nominal beam line (PCA) into two separate
track candidates, with each of the candidates refitted individually to
yield a pair of so-called ``split tracks''.

Given the assortment of algorithms available to reconstruct
standalone-muon and tracker tracks, various types of global muons
can be produced.  The following ones were used in the studies
described in this paper:
\begin{itemize}
\item \textit{LHC-like global muons}, formed from ppCKF tracker tracks
 and ppSTA standalone-muon tracks.  These are
 tracks found by the standard algorithm aimed at reconstructing muons
 produced in beam collisions at the LHC.  Only cosmic muons crossing
 the detector within a few centimeters of the nominal interaction point
 are selected to be reconstructed by this algorithm.
\item \textit{1-leg global muons}, formed from 1-leg CosmicCKF tracker tracks
 and CosmicSTA standalone-muon tracks.  These muons typically consist of
 a single track in the entire tracker sandwiched between two standalone-muon
 tracks, and yield the best estimate of the parameters of the muon.
\item \textit{split global muons}, each formed from a split tracker track
 and a CosmicSTA stand\-alone-muon track.  A cosmic muon traversing the
 core of the detector typically yields a pair of split global muons.
 Comparison of these tracks, fitted independently, provides a measure
 of muon reconstruction performance,
 while the splitting mechanism ensures that they indeed originate from
 the same muon.
\item \textit{2-leg global muons}, each formed from a 2-leg CosmicCKF tracker
 track and a CosmicSTA standalone-muon track.  Since the two 2-leg muon tracks
 typically found for each tracker-pointing cosmic muon are treated
 independently at all stages of reconstruction, they provide fully unbiased
 measurements of reconstruction performance, though care must
 be taken to ensure that they were produced by the same muon.
\end{itemize}

\begin{figure}[htb]
  \centering
  \vspace*{-1.5cm}
  \includegraphics[width=1.0\textwidth]{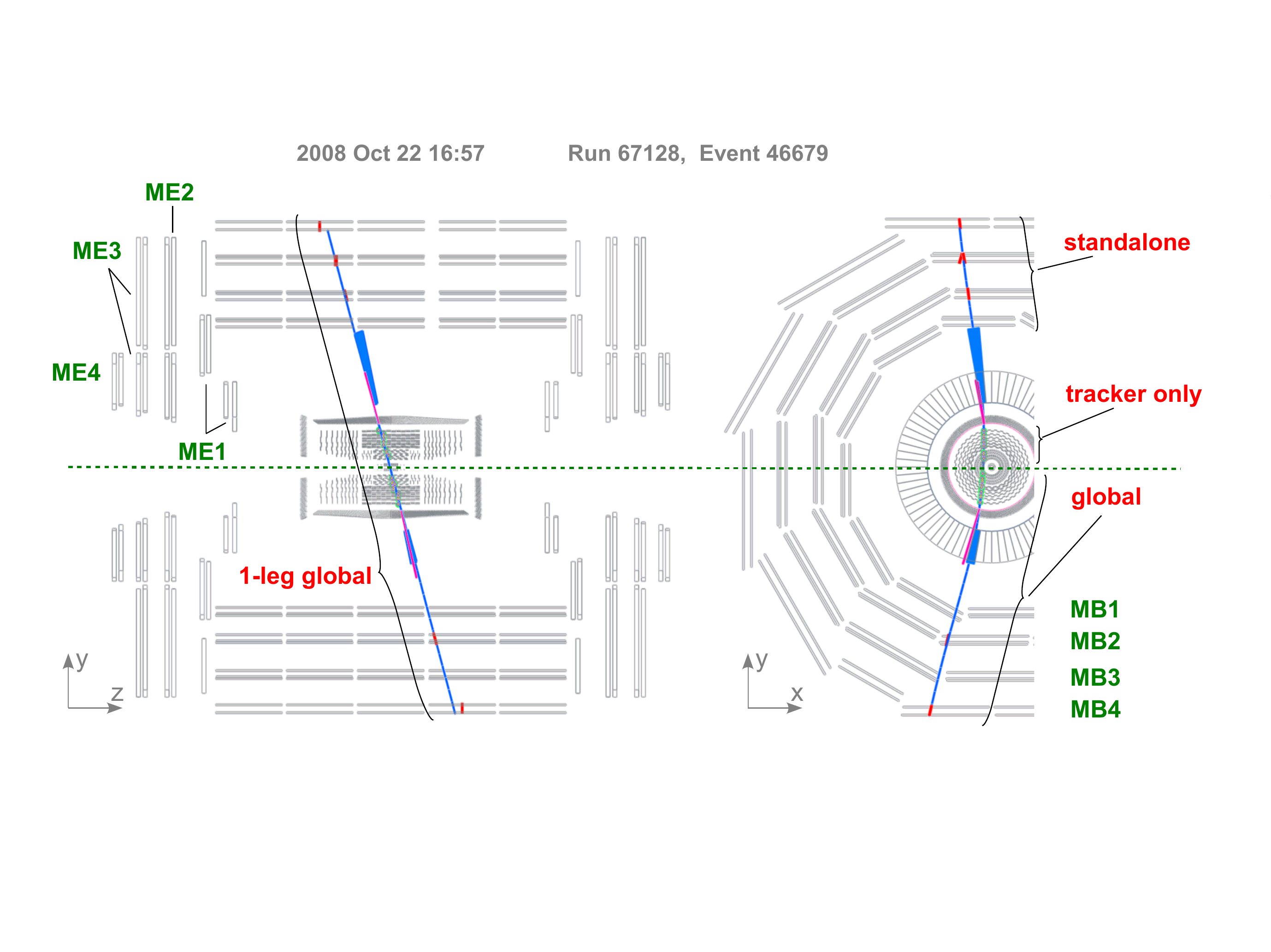}
  \vspace*{-2.9cm}
  \caption{Event display of a cosmic muon crossing CMS: the side
   view (left) and a part of the transverse view (right).  ``MB'' and
   ``ME'' labels indicate positions of the muon barrel and the muon
   endcap stations, respectively.  The solid blue curve represents a
   1-leg global muon reconstructed using silicon tracker and muon
   system hits in the whole detector.  Small green circles indicate hits
   in the silicon tracker.  Short red stubs correspond to fitted track
   segments in the muon system; as the $z$ position is not measured
   in the outer barrel station, the segments in it are drawn at the
   $z$ center of the wheel, with their directions perpendicular to the
   chamber.  Energy deposits in the electromagnetic
   and hadron calorimeters are shown as (thin) magenta and (thick)
   blue bars, respectively.}
  \label{fig:event_display}
\end{figure}

An example of an event display of a cosmic muon crossing CMS in
Fig.~\ref{fig:event_display} illustrates the main topological
differences between tracker tracks, standalone muons, 1-leg global
muons, and other types of global muons (LHC-like, split, and 2-leg).
The CMS coordinate system is right-handed, with the origin at the
nominal collision point, the $x$ axis pointing to the centre of the
LHC, the $y$ axis pointing up (perpendicular to the LHC plane), and
the $z$ axis along the anticlockwise-beam direction.  The pseudorapidity
$\eta$ is defined as $\eta = -$ln tan $(\theta/2)$, where cos
$\theta = p_z/p$, and has the same sign for both legs of a muon.
The radius $r$ is the distance from the $z$ axis; the azimuthal angle
$\phi$ is the angle from the positive $x$ axis measured in the $x$-$y$
plane.

\subsection{Muon reconstruction refinements} \label{sec:tevmus}
As the muon traverses the steel yoke, multiple scattering and radiative
processes can significantly degrade the precision with which the muon
trajectory is measured, so using all available muon system hits in the
track fit --- the approach chosen for global muons
--- is not always the best choice.  An alternative approach studied on CRAFT
data consists of refitting the global-muon track ignoring hits in all
muon stations except the innermost one containing hits.  This approach
is called the ``tracker plus the first muon station'' (TPFMS)
fit.  Another approach, dubbed the ``picky'' fit, imposes tight cuts
on the compatibility of hits with the muon trajectory in those muon stations
which appear to contain electromagnetic showers (i.e., contain a large
number of hits).

Momentum resolution for high-\pt muons can potentially be further
improved by choosing, on a track-by-track basis, between fits including
muon hits and the tracker-only fit, depending on
the fit output.  Three such approaches (``selectors'') were studied in CRAFT:
\begin{itemize}
\item The \textit{sigma-switch method}, in which one chooses the global-muon
 fit if the global and tracker-only fit results for the ratio of the
 charge $q$ to the momentum $p$ of a muon, $q/p$, are within
 2$\sigma_{q/p}$ of the tracker-only fit from each other, and if the \pt values
 found by both fits are above 200~GeV/$c$; one chooses the tracker-only fit
 otherwise.
\item The \textit{truncated muon reconstructor} (TMR), whereby one chooses
 between the TPFMS and tracker-only fits on a track-by-track basis, using
 goodness-of-fit variables for each fit.
\item \textit{Tune P}, which is similar to TMR, but includes the result
 of the ``picky'' fit in the selection.
\end{itemize}

\subsection{Muon identification} \label{sec:trackermus}
An approach complementary to global muon reconstruction,
referred to as muon identification, consists of considering all
tracker tracks to be potential muon candidates and 
checking this hypothesis by looking for compatible signatures in the
calorimeters and the muon system.

Tracker tracks for which at least one matched segment in the muon
system is found are called ``tracker muons''.
Tracker muons have a rather low purity and necessitate further
selection requirements before they can be considered viable muon
candidates.  Two sets of such requirements, compatibility-based and
cut-based, are currently defined:
\begin{itemize}
\item In the compatibility-based selection, two ``compatibility''
 variables are constructed, one based on calorimeter information and
 the other based on information from the muon system.  A tracker muon
 is considered to be a muon candidate if the value of a linear
 combination of these variables is larger than a pre-defined
 threshold.  Two versions of the selection, with a lower
 (\textit{CompatibilityLoose}) and a higher
 (\textit{CompatibilityTight}) threshold, are available.
\item In the cut-based selection, cuts are applied on the number of
 matched muon segments and on their proximity to the extrapolated
 position of the tracker track.  In the \textit{LastStation} method,
 one makes use of the fact that the penetration depth of muons is
 larger than that of hadrons by requiring that there be well-matched
 segments in at least two muon stations, one of them being in the outermost
 station.  Two versions of the LastStation method exist, with
 track-to-segment proximity cuts in only $x$ (\textit{LastStationLoose})
 or in both $x$ and $y$ (\textit{LastStationTight})
 projections.  In a less stringent \textit{OneStation} method, a
 well-matched segment can be located in any muon station.  Track-to-segment
 matching is performed in a local (chamber) coordinate system,
 where local $x$ is the best-measured coordinate (in the $r$-$\phi$ plane)
 and local $z$ is the coordinate perpendicular to the chamber and pointing
 towards the center of CMS.
\end{itemize}

More details on the muon identification algorithms can be found in
Section~9.2 of Ref.~\cite{PTDR1}.

\section{General Comparisons between Data and Simulation} \label{sec:data-sim}
The presentation of the results starts with some general
data-simulation comparisons.  Section~\ref{sec:data-sim-kine} shows
how well the Monte Carlo simulation reproduces the main kinematic
properties of cosmic muons.  Data-simulation comparisons for a few
basic distributions for standalone and global muons are shown in
Section~\ref{sec:data-sim-staglb}.  Section~\ref{sec:data-sim-tracker}
describes the results obtained for various building blocks of the muon
identification algorithm.

\subsection{Kinematic properties of cosmic muons} \label{sec:data-sim-kine}
Since the generation of cosmic muons with the CMSCGEN package does not
include the production of multiple muons (muon showers), only events
with a single muon crossing the
detector were selected for the comparisons below; furthermore, the
reconstructed momentum of the muon at the entry point to the CMS detector was
required to be larger than 10~GeV/$c$, the typical energy loss for a
cosmic muon traversing the entire detector.

\begin{figure}[htb!]
  \centering
  \includegraphics[width=0.5\textwidth]{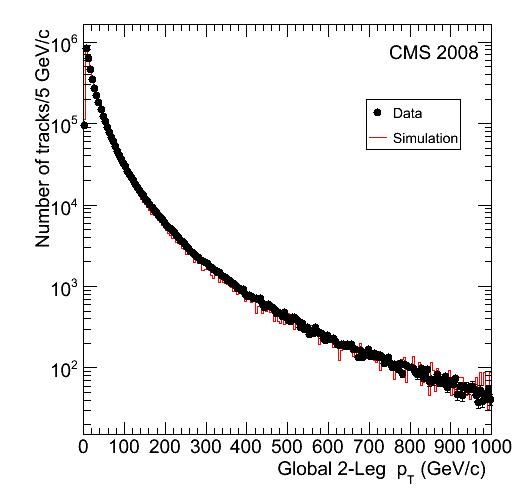}
  \put(-50,80){\bf\large a)}
  \includegraphics[width=0.5\textwidth]{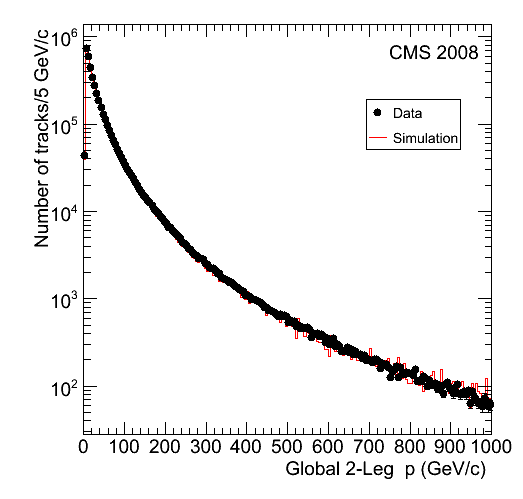}
  \put(-50,80){\bf\large b)}\\
  \includegraphics[width=0.5\textwidth]{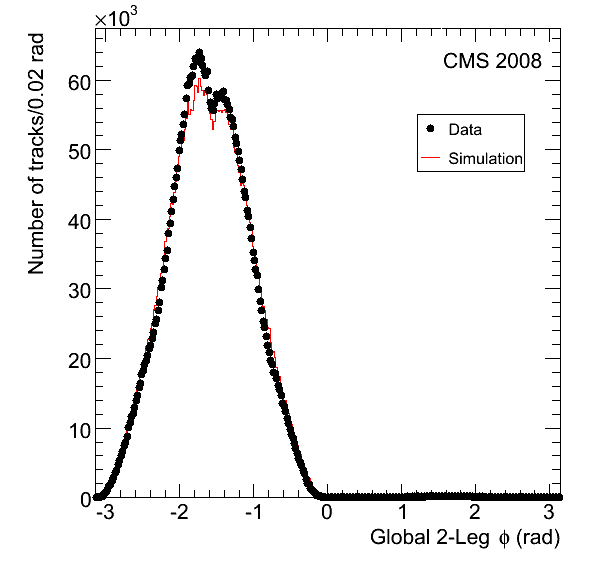}
  \put(-50,80){\bf\large c)}
  \includegraphics[width=0.5\textwidth]{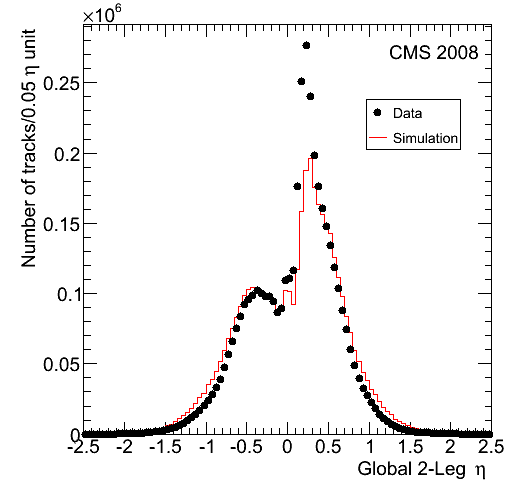}
  \put(-50,80){\bf\large d)}
  \caption{Distributions of a) the transverse momentum \pt, b) the
  momentum $p$, c) the azimuthal angle $\phi$, and d) the
  pseudorapidity $\eta$ of 2-leg global muons at the point of
  closest approach to the nominal beam line, for the data (points)
  and for the Monte Carlo simulation (histogram).  The MC
  distributions are normalized to the number of events in the data.}
  \label{fig:data_mc_kine}
\end{figure}

The comparison between the measured and the predicted distributions
of the transverse momenta of cosmic muons is shown in
Fig.~\ref{fig:data_mc_kine}a, for 2-leg global muons in the
tracker-pointing dataset.  The agreement is good up to very high
\pt values.  Figure~\ref{fig:data_mc_kine}b shows the momentum distributions
for cosmic muons in the same dataset.  Data
and Monte Carlo spectra agree to better than 10\% over 4 orders
of magnitude.  The distributions of the azimuthal angle
$\phi$ of the direction of the track at the point of its closest approach
to the nominal beam line are shown in Fig.~\ref{fig:data_mc_kine}c.
The double-peak structure is due to the superposition of the distributions
for negatively charged and positively charged muons bent in opposite
$\phi$ directions by the CMS magnetic field.  The Monte Carlo simulation
reproduces the shape of the distribution very well, only slightly
underestimating a fraction of near-vertical downward-going
($\phi \sim -\pi/2$) muons.
Finally, Fig.~\ref{fig:data_mc_kine}d shows the distributions of
the pseudorapidity $\eta$.  An excess of events at positive $\eta$
values is due to an extra contribution from muons reaching the core of the
detector through the main access shaft located at negative $z$.
There are more muons from the shaft in data than predicted by the MC
simulation: the simulation of cosmic muons used a simplified description
of the materials surrounding the CMS cavern, notably of a concrete plug
covering the shaft and
of the material between the surface and the detector cavern, leading to an
underestimated flux of muons from the shaft relative to the flux
of muons traversing the rock.
Overall, however, the Monte Carlo simulation reproduces the main kinematic
distributions of cosmic muons fairly well and therefore can be used
in the studies of the muon reconstruction performance described below.

\subsection{Basic distributions for standalone and global muons}
\label{sec:data-sim-staglb}
\begin{figure}[htb!]
  \centering
  \includegraphics[width=0.5\textwidth]{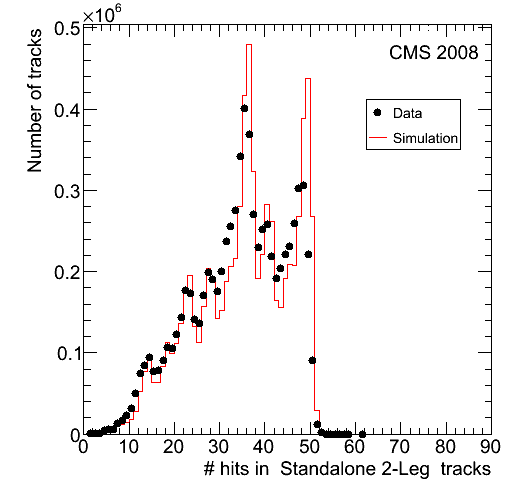}
  \put(-50,80){\bf\large a)}
  \includegraphics[width=0.5\textwidth]{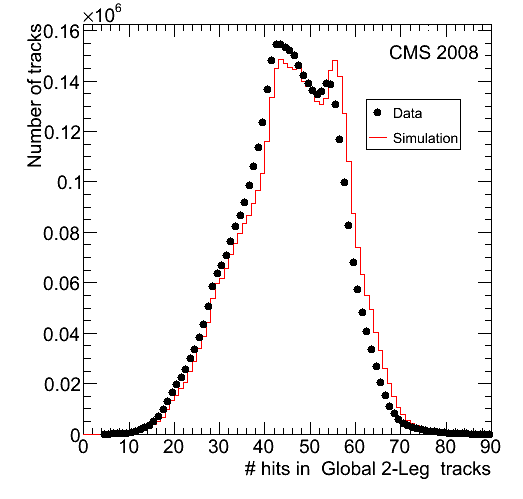}
  \put(-50,80){\bf\large b)}
  \caption{Distributions of the total number of hits per track for
  the data (points) and for the Monte Carlo simulation (histogram),
  for a) CosmicSTA standalone muons and b) 2-leg global
  muons.  The MC distributions are normalized to the number of
  events in the data.}
  \label{fig:data_mc_nhits}
\end{figure}

The same sample of events was used to examine a few other basic
distributions for standalone and global muons.
Figure~\ref{fig:data_mc_nhits} shows the comparison between the data
and the Monte Carlo simulation for the
total number of hits per track, for CosmicSTA standalone and 2-leg global
muons in the
tracker-pointing dataset.  The most prominent peaks correspond to the
number of hits in the tracks spanning 3 and 4 barrel muon stations.  The number
of hits per track in data is on average slightly lower than that in
the MC simulation.  These differences are discussed in
Refs.~\cite{CMS_CFT_09_012, CMS_CFT_09_002}; as shown below, their
impact on the performance of the muon reconstruction is very small,
thanks to the high redundancy of measurements in the tracker and the
muon system.

\begin{figure}[htb!]
  \centering
  \includegraphics[width=0.5\textwidth]{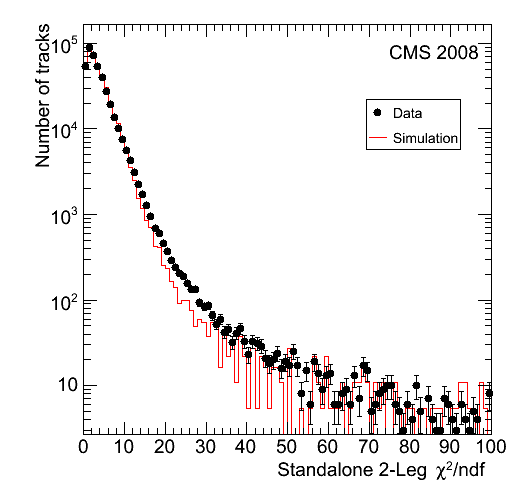}
  \put(-50,80){\bf\large a)}
  \includegraphics[width=0.5\textwidth]{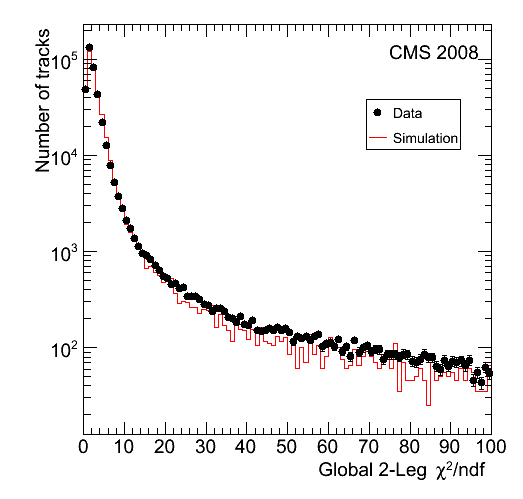}
  \put(-50,80){\bf\large b)}
  \caption{Distributions of the $\chi^2/\text{ndf}$ of the track
  fit for the data (points) and for the Monte Carlo simulation
  (histogram), for a) CosmicSTA standalone muons and b) 2-leg global
  muons.  The MC distributions are normalized to the number of
  events in the data.}
  \label{fig:data_mc_chi2}
\end{figure}

Figure~\ref{fig:data_mc_chi2} shows the comparison between the
data and the simulation for the $\chi^2/\text{ndf}$ of the
track fit for events in the super-pointing dataset.  The tails of
the distributions are larger in the data, indicating that the errors
are underestimated; this issue is discussed further in Section~\ref{sec:pres}.

\subsection{Basic distributions for tracker muons} \label{sec:data-sim-tracker}
A crucial step of the muon identification approach described in
Section~\ref{sec:trackermus} consists of propagating the tracker track to
the calorimeters and to the muon chambers, and associating the
propagated trajectory with the energy depositions and the muon
segments, respectively.  The propagation takes into account the
magnetic field, the average expected energy losses, and multiple
scattering in the detector materials.

\begin{figure}[thb!]
  \centering
  \includegraphics[width=0.5\textwidth]{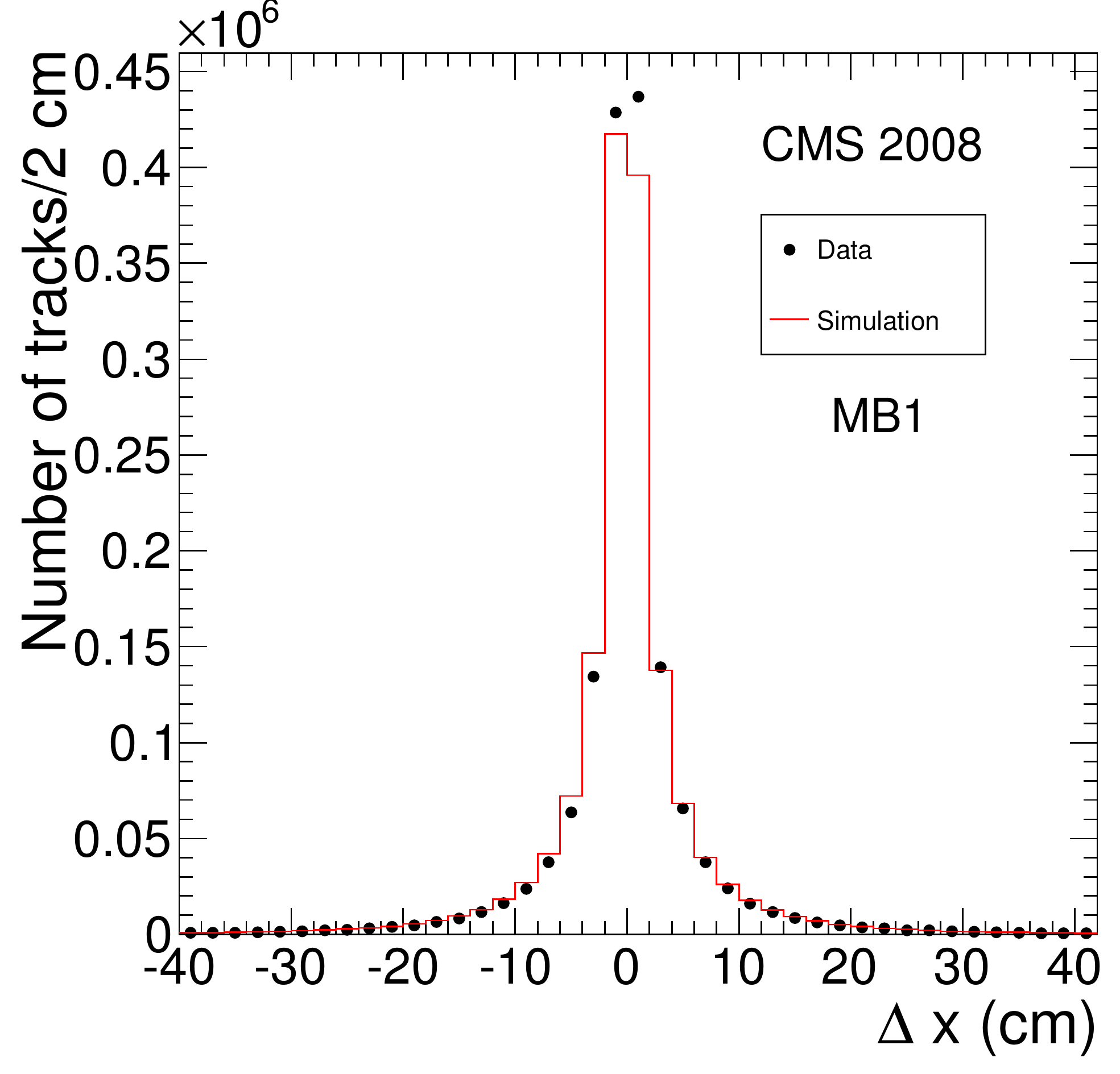}
  \put(-40,60){\bf\large a)}
  \includegraphics[width=0.5\textwidth]{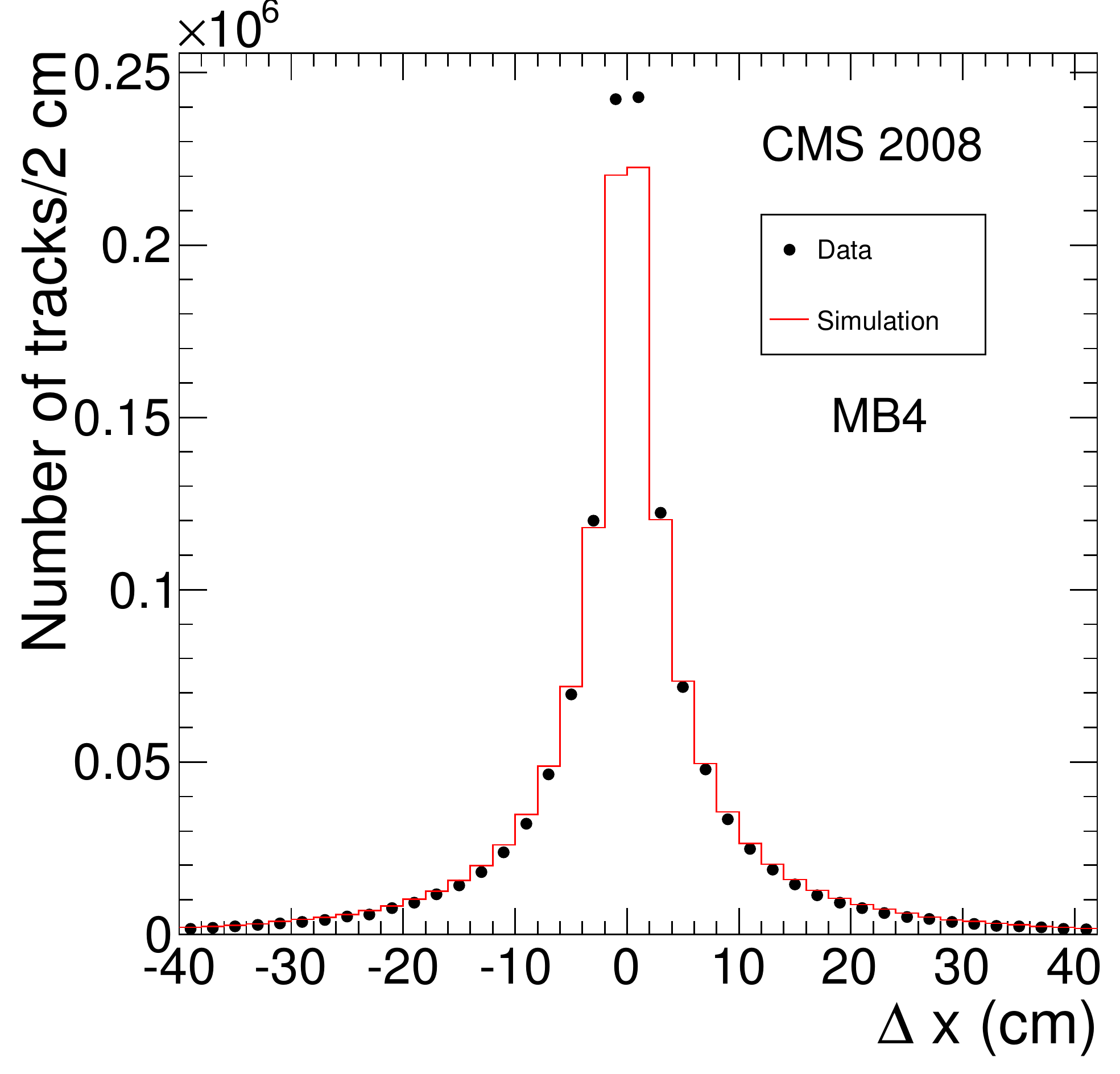}
  \put(-40,60){\bf\large b)}\\
  \includegraphics[width=0.5\textwidth]{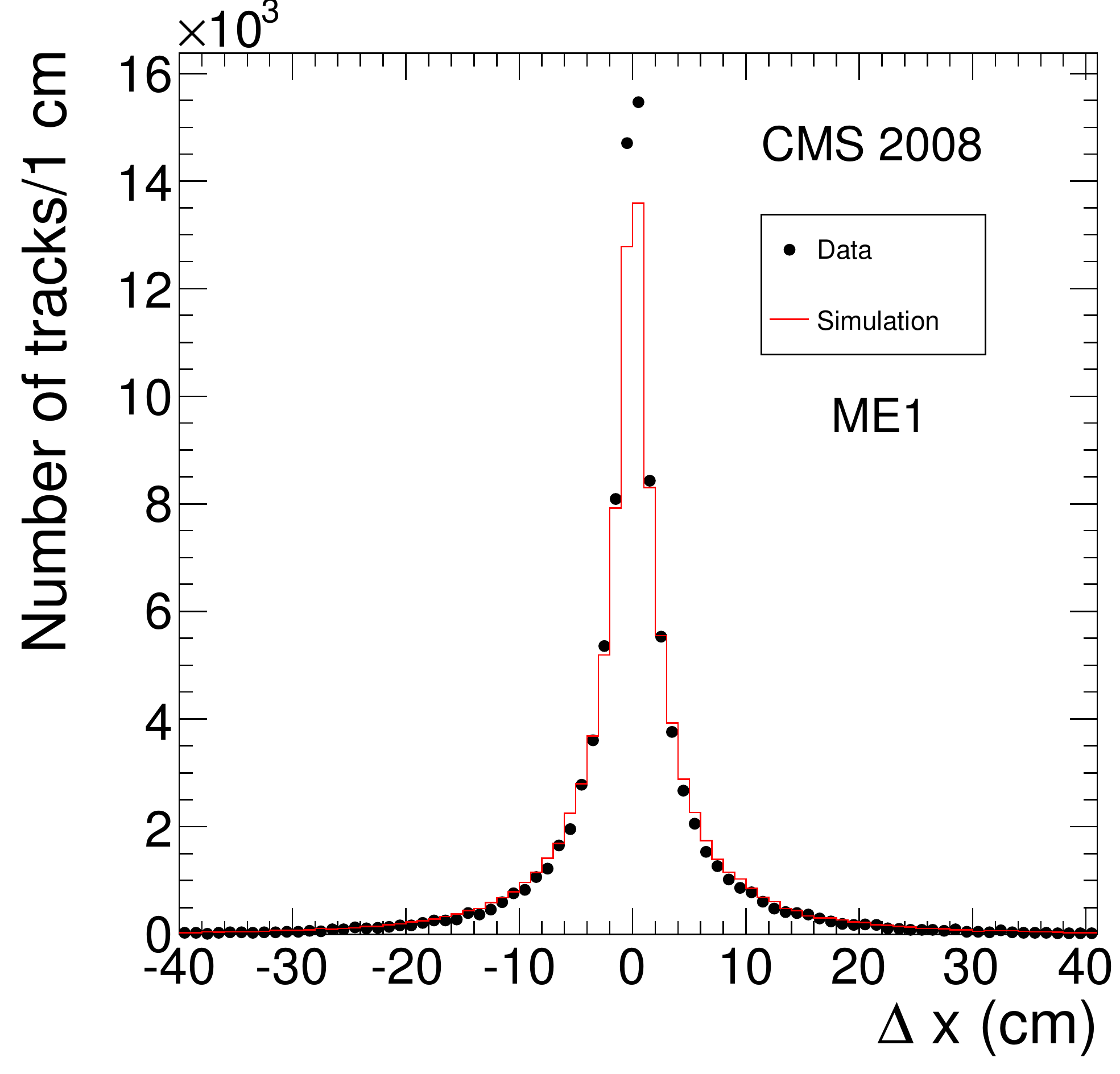}
  \put(-40,60){\bf\large c)}
  \includegraphics[width=0.5\textwidth]{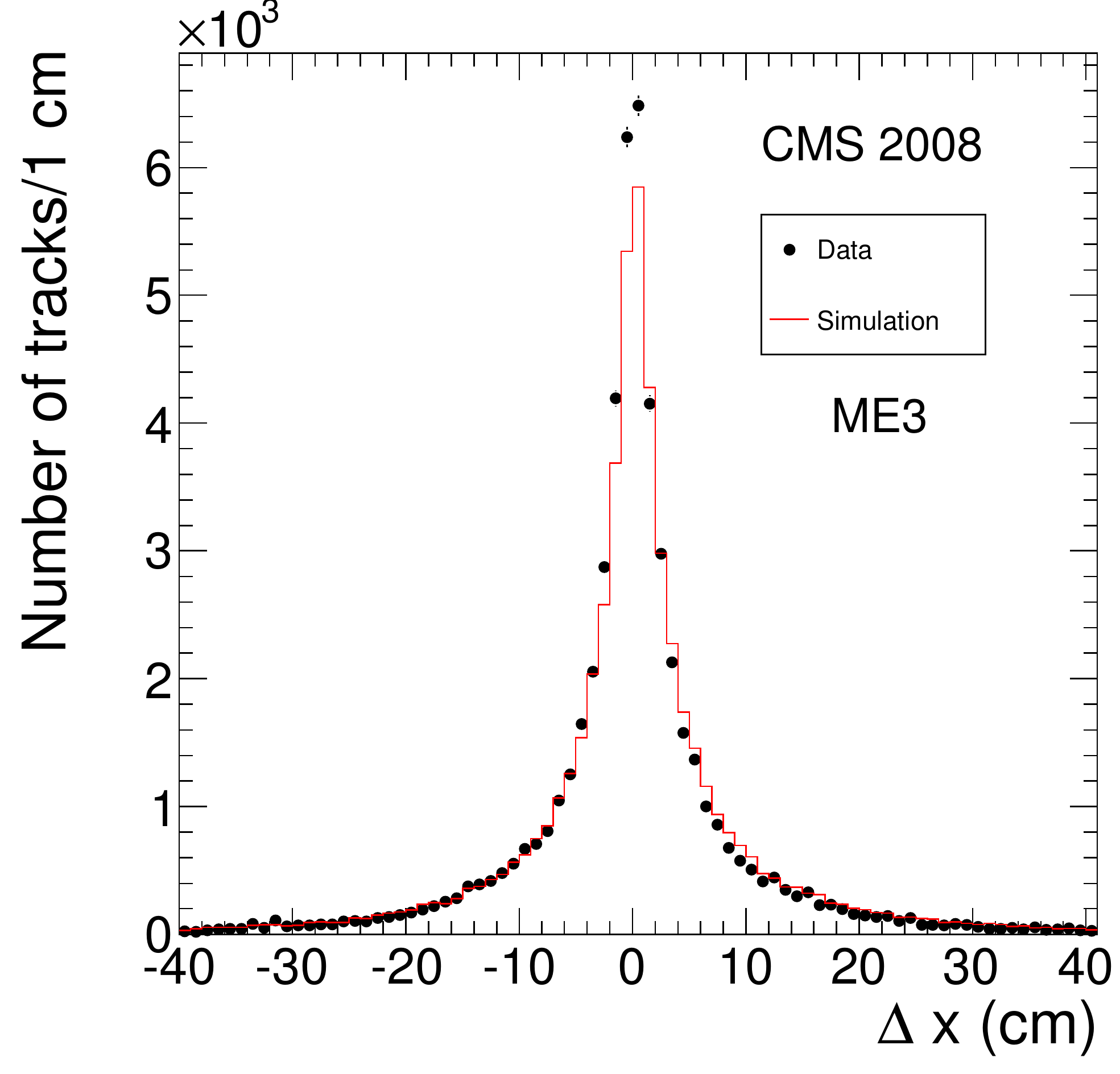}
  \put(-40,60){\bf\large d)}
 \caption{Distributions of residuals of the local $x$ position for the
  track-to-segment match in the data (points) and in the
  Monte Carlo simulation (histogram) for a) MB1 chambers; b) MB4 chambers;
  c) ME1 chambers; d) ME3 chambers.  The MC distributions are normalized
  to the number of events in the data.}
  \label{fig:resids_dt_trackermus}
\end{figure}

The accuracy of the propagation and the performance of the
track-to-segment match were studied using events in the tracker-pointing
dataset.  Every CosmicTF track with \pt $>$ 1.5~GeV/$c$ or
$p >$ 3~GeV/$c$ at the point of its closest approach to the nominal
beam line (PCA) was propagated to the muon stations, and a search
for the nearest muon segment reconstructed in each station was
performed.  For each segment found, the normalized residuals (pulls)
for position and direction were calculated; the pull is defined as the
difference between the position (or direction) of the extrapolated track
and the position (or direction) of the nearest segment, divided by their
combined uncertainty.
The track and the segment were considered to be matched if the distance
between them in local $x$ is less than 3~cm or if the value of the pull
for local $x$ is less than 4.

\begin{figure}[thb!]
  \centering
  \includegraphics[width=0.5\textwidth]{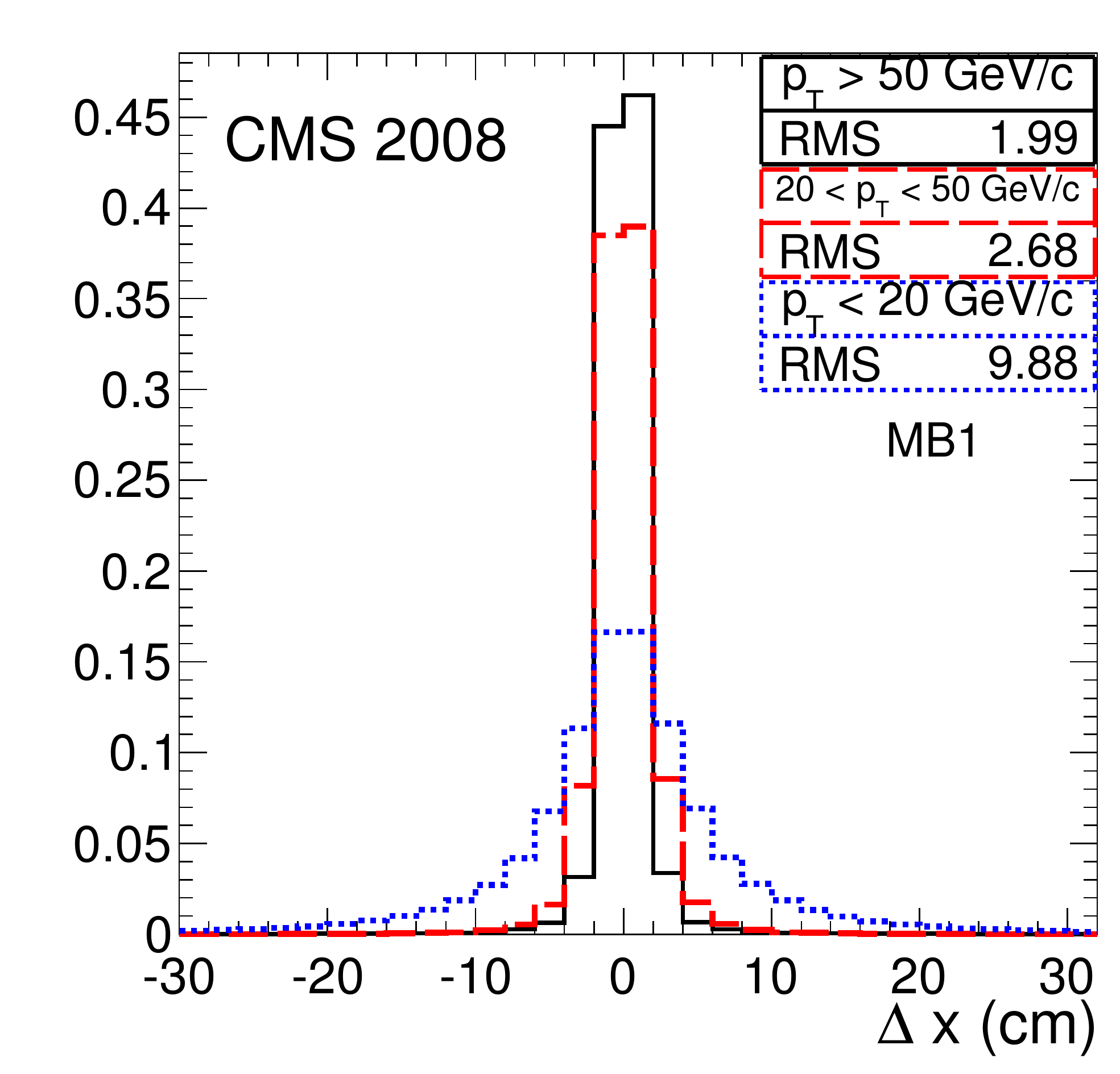}
  \put(-40,60){\bf\large a)}
  \includegraphics[width=0.5\textwidth]{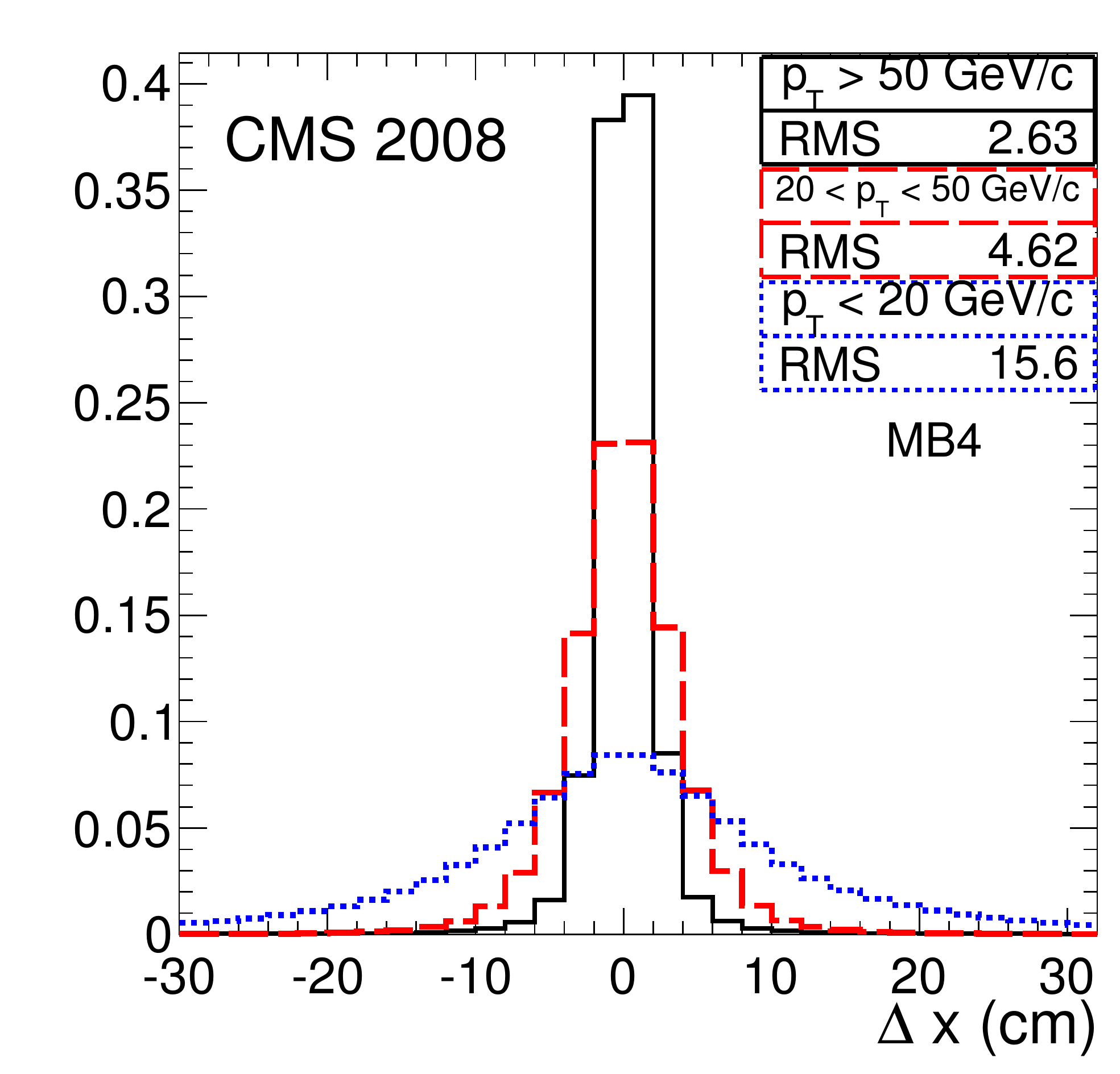}
  \put(-40,60){\bf\large b)}\\
  \includegraphics[width=0.5\textwidth]{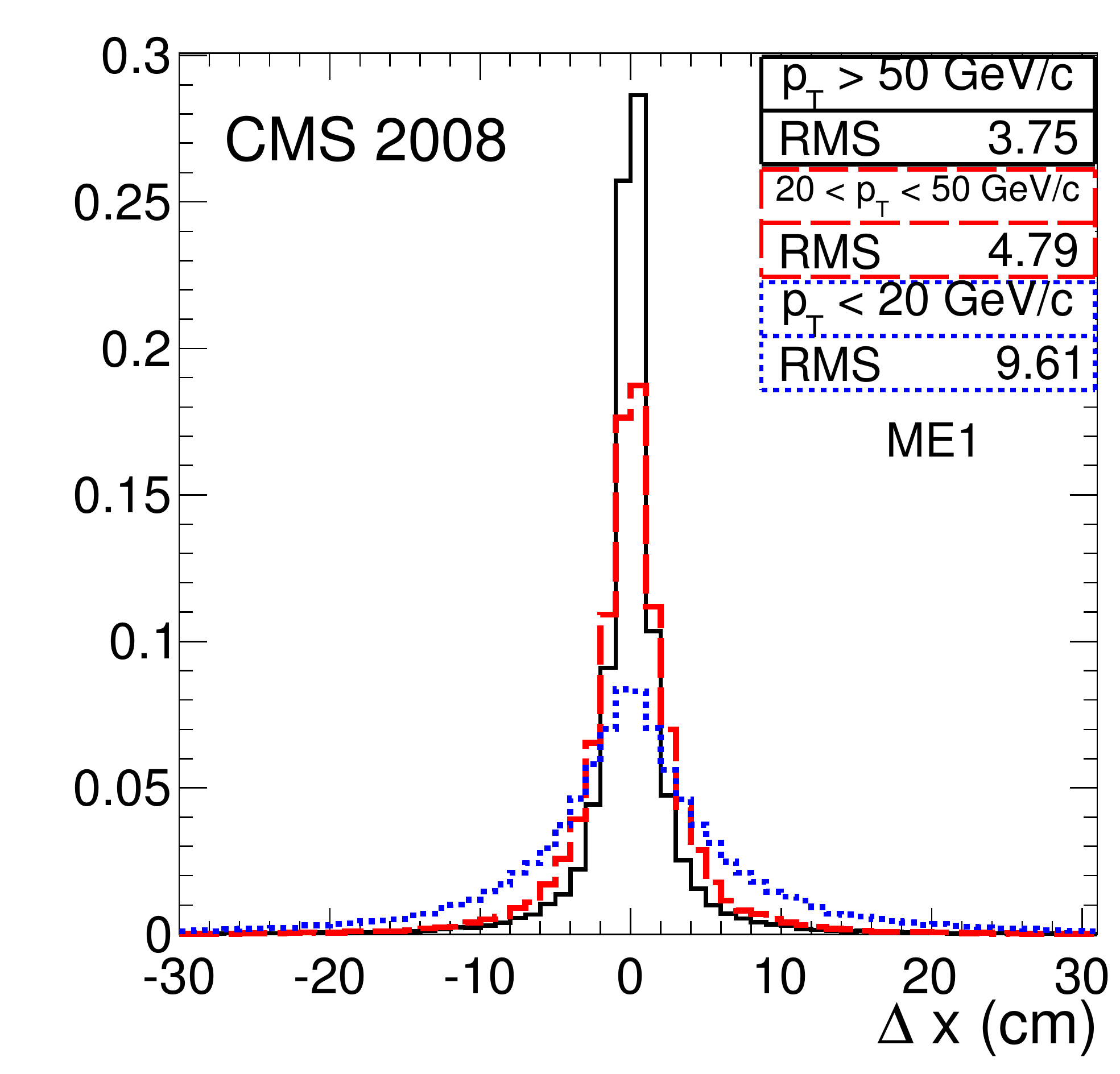}
  \put(-40,60){\bf\large c)}
  \includegraphics[width=0.5\textwidth]{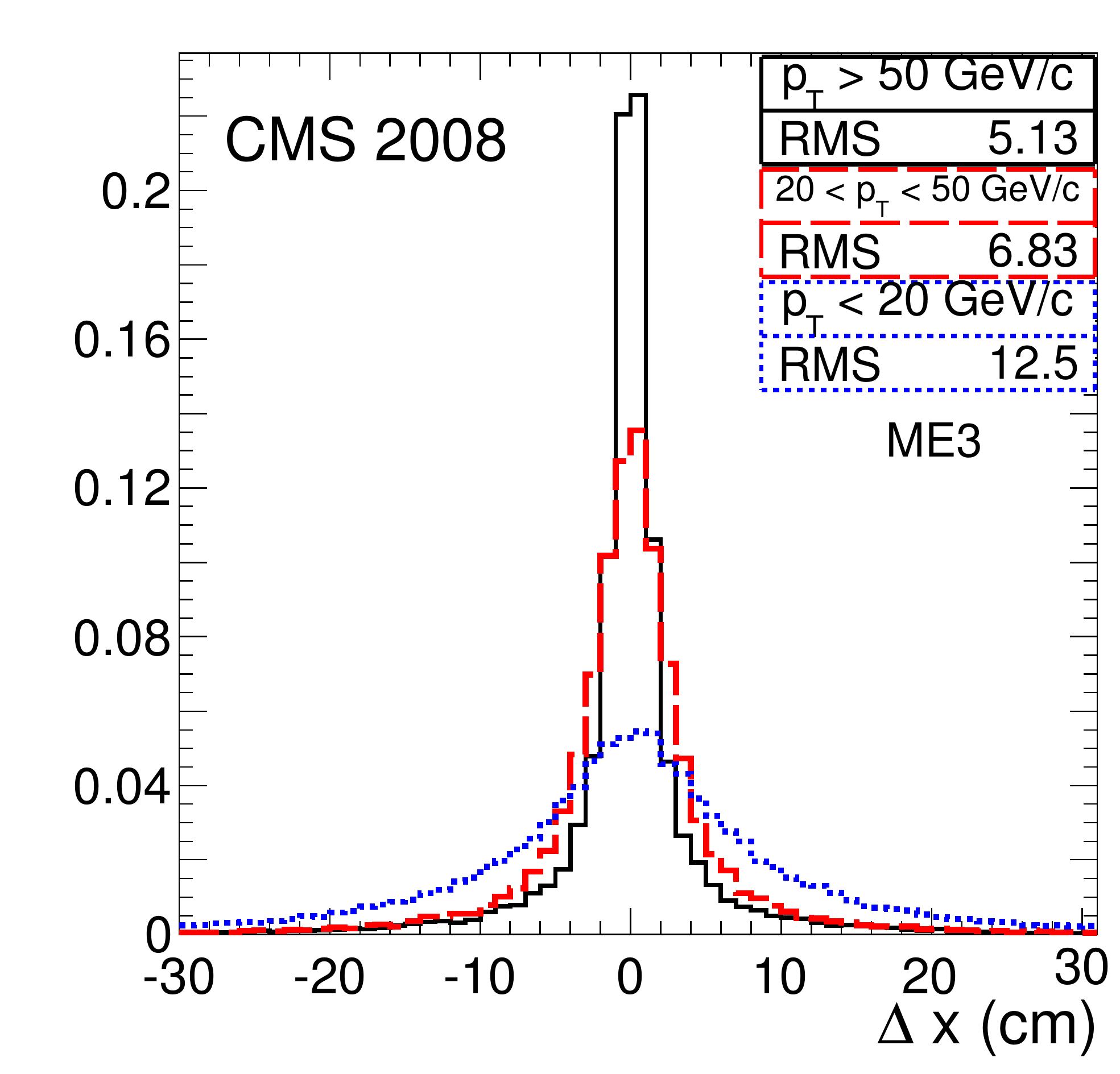}
  \put(-40,60){\bf\large d)}
  \caption{Distributions of residuals of the local $x$ position for
   the track-to-segment match in the data, shown separately in three
   \pt bins: less than 20~GeV/$c$ (dotted histograms), 20--50~GeV/$c$
   (dashed histograms), and above 50~GeV/$c$ (solid
   histograms).  The four panels show residuals for different chambers:
   a) MB1; b) MB4; c) ME1; d) ME3.  Each histogram is
   normalized to unit area; histograms and boxes
   with statistics are matched in style and color.}
  \label{fig:ptbins_dt_trackermus}
\end{figure}

The comparison between the measured and the predicted distributions of
the distance in local $x$ between the extrapolated track position and the
position of the segment for successful track-to-segment matches is shown
in Fig.~\ref{fig:resids_dt_trackermus}.  As expected, the width of the
distributions increases with the distance over which the track is
extrapolated, from the innermost to the outermost muon stations
(from MB1 to MB4 and from ME1 to ME3 in the DT and the CSC systems,
respectively; see Fig.~\ref{fig:event_display}).  This effect is well
reproduced by the Monte Carlo
simulation.  Figure~\ref{fig:ptbins_dt_trackermus} shows the same
residuals plotted separately in three bins of tracker-track \pt: less than
20~GeV/$c$, \mbox{20--50}~GeV/$c$, and above 50~GeV/$c$.  No bias is observed
in any of the \pt bins.  As expected, the width of the residuals decreases
with increasing \pt, because of smaller multiple-scattering effects.

\begin{figure}[htb!]
  \centering
  \includegraphics[width=0.5\textwidth]{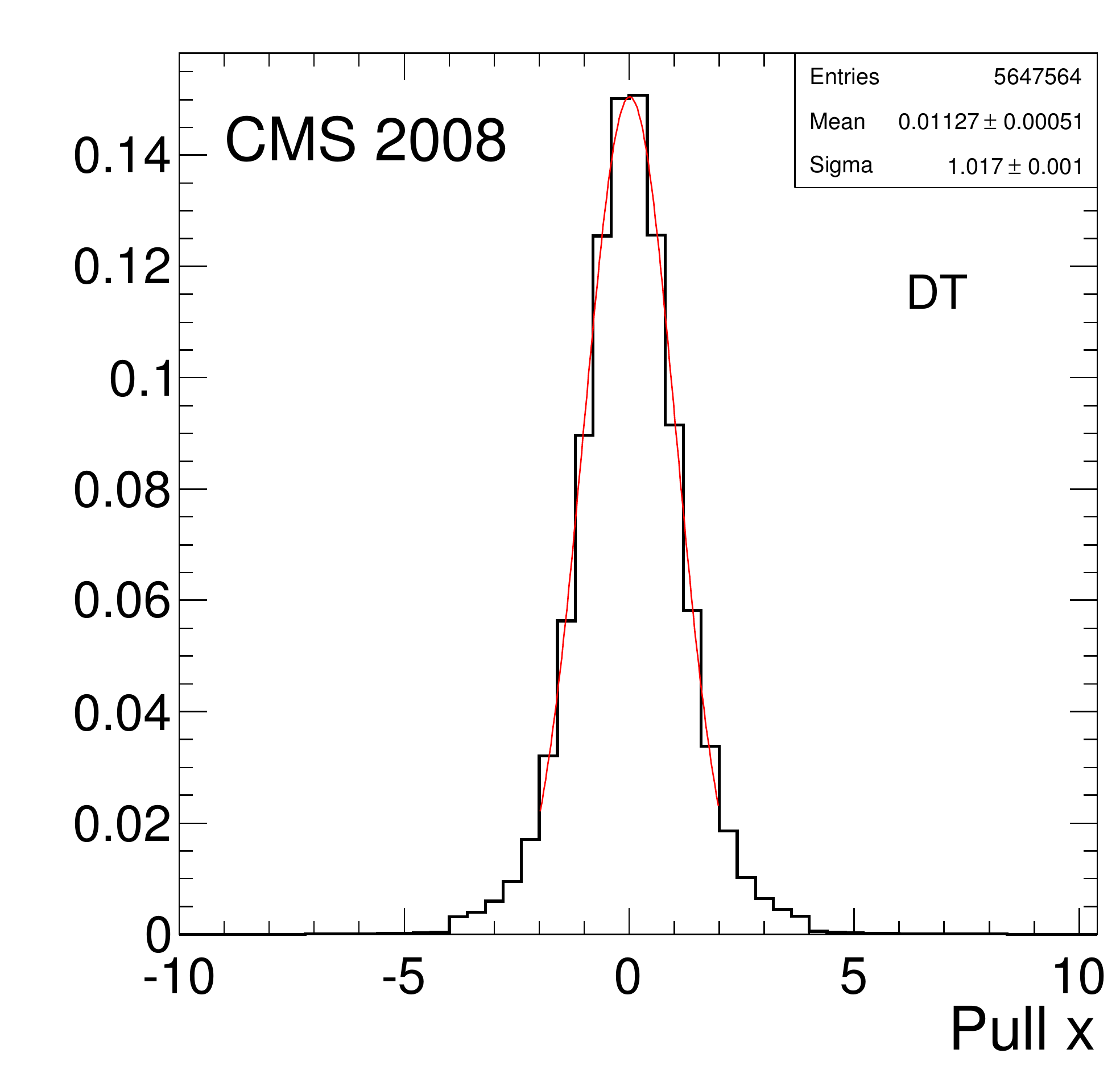}
  \put(-40,60){\bf\large a)}
  \includegraphics[width=0.5\textwidth]{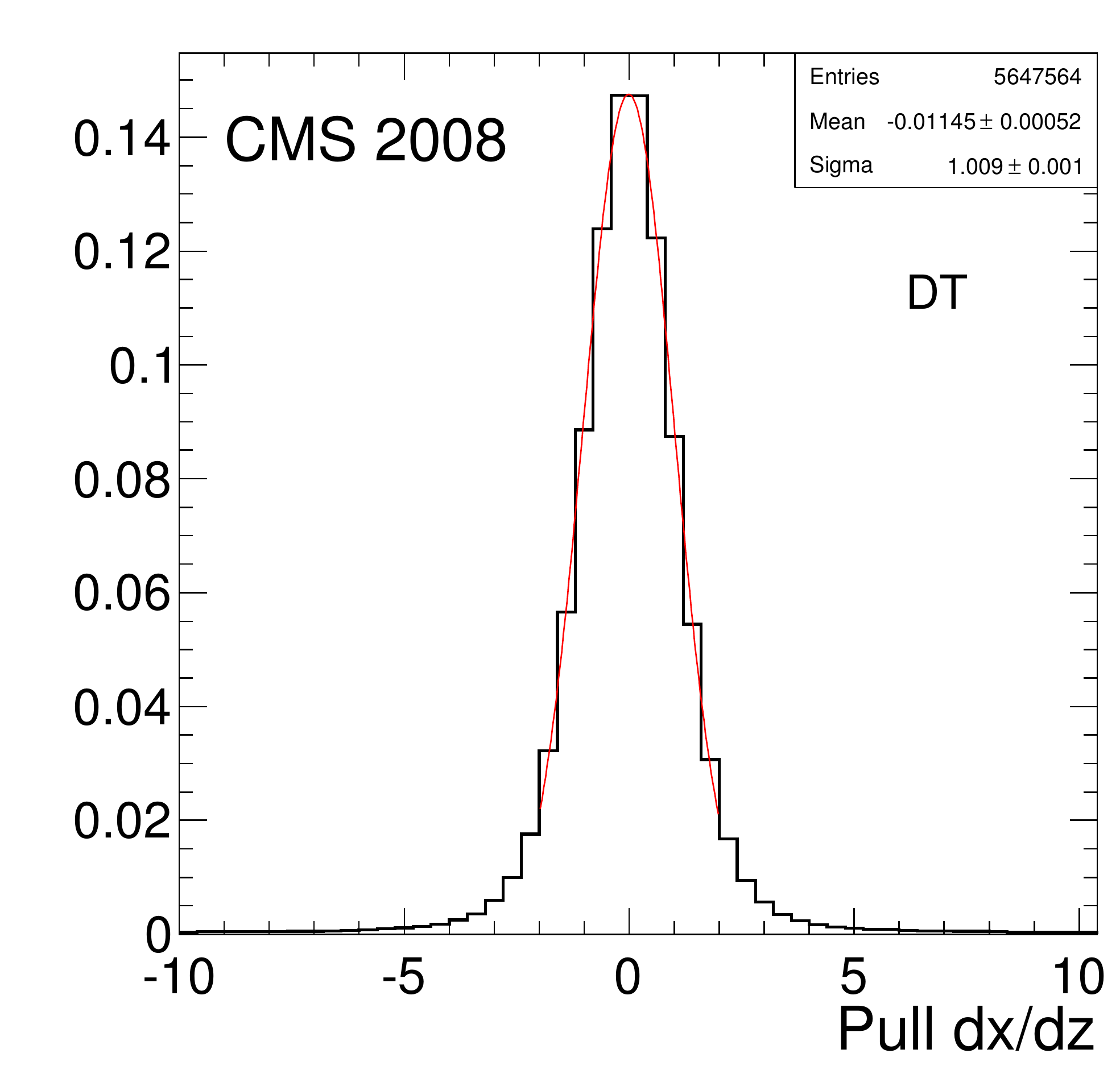}
  \put(-40,60){\bf\large b)}\\
  \includegraphics[width=0.5\textwidth]{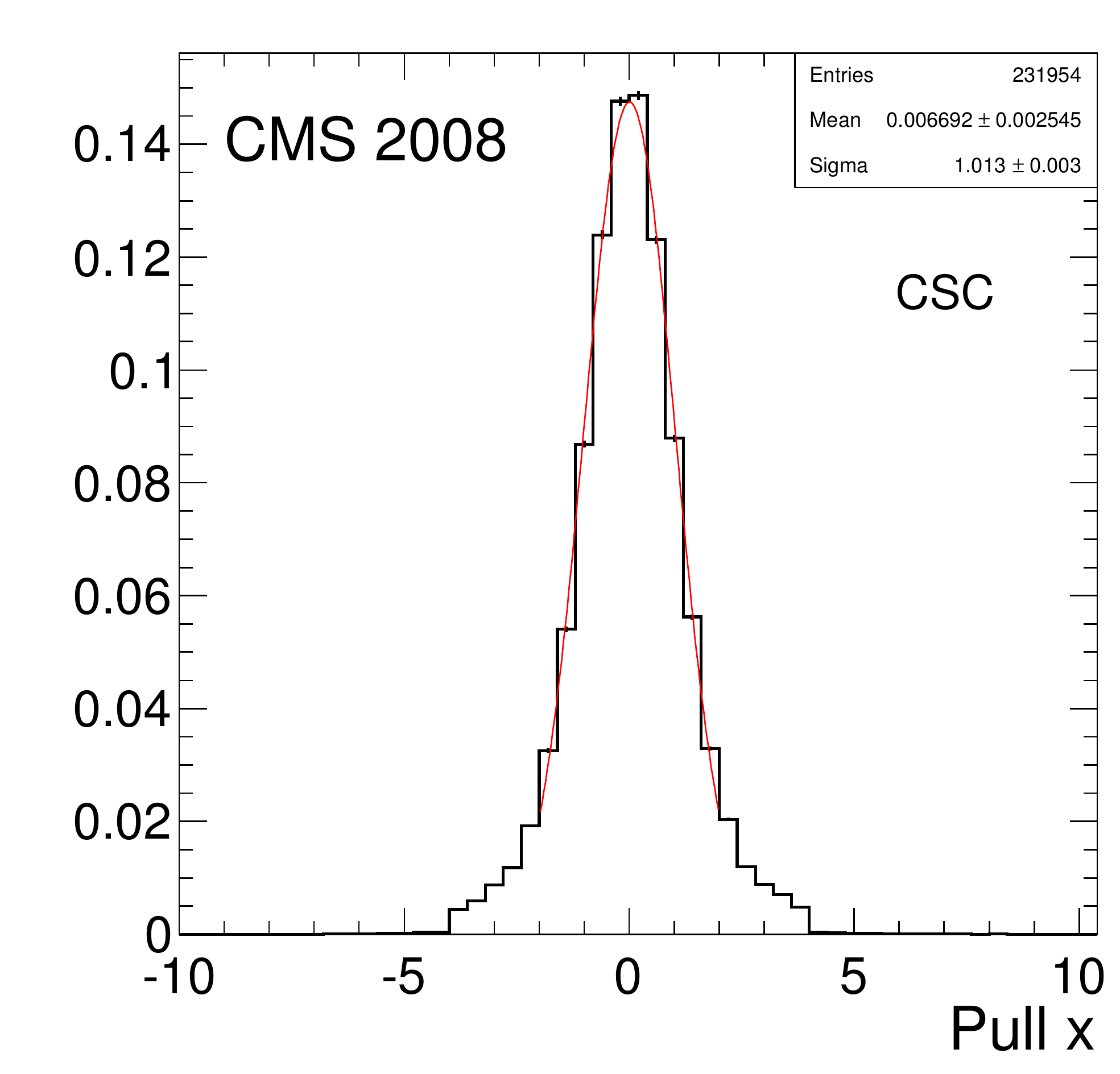}
  \put(-40,60){\bf\large c)}
  \includegraphics[width=0.5\textwidth]{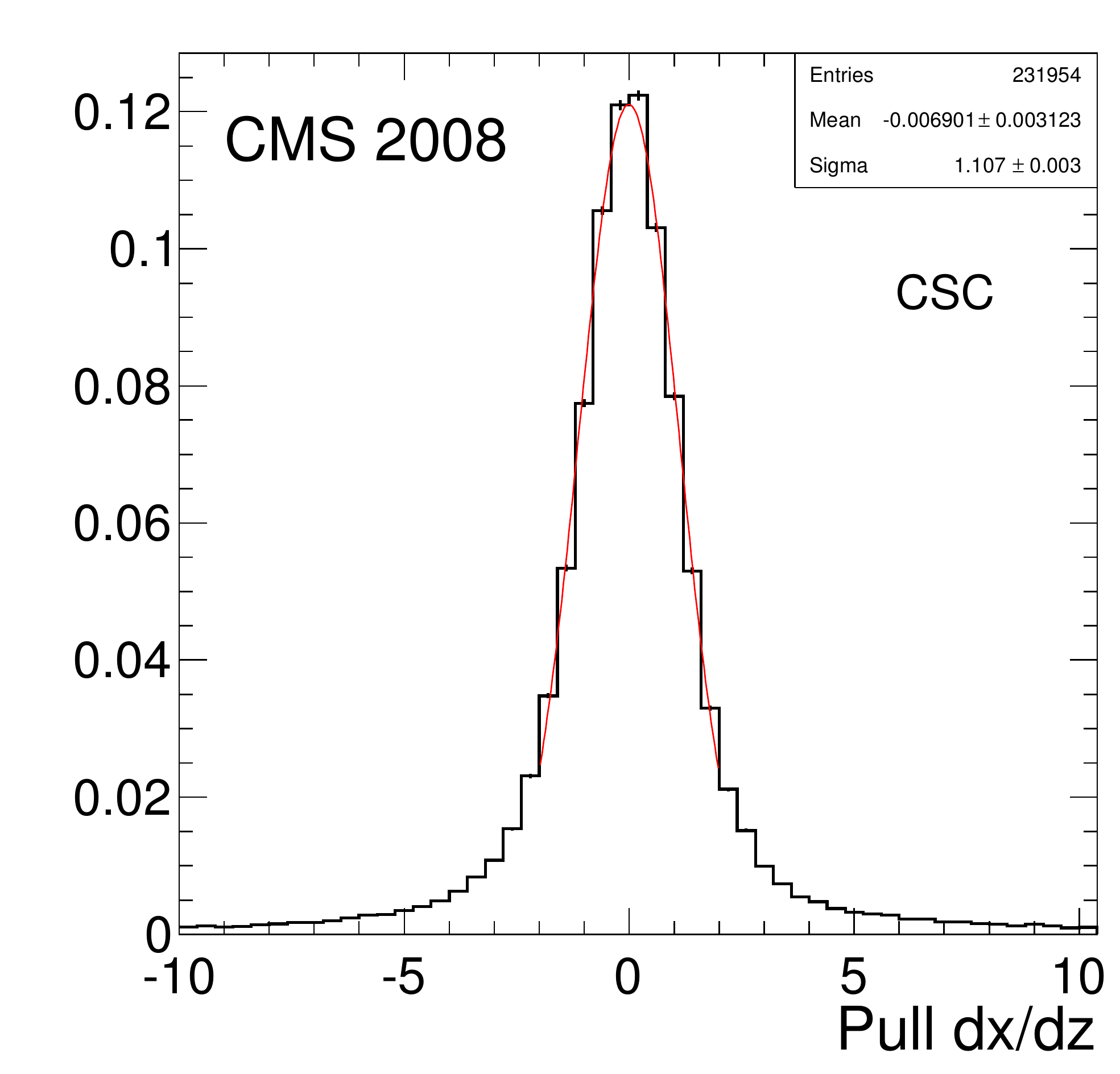}
  \put(-40,60){\bf\large d)}
  \caption{Distributions of pulls of the local $x$ position and of the
  local ${\rm d}x/{\rm d}z$ direction of the track-to-segment match:
  a) pull of $x$ in the DT system; b) pull of ${\rm d}x/{\rm d}z$ in the DT
  system; c) pull of $x$ in the CSC system; d) pull of ${\rm d}x/{\rm d}z$
  in the CSC system.  Each plot is a combined distribution of pulls
  in all MB or ME stations.  All histograms are normalized to unit area;
  the superimposed curves are the results of
  Gaussian fits in the range from $-$2 to 2.}
  \label{fig:pulls_trackermus}
\end{figure}

Figure~\ref{fig:pulls_trackermus} shows the distributions of pulls
of the local $x$ position and of the local ${\rm d}x/{\rm d}z$ direction
in the DT and CSC systems.  The widths of these and other
pulls were found to be close to unity and no large biases were observed,
thus demonstrating that the propagation works as expected and that the
uncertainties are estimated correctly.

\begin{figure}[htb]
  \centering
  \vspace*{-0.2cm}
  \includegraphics[width=0.7\textwidth]{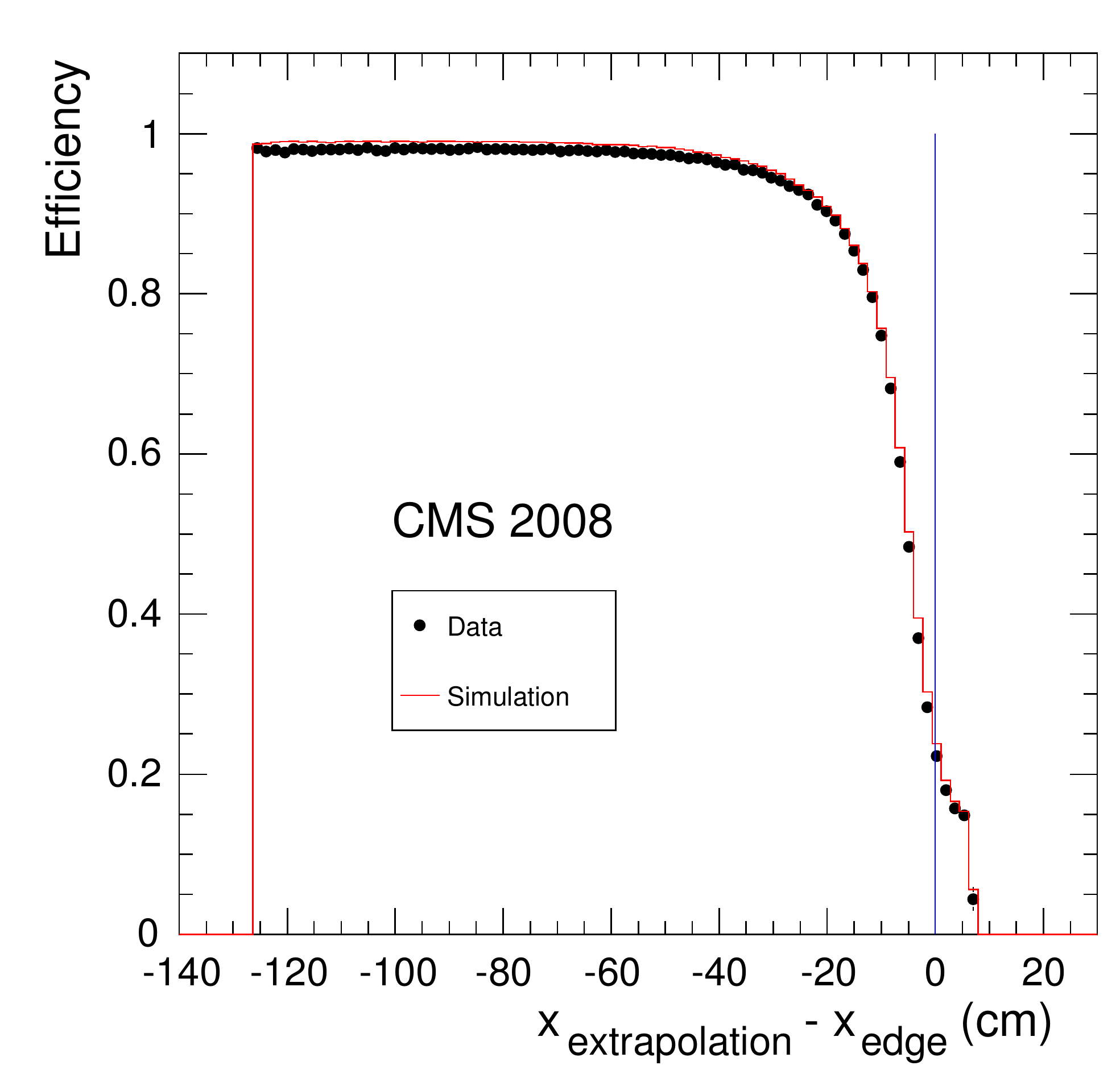}
  \caption{Efficiency of a successful track-to-segment match in
   the DT system, averaged over all DT chambers, as a function of the
   distance between the propagated
   track position and the nearest chamber edge, for the data (points)
   and the Monte Carlo simulation (histogram).}
  \label{fig:eff_vs_dist_trackermus}
\end{figure}

Figure~\ref{fig:eff_vs_dist_trackermus} shows the efficiency of a
successful track-to-segment match, averaged over all DT chambers, as a
function of the distance between the propagated track position and the nearest
chamber edge.  The distance to the chamber edge is defined to be
negative when the extrapolated position of the track is inside the
nominal chamber volume, and to be positive otherwise.  One can see
that the efficiency of finding a muon segment well inside the chamber is
close to unity.  The inefficiency observed near the edge of the chamber is
explained by the increased probability that a given muon (mostly with low
momentum) passed outside the chamber, considering the extrapolation
uncertainty; for higher-\pt muons the efficiency drop-off is steeper
and begins nearer to the chamber edge.  The slope of this efficiency drop
is consistent with that expected from the Monte Carlo simulation.

\begin{figure}[htbp]
  \centering
  \includegraphics[width=0.5\textwidth]{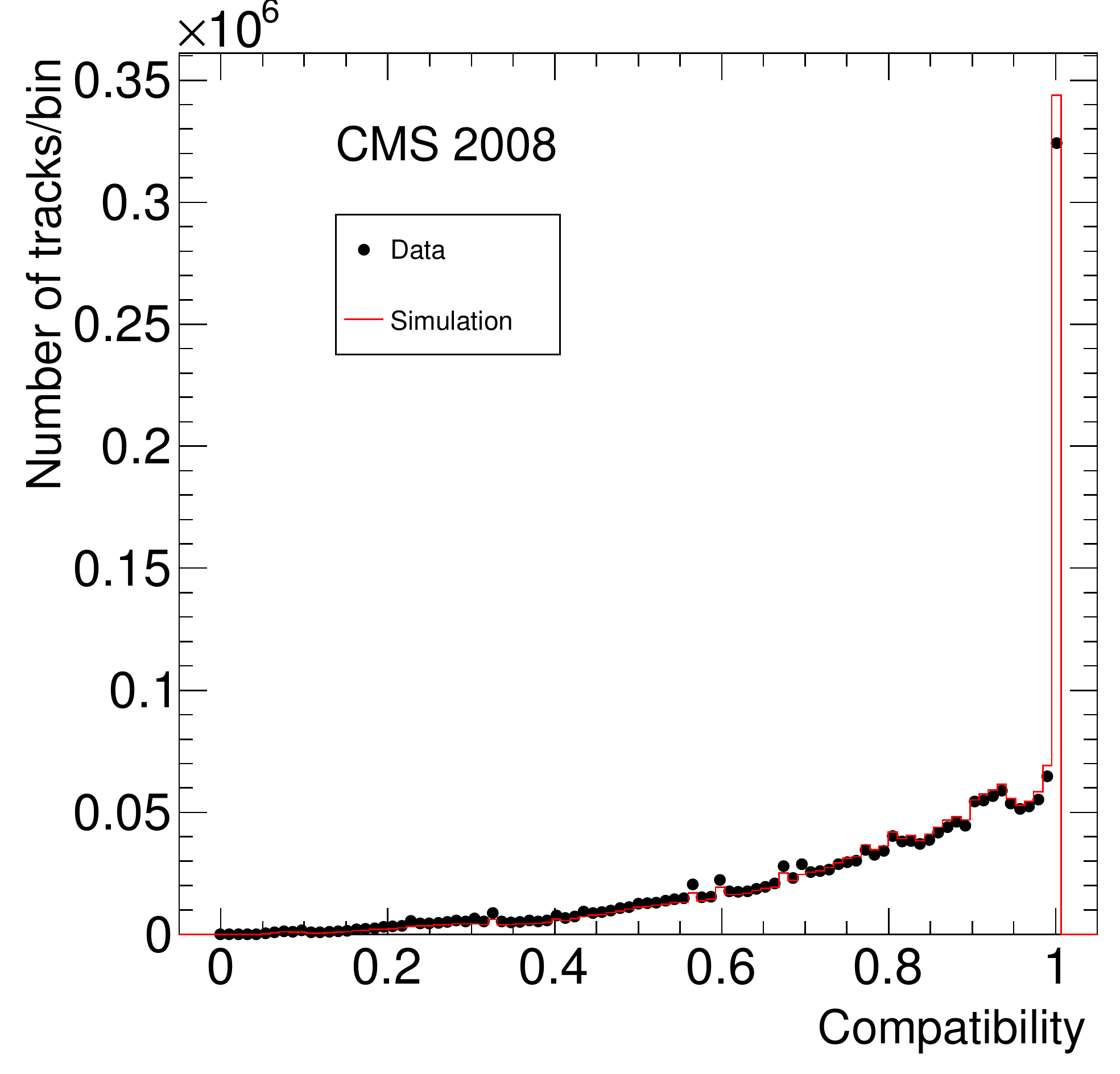}
  \put(-160,60){\bf\large a)}
  \includegraphics[width=0.5\textwidth]{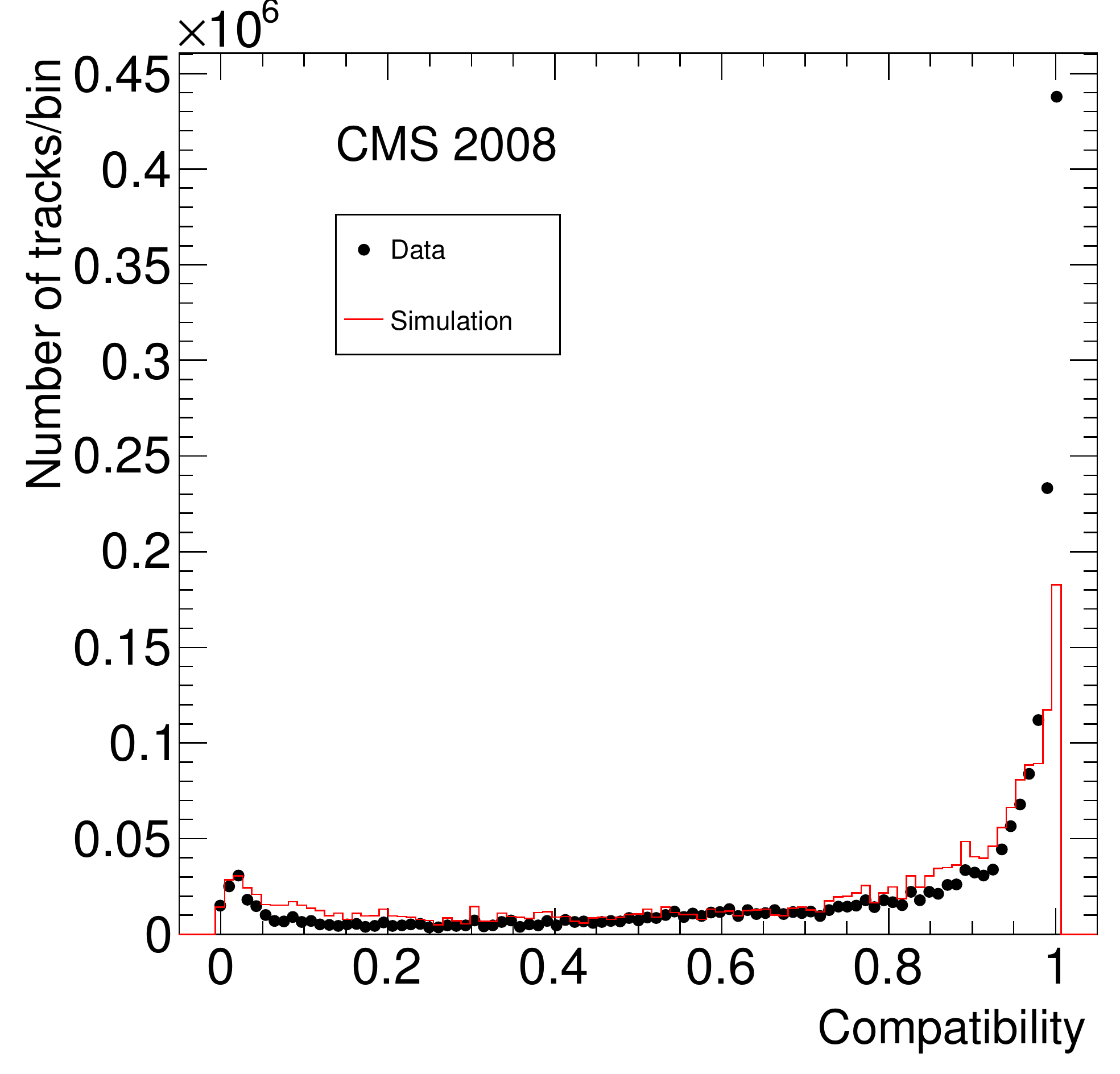}
  \put(-160,60){\bf\large b)}\\
 \caption{Compatibility of the muon hypothesis with a) segments reconstructed
  in the muon chambers and b) energy deposits in the calorimeters,
  for the data (points) and the simulation (histogram).}
  \label{fig:compat_trackermus}
\end{figure}

The ratio of the number of successful track-to-segment matches to the
total number of possible ones for a given track, and detailed information on
how well the extrapolated track and the segments match, are combined
into a single variable used to quantify the compatibility of a given
track with the hypothesis that it is from a muon.  Such a segment-based
compatibility
variable is constructed to be in the interval from 0 to 1, with a higher
value indicating a higher probability for the track to be from a muon.  The
distributions of this variable in the data and in the Monte Carlo
simulation are shown in Fig.~\ref{fig:compat_trackermus}a.  As
expected, most of the cosmic muons have large values of compatibility.
The Monte Carlo simulation reproduces the shape of the measured distribution
very well.  Figure~\ref{fig:compat_trackermus}b shows a similar
muon-compatibility variable
constructed from the energy depositions in the electromagnetic and
hadron calorimeters~\cite{PTDR1}.  Again, the distribution in the data
behaves as expected.
Since the compatibility algorithms were built and optimized
for muons produced in pp collisions but were applied to cosmic muons
in this study without any modifications, their current performance
is not expected to be optimal.
For example, the small enhancement at zero in Fig.~\ref{fig:compat_trackermus}b
is produced by muons crossing calorimeters sideways and depositing
more energy than expected for a muon coming from the interaction point;
this effect is well described by the cosmic-muon simulation.
The efficiency of the muon identification algorithms using these segment-based
and calorimeters-based compatibility variables is discussed in the next
section.

\section{Reconstruction and Identification Efficiency} \label{sec:eff}
This section reports on the efficiency of muon reconstruction
and identification algorithms, measured using two different
approaches: by independently reconstructing the two halves of a cosmic-muon
track in opposite detector hemispheres (Section~\ref{sec:eff_Ivan}) and
by searching for a track in the muon system corresponding to a track
reconstructed in the silicon tracker (Section~\ref{sec:eff_Chang}).

\subsection{Efficiency measurements with tracks in opposite detector
 hemispheres} \label{sec:eff_Ivan}
The efficiency of various muon reconstruction and identification algorithms
was measured
by selecting events with a good-quality global muon reconstructed in
one hemisphere of the detector (top or bottom) and examining whether
there is a corresponding track in the opposite hemisphere, in the region
of $|\Delta \phi| <$ 0.3 and $|\Delta \eta| <$ 0.3 around the direction of
the reference global-muon track.
Since this method of measuring muon efficiency is sensitive to the
efficiency of the silicon tracker, it was only applied to the runs from the last
part of CRAFT (``period B''), in which all parts of the tracker were
correctly synchronized with the rest of CMS~\cite{CMS_CFT_09_002}.
To ensure that the muon traversed the whole detector,
the \pt of the reference global-muon track at the point of its closest
approach to the nominal beam line was required to be larger than 10~GeV/$c$.

Two groups of muon reconstruction and identification algorithms
were considered: the dedicated cosmic-muon and, most importantly,
the standard algorithms developed for muons produced in pp collisions.
Efficiencies of the cosmic-muon algorithms were evaluated on a sample
of muons with a topology similar to that of muons produced in
beam collisions (i.e., with muon trajectories pointing to the nominal
interaction point).  Such events were selected by requiring that the
distance between the point of closest approach of the reference track
to the beam line and the nominal position of pp interactions did
not exceed 10~cm in the direction perpendicular to the beam axis ($r$)
and 20~cm along the nominal beam line ($z$).  Efficiencies of the
standard algorithms were measured on a smaller subsample of muons
selected by applying even tighter impact-parameter cuts to reference
tracks: $r <$ 4~cm (the beam-pipe radius) and $|z| <$ 10~cm
($\sim 3\sigma$ boundary of the collision region at start-up).
The total numbers of reference tracks available for this study were
4530 and 1028 for the dedicated cosmic-muon and standard algorithms,
respectively.

\begin{figure}[hbtp]
  \begin{center}
      \vspace*{-0.1cm}
      \rotatebox{90}{\includegraphics[width=0.359\textwidth]{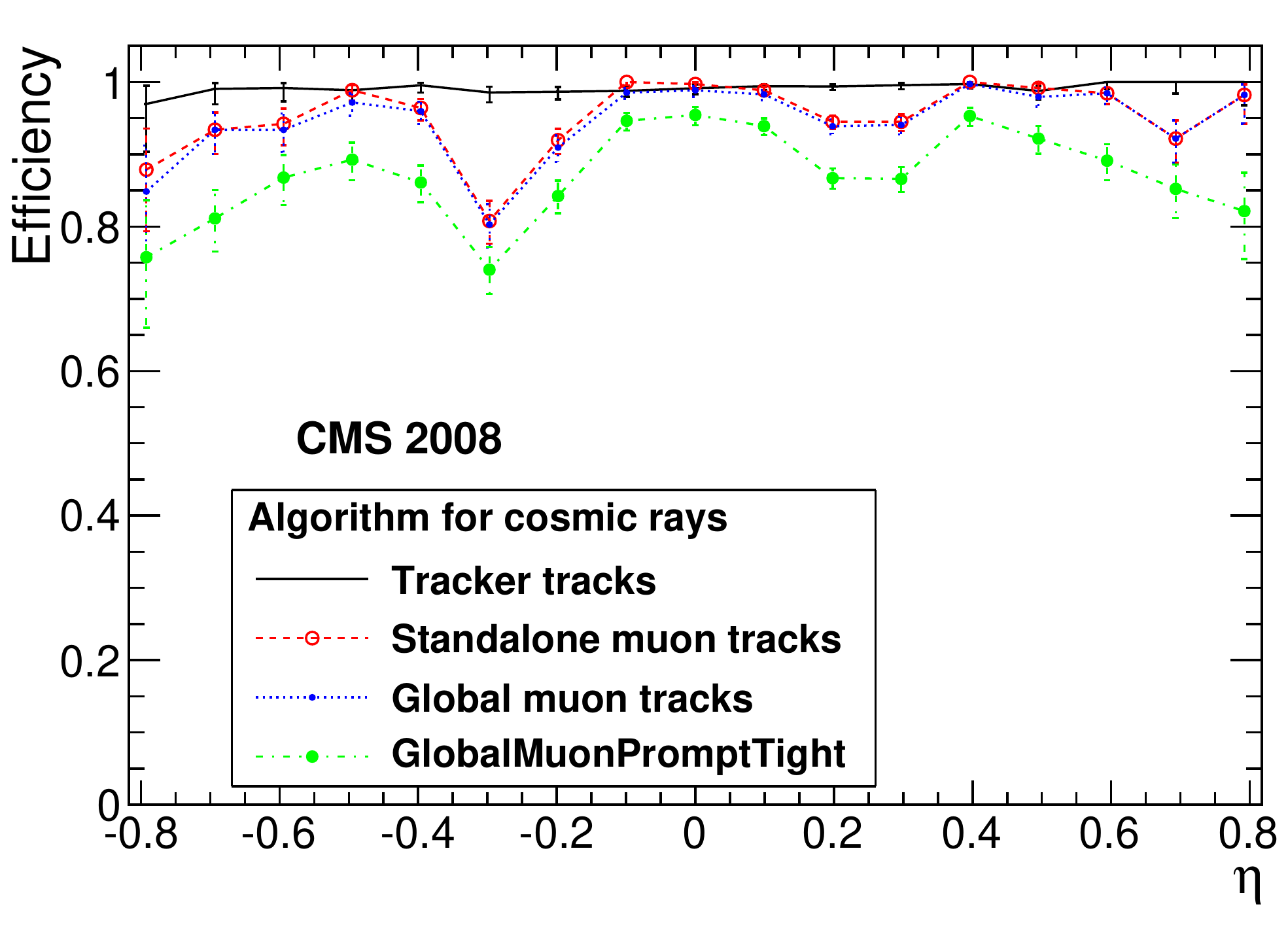}}
      \put(-40,40){\bf\large a)}
      \rotatebox{90}{\includegraphics[width=0.359\textwidth]{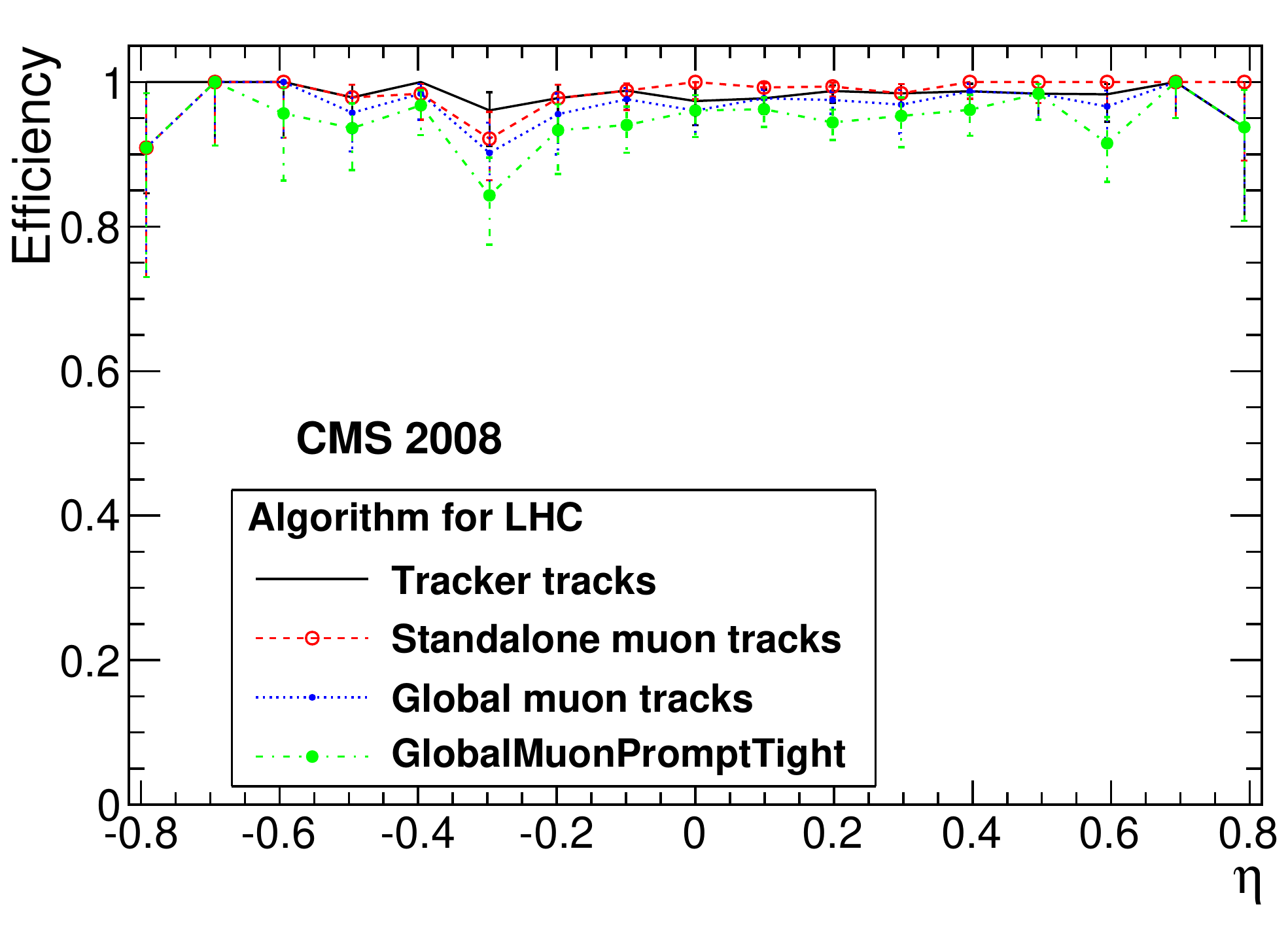}}
      \put(-40,40){\bf\large b)}
      \caption{Muon reconstruction efficiencies as a function of $\eta$
       of the reference track, for a) cosmic-muon algorithms and b)
       algorithms developed for muons produced in beam collisions at the
       LHC. Plot a) shows efficiencies of 2-leg CosmicCKF tracker tracks
       (solid line), CosmicSTA standalone muons (open circles),
       2-leg global muons (small filled circles), and
       2-leg global muons with an additional $\chi^2$
       cut applied (large filled circles).  Plot b) shows efficiencies of
       ppCKF tracker tracks (solid line), ppSTA standalone muons (open circles),
       LHC-like global muons (small filled circles), and
       LHC-like global muons with an extra $\chi^2$ cut
       (large filled circles).}
     \label{fig:eff_vs_eta}
  \end{center}
\end{figure}

Figure~\ref{fig:eff_vs_eta} shows the efficiencies to reconstruct a)
2-leg and b) LHC-like global muons and their constituents as a
function of the pseudorapidity of the reference tracks.  Integrated over the
barrel region of the detector ($|\eta| <$ 0.8), the average efficiency for
2-leg global muons produced by the dedicated cosmic-muon
reconstruction algorithm (Fig.~\ref{fig:eff_vs_eta}a) was found to be
(95.4 $\pm$ 0.3)\%.  The main source of efficiency loss is
an inefficiency of about 4\% in the cosmic standalone muon reconstruction,
mostly in the gaps between the barrel wheels.  The efficiencies of the
CosmicCKF tracker-track reconstruction and of the tracker-track to
standalone-muon matching are both larger than 99\%.
The efficiency of the standard global muon reconstruction
algorithm in the barrel region (Fig.~\ref{fig:eff_vs_eta}b), evaluated
on a sample of collision-like cosmic muons, was measured to be (97.1
$\pm$ 0.6)\%.
The small inefficiency stems mainly from the component tracks of global muons:
for events in which both the tracker track and the standalone-muon track
are found, the efficiency to reconstruct the global muon is (99.7 $\pm$ 0.1)\%.
Figure~\ref{fig:eff_vs_eta} also shows the efficiency for the global muons
with an additional requirement applied to the normalized
$\chi^2$ of the fit, $\chi^2/\text{ndf} <$ 10;
this cut is expected to strongly suppress hadronic punch-throughs
and muons from decays of $\pi$- and $K$-mesons in collision events.
The results for LHC-like global muons in Fig.~\ref{fig:eff_vs_eta}b
confirm that the proposed cut value leaves the efficiency for
prompt muons almost intact: the corresponding decrease in efficiency
is on the order of 2\%.

\begin{figure}[hbtp]
  \begin{center}
      \rotatebox{90}{\includegraphics[width=0.359\textwidth]{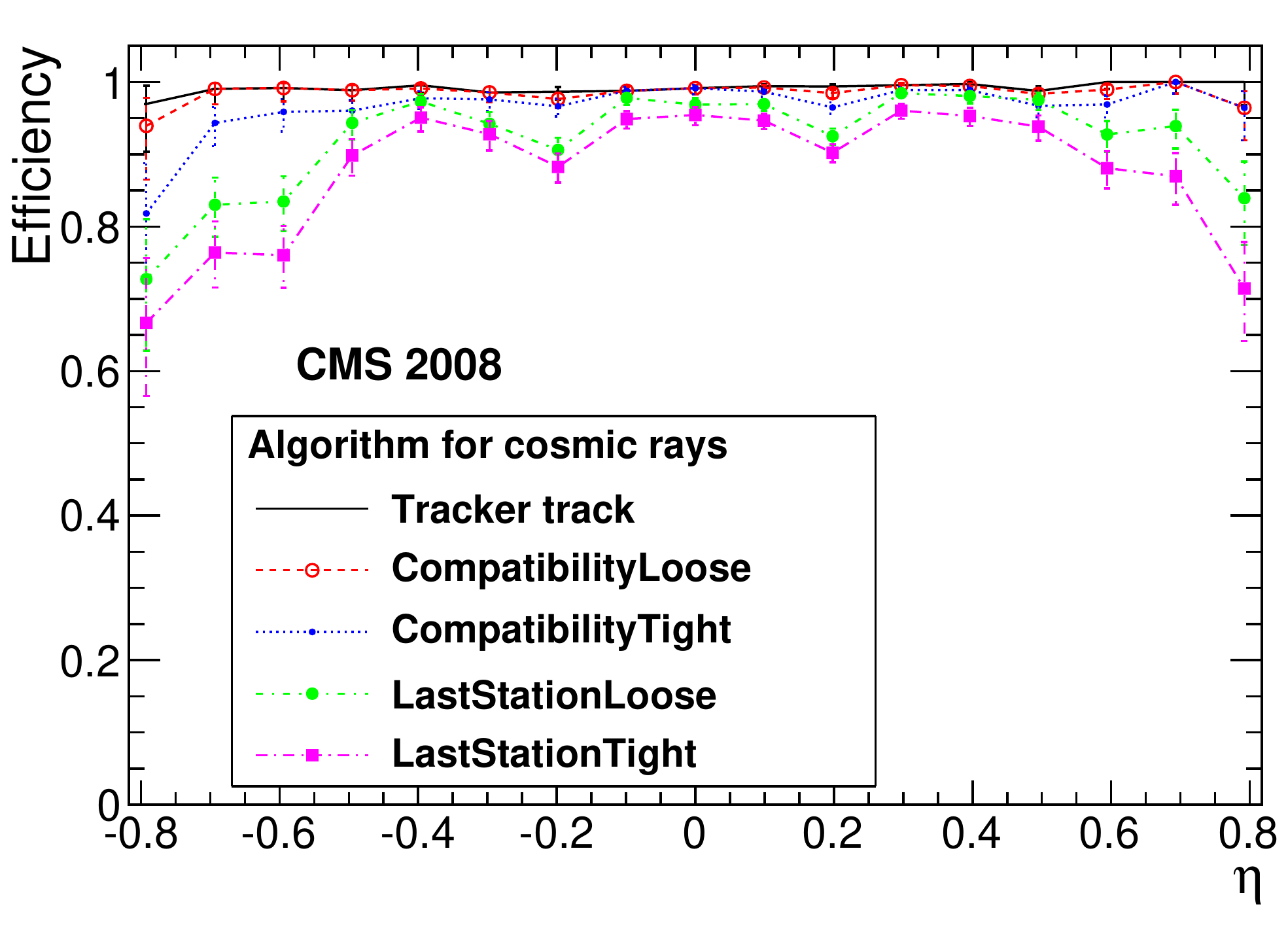}}
      \put(-40,40){\bf\large a)}
      \rotatebox{90}{\includegraphics[width=0.359\textwidth]{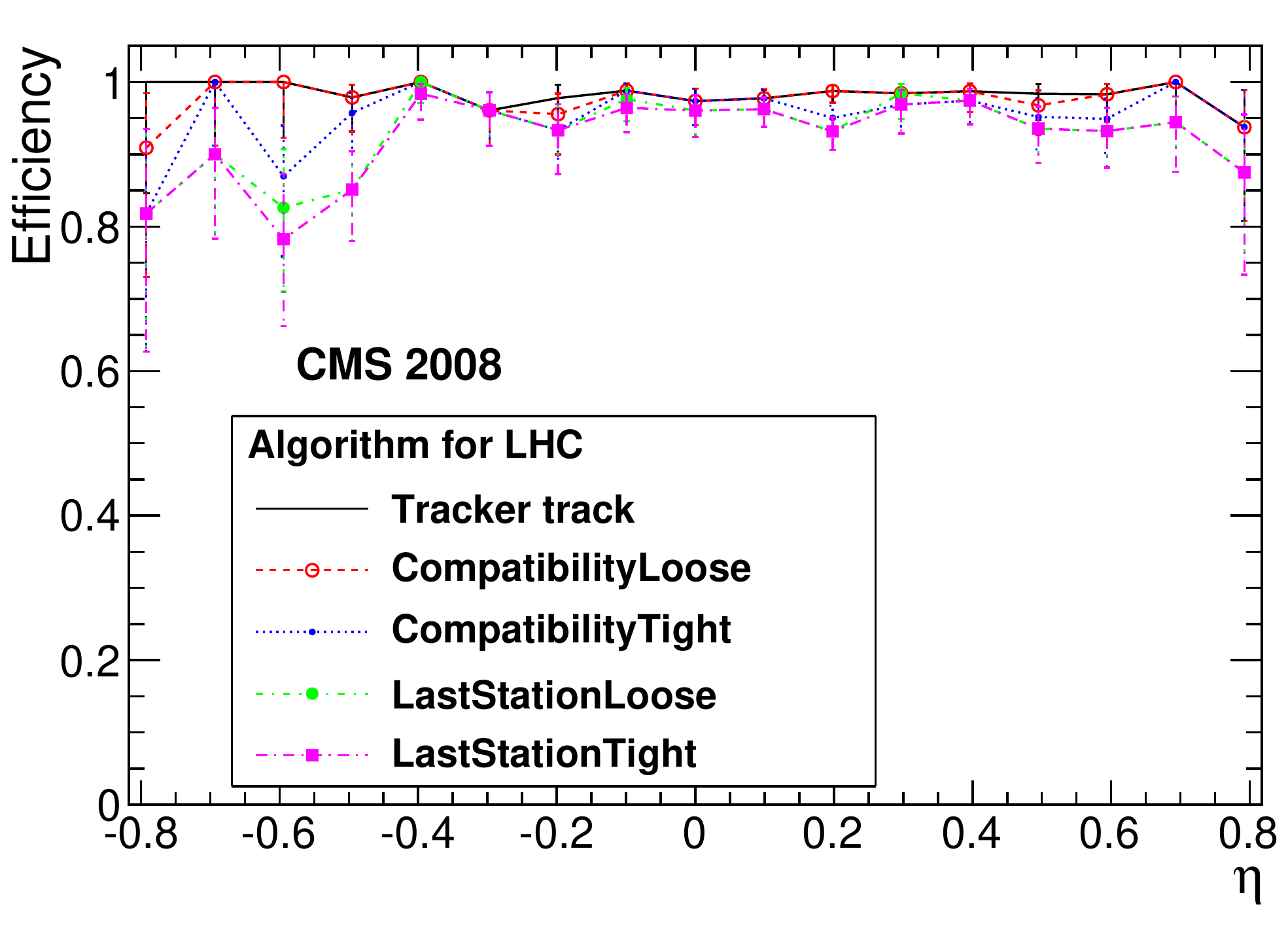}}
      \put(-40,40){\bf\large b)}
      \caption{Muon identification efficiencies as a function of $\eta$
       of the reference track, for a) cosmic-muon algorithms and b)
       algorithms developed for muons produced in beam collisions.
       The efficiencies for
       loose (CompatibilityLoose) and tight (CompatibilityTight)
       versions of the compatibility-based selection of tracker muons
       are shown in open and small filled circles,
       respectively.  The efficiencies for loose (LastStationLoose)
       and tight (LastStationTight) versions of the cut-based
       selection of tracker muons are shown in large filled circles
       and in squares, respectively.  For comparison,
       the efficiency for tracker tracks (upper line) is also shown.}
       \label{fig:eff_vs_eta_trackermus}
  \end{center}
\end{figure}

The efficiencies for the loose and tight versions of the
compatibility-based and cut-based muon identification algorithms are
compared with the efficiencies of tracker tracks in
Fig.~\ref{fig:eff_vs_eta_trackermus} as a function of $\eta$ of the
reference track.  For both cosmic-muon and standard track reconstruction
methods, the efficiency of CompatibilityLoose tracker muons is very
similar to that of the tracker tracks: the overall efficiency
reduction caused by the CompatibilityLoose selection does not exceed 0.3\%.
The loss of efficiency due to the CompatibilityTight selection
criteria is also small, of the order of 2\%.  The average efficiencies of the
loose and tight versions of the LastStation variant of the cut-based
selection are all above 90\%.  All measured efficiency values are
summarized in Table~\ref{tab:effs_dataVmc}.  To evaluate a possible bias
from correlations between reference and probe tracks, average efficiencies
were calculated in two ways: by dividing the number of probe tracks
found by the number of reference tracks, and as an arithmetic mean of
efficiencies in $\eta$ bins, neglecting their statistical uncertainties.
As can be seen in Table~\ref{tab:effs_dataVmc}, the efficiencies obtained
by the two methods agree within 1--2\% in most cases.

\begin{table}[htb]
\centering
\caption{Summary of muon reconstruction and identification efficiencies (in \%)
 for cosmic-muon algorithms and algorithms developed for muons produced in beam
 collisions at the LHC, for muons in the region $|\eta| <$~0.8.  Errors
 represent statistical uncertainties only.  Numbers in parentheses show
 efficiencies calculated by a simple (non-weighted) averaging of the
 efficiencies in $\eta$ bins.}
\begin{tabular}{|c|c|c|c|c|} \hline
Algorithm & \multicolumn{2}{c|}{Cosmic-muon algorithms} &
            \multicolumn{2}{c|}{Beam collision algorithms} \\
                   &       Data     &     Simulation       &      Data      &      Simulation       \\ \hline
\multicolumn{5}{|c|}{Reconstruction algorithms}\\ \hline
Tracker-only       & $99.2 \pm 0.2$ ($99.2$) & $99.9 \pm 0.1$ ($99.9$) & $98.3 \pm 0.5$ ($98.3$) & $99.1 \pm 0.3$ ($98.8$)\\
Standalone muon    & $96.1 \pm 0.3$ ($95.2$) & $91.5 \pm 0.3$ ($92.7$) & $98.8 \pm 0.4$ ($98.7$) & $96.2 \pm 0.5$ ($96.8$)\\
Global muon        & $95.4 \pm 0.3$ ($94.5$) & $91.3 \pm 0.3$ ($92.5$) & $97.1 \pm 0.6$ ($96.9$) & $95.0 \pm 0.5$ ($95.5$)\\ \hline
\multicolumn{5}{|c|}{Identification algorithms}\\ \hline
CompatibilityLoose & $98.9 \pm 0.2$ ($98.5$) & $98.8 \pm 0.1$ ($98.7$) & $98.1 \pm 0.5$ ($97.8$) & $97.9 \pm 0.4$ ($97.4$)\\
CompatibilityTight & $97.6 \pm 0.2$ ($96.5$) & $97.2 \pm 0.2$ ($97.0$) & $96.4 \pm 0.7$ ($95.9$) & $96.6 \pm 0.5$ ($96.2$)\\
LastStationLoose   & $94.7 \pm 0.4$ ($92.0$) & $94.6 \pm 0.3$ ($94.8$) & $94.6 \pm 0.8$ ($93.3$) & $93.2 \pm 0.6$ ($93.1$)\\
LastStationTight   & $91.7 \pm 0.4$ ($87.8$) & $84.9 \pm 0.4$ ($84.1$) & $94.2 \pm 0.8$ ($92.2$) & $92.0 \pm 0.7$ ($91.2$)\\ \hline
\end{tabular}
\label{tab:effs_dataVmc}
\end{table}

\begin{figure}[hbt]
  \begin{center}
      \rotatebox{90}{\includegraphics[width=0.359\textwidth]{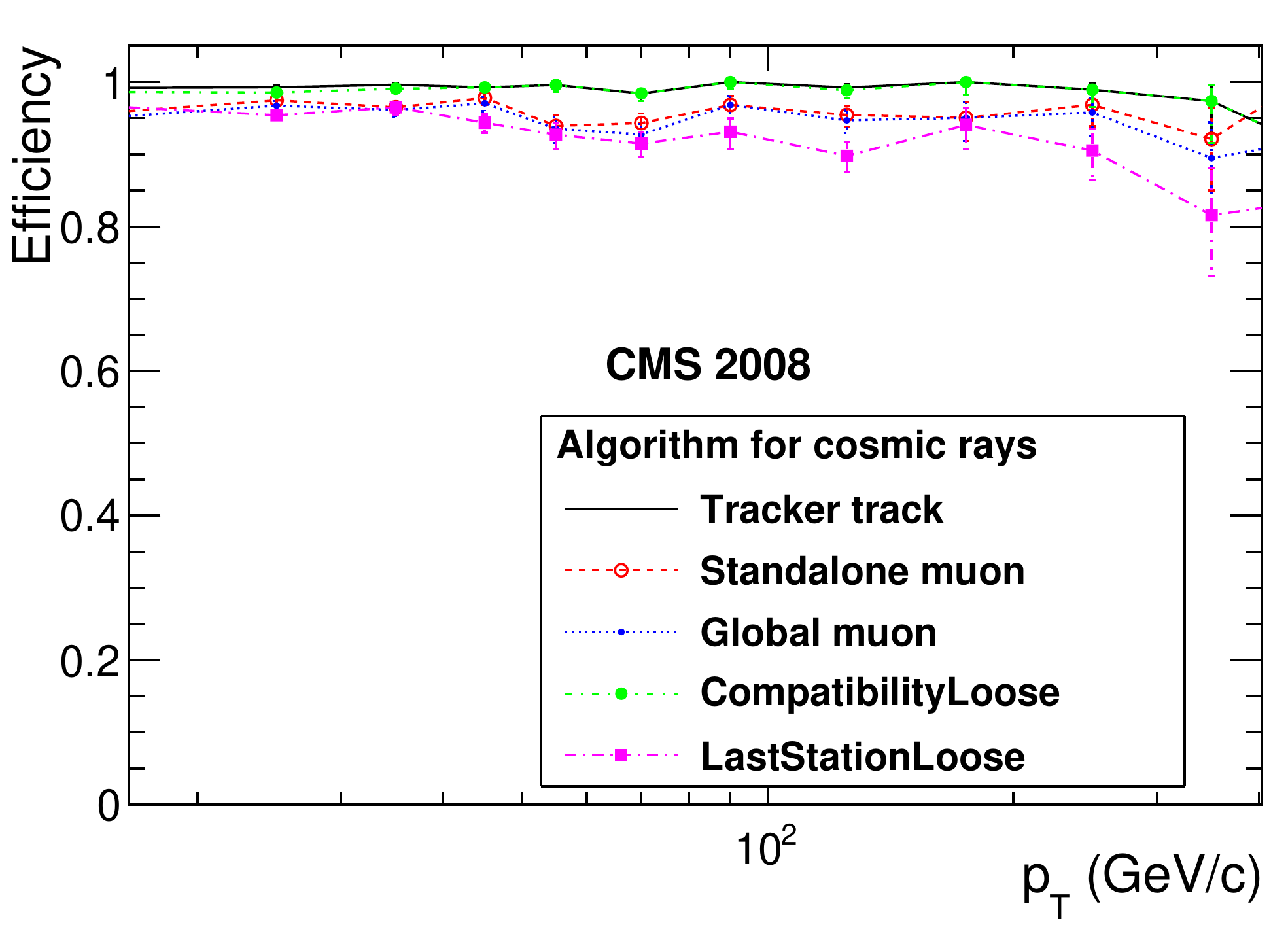}}
      \put(-180,40){\bf\large a)}
      \rotatebox{90}{\includegraphics[width=0.359\textwidth]{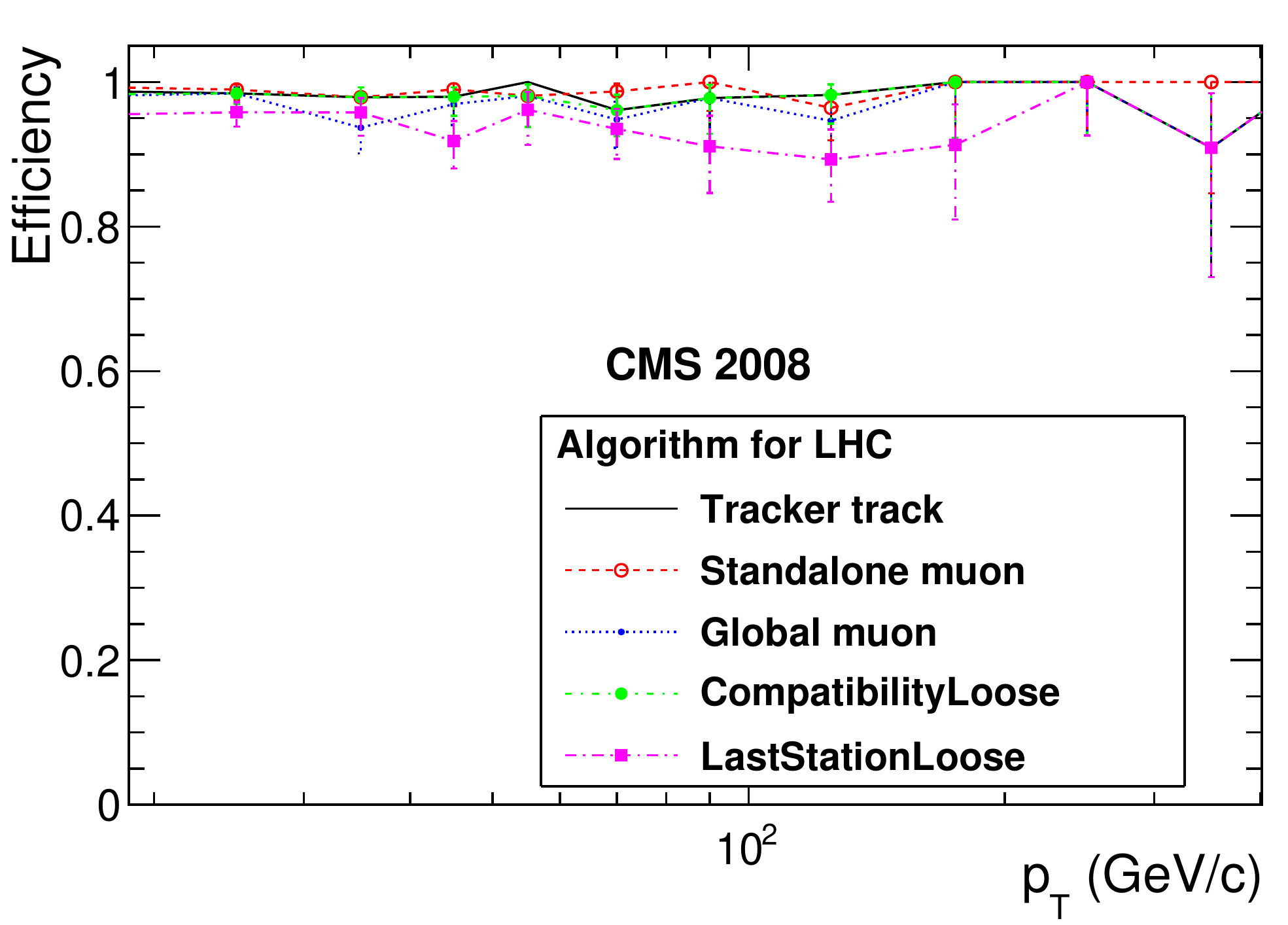}}
      \put(-180,40){\bf\large b)}
      \caption{Muon reconstruction and identification efficiencies in
       the barrel region of the detector ($|\eta| <$ 0.8) as
       a function of \pt of the reference track, for a) cosmic-muon
       algorithms and b) algorithms developed for muons produced in
       beam collisions, for
       tracker tracks (solid line), standalone muons (open
       circles), global muons (small filled circles), and the loose versions
       of the compatibility-based and cut-based muon identification
       algorithms (large filled circles and squares, respectively).}
      \label{fig:eff_vs_pt}
  \end{center}
\end{figure}

The dependence of the efficiencies of the various muon reconstruction and
identification algorithms on the \pt of the reference muons at the PCA is shown
in Fig.~\ref{fig:eff_vs_pt}.  None of the studied algorithms show a
strong \pt dependence in the range above 10~GeV/$c$, as expected.

\begin{figure}[hbt]
  \begin{center}
      \rotatebox{90}{\includegraphics[width=0.359\textwidth]{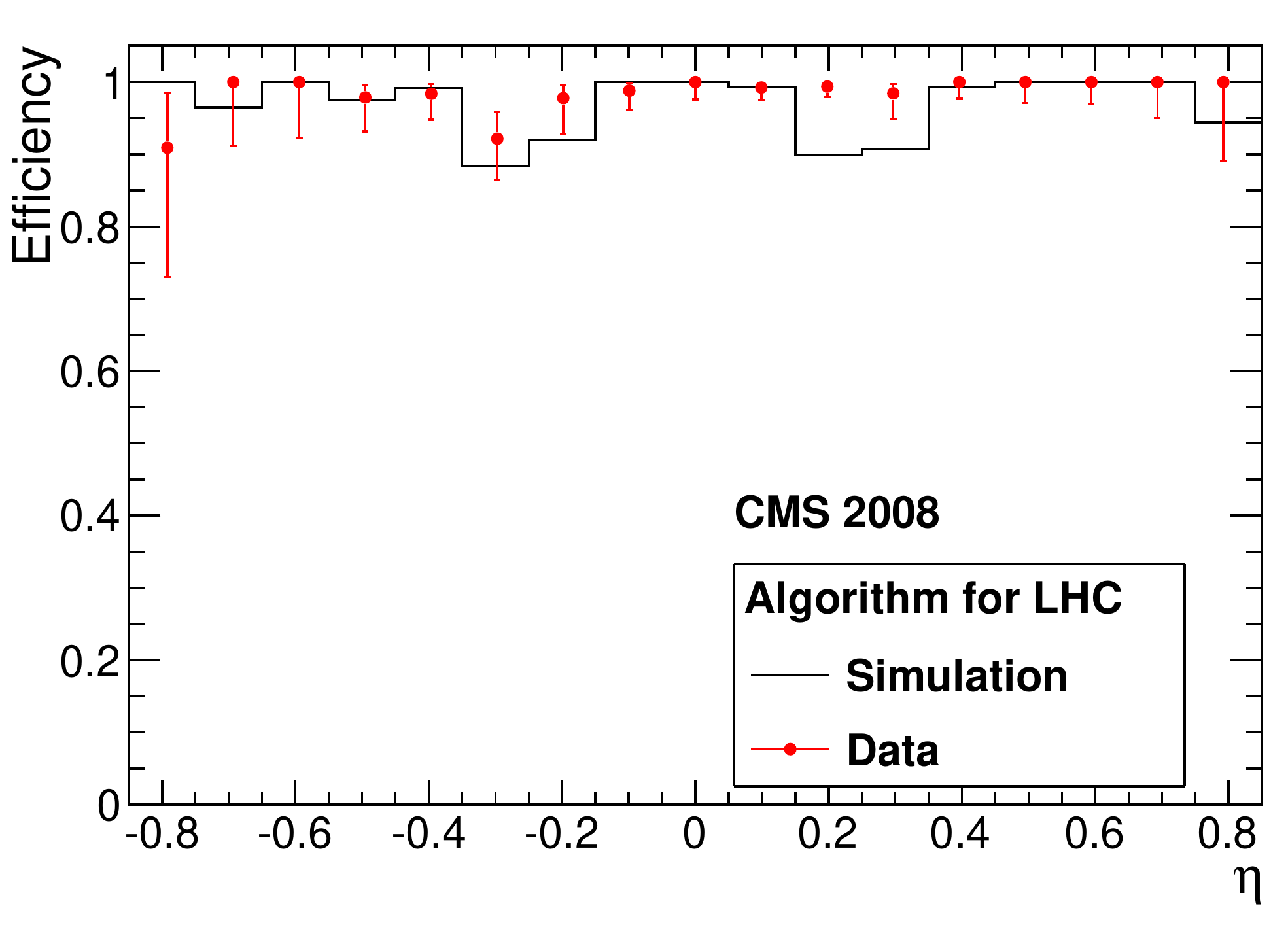}}
      \put(-180,40){\bf\large a)}
      \rotatebox{90}{\includegraphics[width=0.359\textwidth]{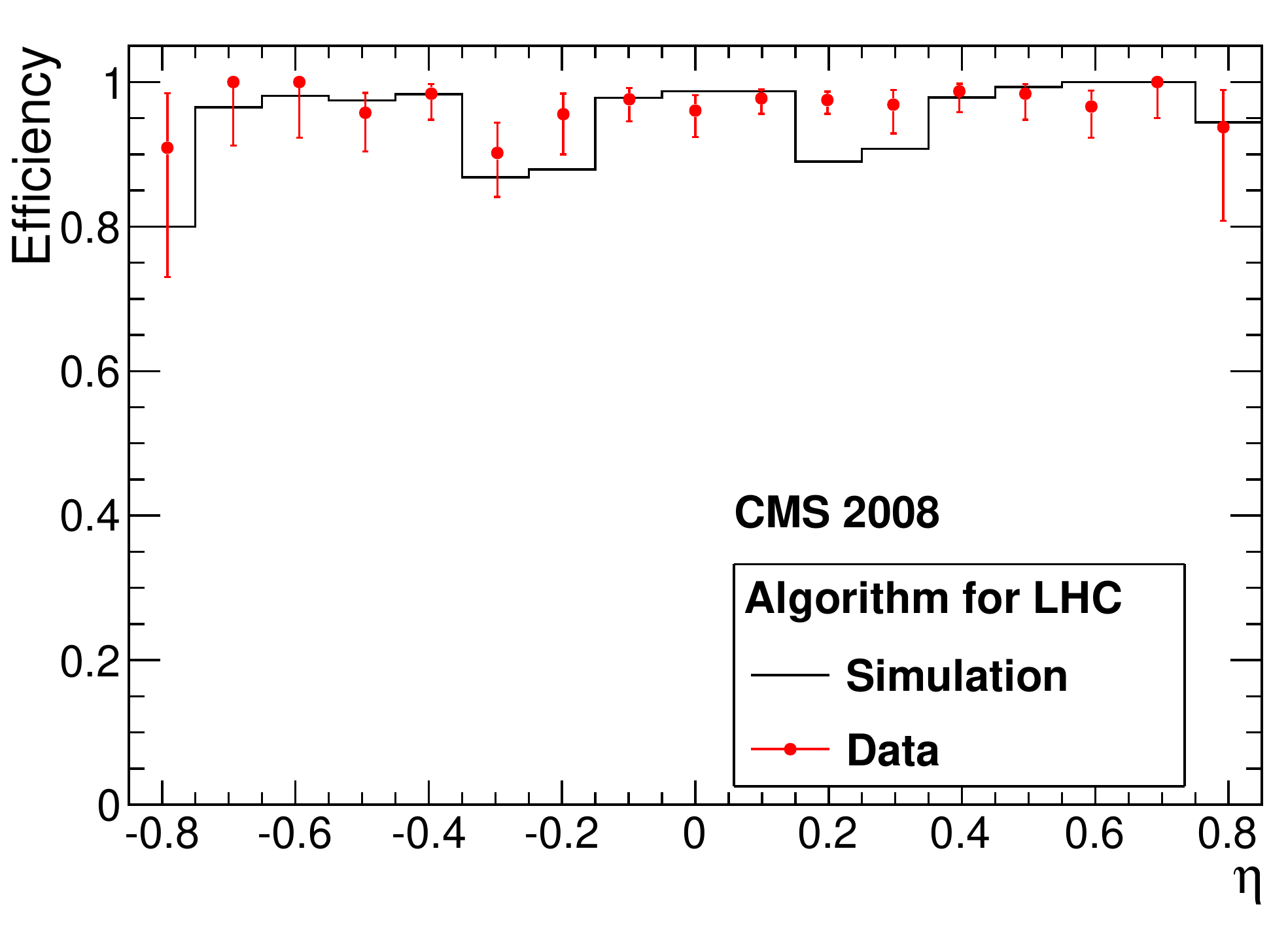}}
      \put(-180,40){\bf\large b)}
      \caption{Reconstruction efficiencies for LHC-like a) standalone
       (ppSTA) and b) global muons as a function of $\eta$ of the
       reference track.  Efficiencies in data are shown as points with
       error bars; efficiencies predicted by the Monte Carlo simulation are
       depicted as histograms.  Statistical uncertainties for simulated
       points are similar to those in data.}
      \label{fig:eff_vs_eta_dataVmc}
  \end{center}
\end{figure}

Measured efficiencies were compared with those obtained by applying
the same method of evaluating efficiencies to the
simulated samples of cosmic muons.  Two examples of such comparisons
are displayed in Fig.~\ref{fig:eff_vs_eta_dataVmc}, showing the
$\eta$ dependence of efficiencies for standalone
and global muons reconstructed by the standard algorithms.  MC efficiencies
integrated over the barrel region of the detector are compared in
Table~\ref{tab:effs_dataVmc} to efficiencies measured in the data.
In general, the results are in good agreement.  In several cases,
the measured efficiencies are slightly larger than the
predicted ones: in particular, this is the case for standalone
muons and for muons selected by the tight version of the cut-based
identification algorithm.  The measured value of the key efficiency,
namely that for LHC-like global muons (and, therefore, for the refits and
selectors described in Section~\ref{sec:tevmus}), exceeds that predicted
by the MC simulation by (2.1 $\pm$ 0.8)\%.

\subsection{Measurements of standalone-muon efficiency with tracker tracks}
\label{sec:eff_Chang}
In addition to the efficiency studies described above, the efficiency
of the standalone muon reconstruction was measured relative to the number
of tracks in the tracker and compared to the efficiency expected from the Monte
Carlo simulation.  Events triggered by the DT or barrel RPC detectors
and containing at least one tracker track reconstructed by the
CosmicTF algorithm were first selected from the tracker-pointing
dataset.  Tracker tracks including
more than 10 hits and with $|\eta| < 0.8$ and $p > 10$~GeV/$c$ were
defined as tags.  The trajectories of these tracks were then propagated to
the outer surface of the CMS detector, and a standalone-muon track
reconstructed by the 1-leg CosmicSTA algorithm (the probe) was searched
for in the nearby region.

\begin{figure}[htb!]
  \centering
  \includegraphics[width=0.5\textwidth]{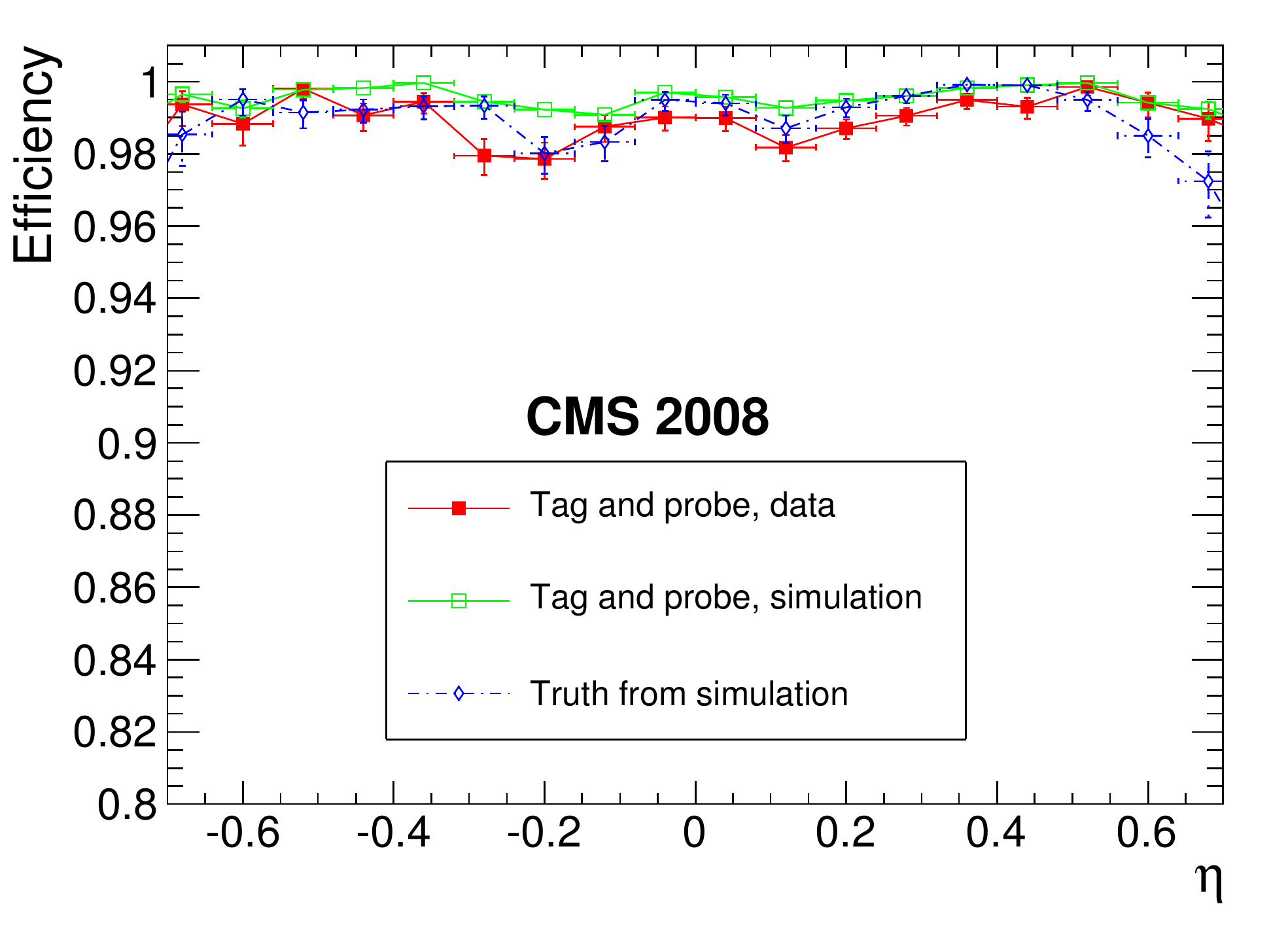}
  \put(-180,40){\bf\large a)}
  \includegraphics[width=0.5\textwidth]{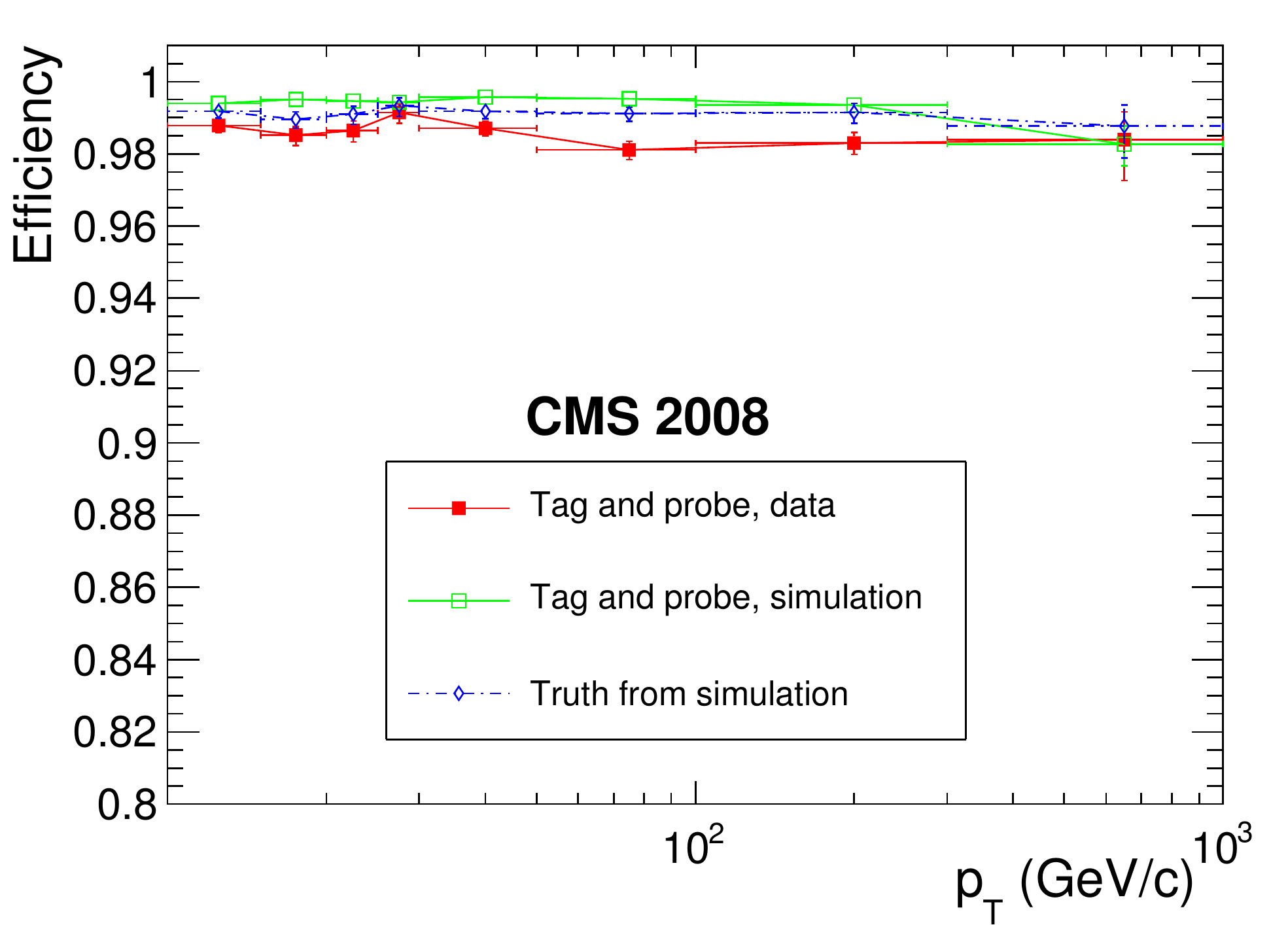}
  \put(-180,40){\bf\large b)}\\
  \includegraphics[width=0.5\textwidth]{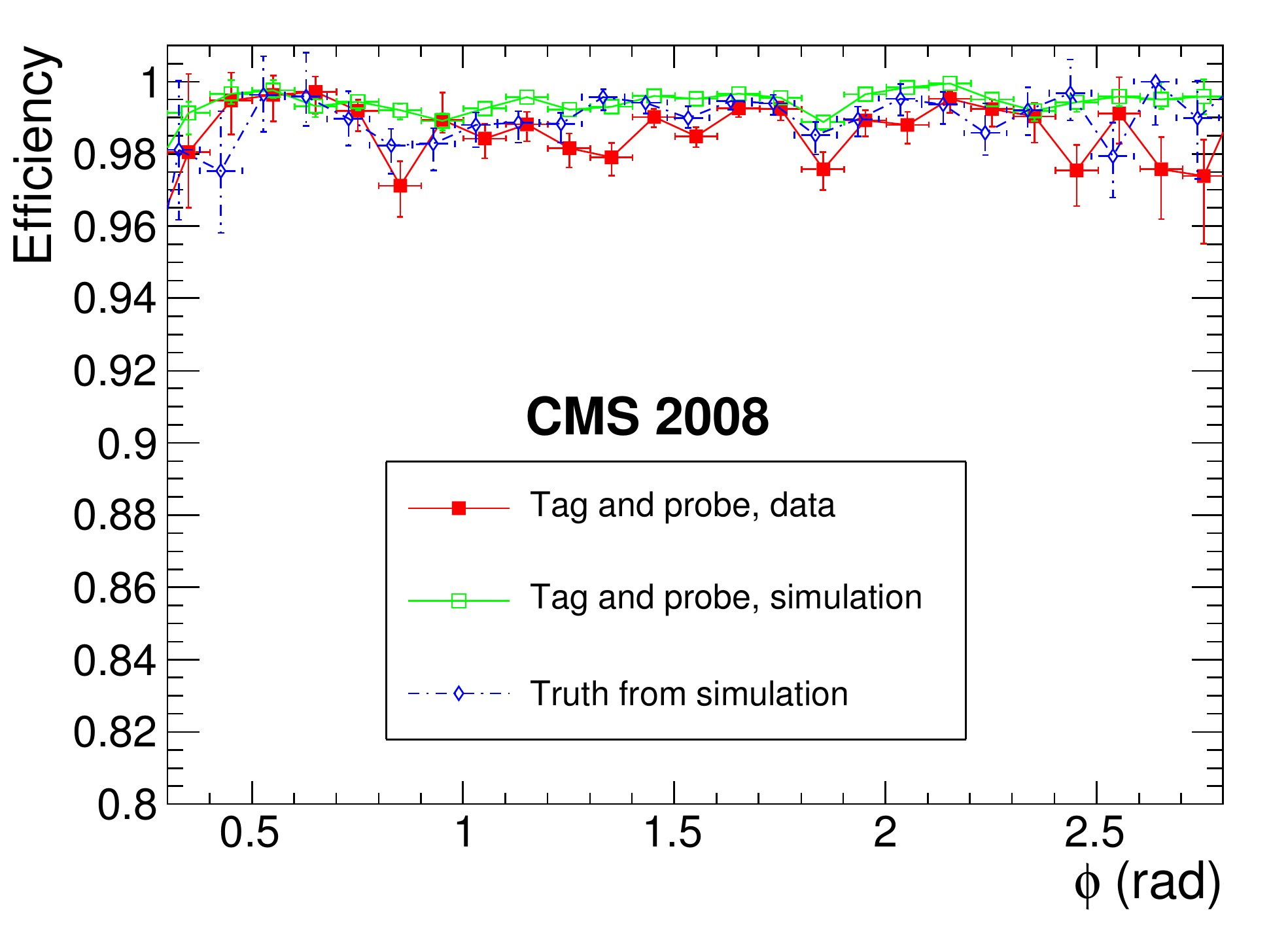}
  \put(-180,40){\bf\large c)}
  \includegraphics[width=0.5\textwidth]{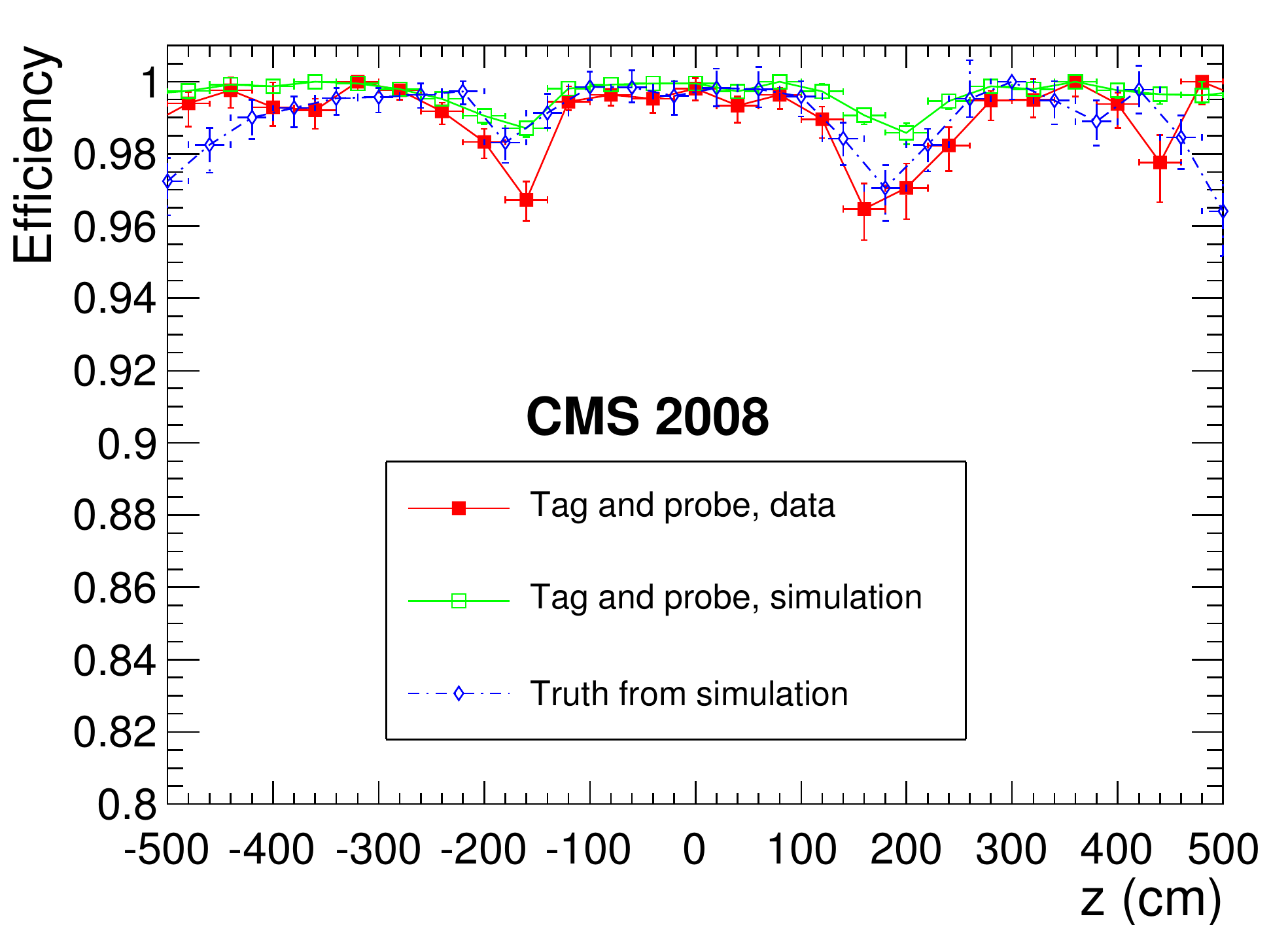}
  \put(-180,40){\bf\large d)}
  \caption{Reconstruction efficiency for 1-leg CosmicSTA standalone muons
   in the data (filled squares) and in the Monte Carlo simulation
   (open squares) as a function of a) $\eta$, b) \pt, c) $\phi$,
   and d) the $z$ coordinate of the muon entry point into the detector.
   Also shown is the efficiency calculated relative to the number of
   generated muons (diamonds).}
  \label{fig:sta_eff}
\end{figure}

The efficiencies measured in the data were compared to those in the Monte Carlo
simulation, calculated in two ways: relative to the number of reconstructed
tracker tracks using the tag-and-probe method, as for the data, and relative
to the number of generated muons.  Simulated events were required to satisfy
the (very loose) \Lone trigger selection criteria used during
CRAFT~\cite{CMS_CFT_09_013}.  Figure~\ref{fig:sta_eff}
shows reconstruction efficiencies for standalone muons as a function of the
tracker-track $\eta$, \pt, azimuthal angle $\phi$, and the $z$ coordinate
of the muon entry point into the detector.  The standalone-muon
efficiency in the data is seen here to be 98--99\%, except for small
regions at the boundaries between the barrel wheels ($z \sim \pm$~200~cm
and $\eta \sim \pm$~0.2).
The efficiency remains high at the largest \pt values studied (of the
order of 500~GeV/$c$).  As expected, no dependence on $\phi$ is observed.
The Monte Carlo simulation reproduces all efficiency distributions to within
1--2\%.  Good agreement with the true efficiency calculated
relative to the generated number of muons confirms the validity of the
method used.

\section{Momentum and Angular Resolutions} \label{sec:pres}

The muon momentum resolution was studied using 2-leg muons.  A pure sample
of muons with a topology similar to that of muons produced in beam collisions
at the LHC was obtained by requiring that each of the muon tracks has at least
1 hit in the pixel detector and at least 8 hits in the silicon-strip detector.
To further suppress contamination from events in which one could
inadvertently compare tracks from different muons, only events with
exactly one pair of tracks were considered.  Since the alignment
of the muon endcaps has not been completed,
the small subset of events with muons in the
CSCs was removed explicitly. Finally, events with muons having hits in
several DT chambers in the horizontal sectors that could not be aligned to
satisfactory precision~\cite{CMS_CFT_09_016} were discarded.  Overall, 23\,458
events were selected.

For each pair of muon tracks in the selected events, the relative
$q/\pt$ residual, $R(q/\pt)$, was calculated as
\begin{equation}
R(q/\pt) = \frac{{(q/\pt)}^\text{upper} - {(q/\pt)}^\text{lower}}{\sqrt{2}{(q/\pt)}^\text{lower}}\quad, \label{eq:r_pt}
\end{equation}
where ${(q/\pt)}^\text{upper}$ and ${(q/\pt)}^\text{lower}$ are the ratios
of the charge sign to the transverse momentum for muon tracks in the upper
and lower detector halves, respectively.  The $\sqrt{2}$ factor accounts
for the fact that the upper and lower tracks are reconstructed
independently and with a similar precision.  The normalized $q/\pt$ residual
(or pull), $P(q/\pt)$, was defined as
\begin{equation}
P(q/\pt) = \frac{{(q/\pt)}^\text{upper} - {(q/\pt)}^\text{lower}}
{\sqrt{\sigma_{(q/\pt)^\text{upper}}^2 + \sigma_{(q/\pt)^\text{lower}}^2}}
\quad,
\end{equation}
where $\sigma_{(q/\pt)^\text{upper}}$ and
$\sigma_{(q/\pt)^\text{lower}}$ are the estimates of $q/\pt$ errors
for the upper and lower muon tracks, respectively.  The values of $q/\pt$
and the corresponding errors were evaluated at the point of closest
approach of each track to the nominal beam line.
As the momentum resolution for standalone muons is expected to be
significantly worse than that obtained using the other muon
reconstruction algorithms~\cite{PTDR1}, the residuals and pulls for
standalone muons were estimated by comparing $q/\pt$ of each standalone
muon reconstructed in the lower detector hemisphere with $q/\pt$ of the
global muon in the same hemisphere (and omitting $\sqrt{2}$ in
Eq.~(\ref{eq:r_pt})).  Since the momentum vector of a tracker muon
is the same as that of the corresponding tracker track, the results for
tracker tracks shown in this and the next section are valid for tracker
muons as well.

\begin{figure}[htb]
  \begin{center}
      \vspace*{-0.6cm}
      \includegraphics[width=1.\textwidth]{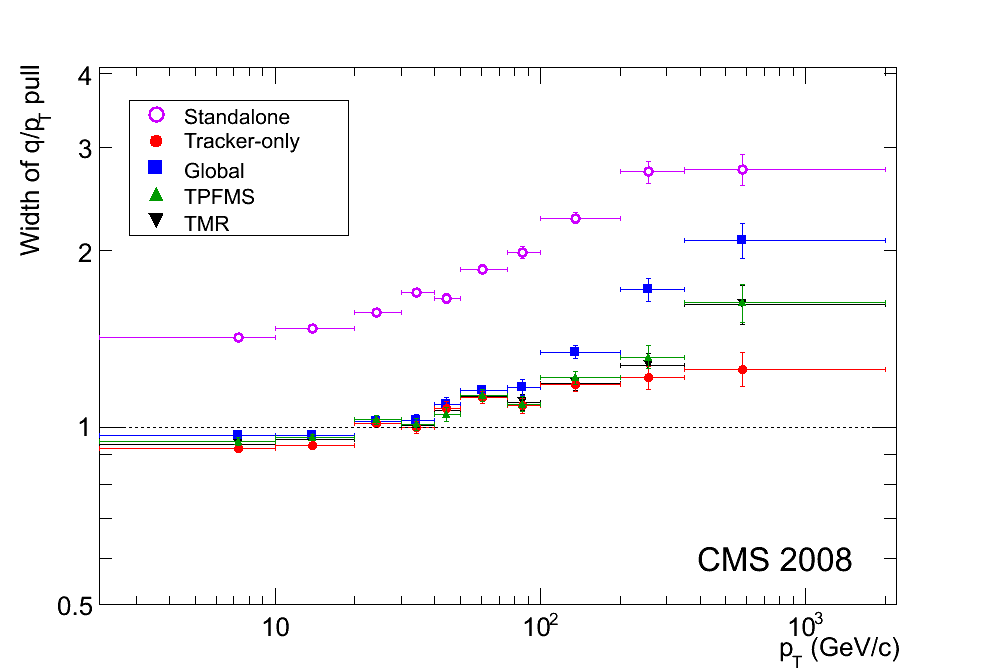}
      \vspace*{-0.5cm}
      \caption{Widths of Gaussian fits to the distributions of the
      normalized residuals, $P(q/\pt)$, for various muon reconstruction
      algorithms, as a function of \pt of the reference track.}
      \label{fig:pull_vs_pt}
  \end{center}
\end{figure}

The widths of the pull distributions were examined to verify the
accuracy of the estimated track-parameter errors.  These estimates
depend, among other things, on the so-called alignment position errors
(APEs) accounting for the precision with which the positions of
different detector components are known~\cite{CMS_CFT_09_003}.  The
available sample of 2-leg muons was subdivided into several subsamples
according to the \pt of the muon track reconstructed in the lower
hemisphere (the reference track), and a Gaussian
fit to the $P(q/\pt)$ distribution for each subset was performed.
The fit range used throughout this section was $\pm$ 2$\cdot$RMS;
various other ranges were tried and only small differences were
observed.  Figure~\ref{fig:pull_vs_pt} shows the widths of
these Gaussian fits as a function of reference-track \pt for muon tracks
reconstructed by various algorithms described
in Section~\ref{sec:algos}.  If all errors were calculated correctly,
these widths should be 1.0.  The widths of the pulls for standalone
muons are greater than unity at all \pt values because the muon APEs,
which were not yet fully implemented, were all set to zero in the
reconstruction.  The widths of other pulls are consistent with unity
in the region of \pt $\lesssim$ 40~GeV/$c$, confirming that
the estimates of errors for the low-\pt region
are accurate.  In the higher-\pt region, the widths of the tracker-only
pulls are larger than 1.0, indicating that the tracker APEs are
underestimated.  As the muon \pt increases, so does the importance of the
muon system in the momentum measurement, and the widths of the pulls
for the combined tracker-muon fits move closer to the widths of the
pulls of the standalone-muon fit.

\begin{figure}[htb]
  \begin{center}
      \vspace*{-0.6cm}
      \includegraphics[width=1.\textwidth]{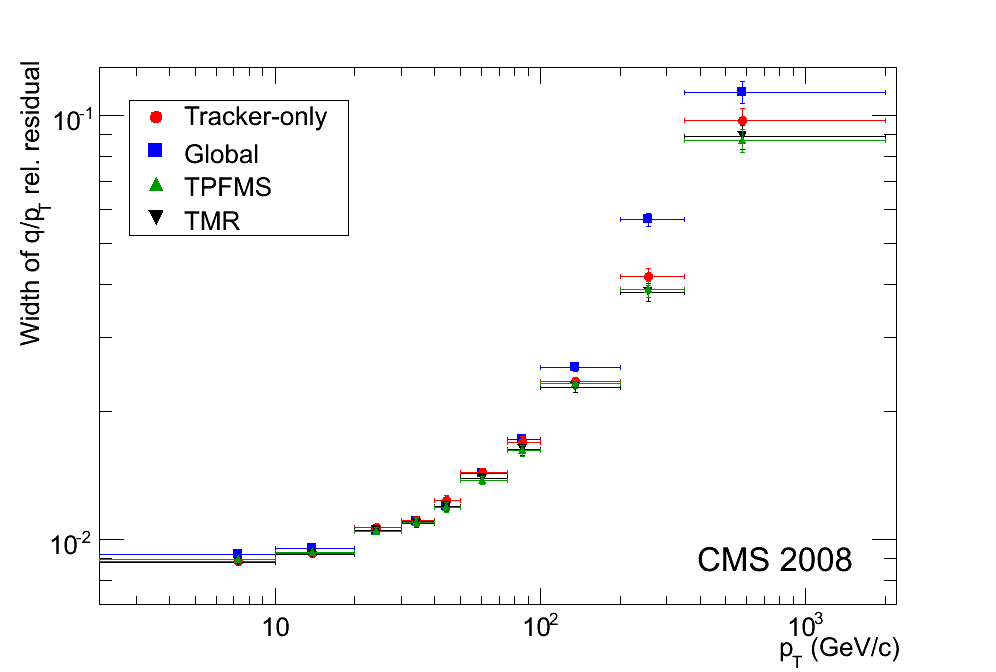}
      \vspace*{-0.5cm}
      \caption{Widths of Gaussian fits to the distributions of the
      relative residuals, $R(q/\pt)$, for various muon reconstruction
      algorithms, as a function of \pt of the reference track.}
      \label{fig:resol_vs_pt}
  \end{center}
\end{figure}

Figure~\ref{fig:resol_vs_pt}
shows the widths of the Gaussian fits to $R(q/\pt)$ distributions
obtained with various muon reconstruction algorithms; these widths are a
measure of the momentum resolution.  In the \pt region below approximately
200~GeV/$c$, the resolution in the muon detector is dominated by
multiple-scattering effects and the inclusion of muon hits does not
improve the resolution beyond that obtained with the tracker-only fits.
At higher \pt, the resolution of the global muon reconstruction
algorithm is currently not as good as that of the tracker tracks;
we expect it to improve once the muon APEs are taken into account
or there is a sufficiently better alignment.  On the other hand,
the resolution in the high-\pt region obtained with
TPFMS and TMR 
is already better than that of global muons and of tracker-only tracks, as
expected.  These and other algorithms described in Section~\ref{sec:tevmus}
improve not only the ``core'' resolution but also the resolution tails,
as can be seen from the summary of the performance of all studied
muon reconstruction algorithms obtained on a sample of muons with
reference-track $\pt >$ 200~GeV/$c$ in Table~\ref{tab:highptres}.
Very similar results (albeit with somewhat larger statistical
uncertainties) were obtained by repeating the analysis on a sample of
split global muons and using the \pt of the original ``unsplit'' track as
the reference \pt.


\begin{table}[tbh!]
\centering
\caption{Summary of figures characterizing $R(q/\pt)$ residuals for the
studied muon reconstruction
algorithms, evaluated on a sample of 567 muons with reference-track $\pt >$
200~GeV/$c$: the width of the Gaussian fit; the value of the RMS truncated at
$\pm$~0.5; the number of events with $R(q/\pt) < -$0.5; the number of
events with $R(q/\pt) >$ 0.5.}
\begin{tabular}{|l|c|c|c|c|} \hline
Fit/selector & Fitted $\sigma$ (\%) & RMS (\%) & $R(q/\pt) < -$0.5 & $R(q/\pt) >$ 0.5 \\ \hline
Tracker-only fit & 5.5 $\pm$ 0.1 & 7.6 $\pm$ 0.2 &       1 &        1 \\
Global fit       & 6.1 $\pm$ 0.2 & 9.5 $\pm$ 0.3 &       8 &       14 \\
TPFMS fit        & 5.2 $\pm$ 0.1 & 6.9 $\pm$ 0.2 &       4 &        3 \\
``Picky'' fit    & 5.5 $\pm$ 0.2 & 6.9 $\pm$ 0.2 &       0 &        0 \\
Sigma switch     & 5.3 $\pm$ 0.1 & 7.4 $\pm$ 0.2 &       1 &        1 \\
TMR              & 5.1 $\pm$ 0.1 & 7.3 $\pm$ 0.2 &       0 &        1 \\
Tune P           & 5.0 $\pm$ 0.1 & 6.5 $\pm$ 0.2 &       0 &        1 \\ \hline
\end{tabular}
\label{tab:highptres}
\end{table}

\begin{figure}[thb!]
  \begin{center}
      \vspace*{-0.2cm}
      \includegraphics[width=0.5\textwidth]{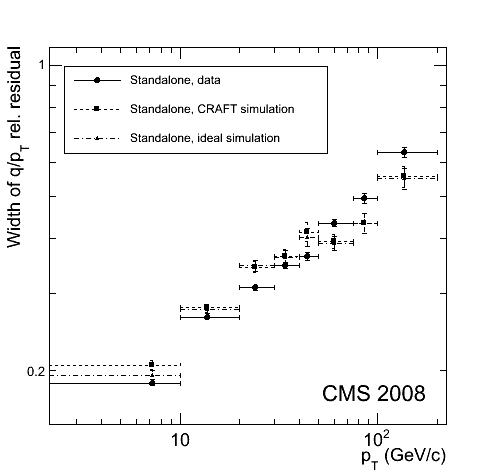}
      \put(-50,60){\bf\large a)}
      \includegraphics[width=0.5\textwidth]{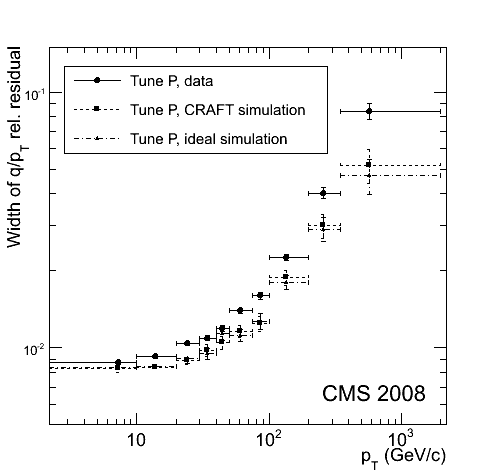}
      \put(-50,60){\bf\large b)}
      \caption{Widths of Gaussian fits to the distributions of the
      relative residuals, $R(q/\pt)$, for a) standalone muons (without the
      beam-spot constraint) and b) muons reconstructed by the ``Tune P''
      method as a function of \pt of the reference track.  The widths
      are compared to two different MC predictions: one assuming
      a CRAFT-based alignment precision and the other an ideal alignment.}
      \label{fig:resol_vs_pt_dataVmc}
  \end{center}
\end{figure}

\begin{figure}[thb!]
  \begin{center}
      \vspace*{-0.2cm}
      \includegraphics[width=0.8\textwidth]{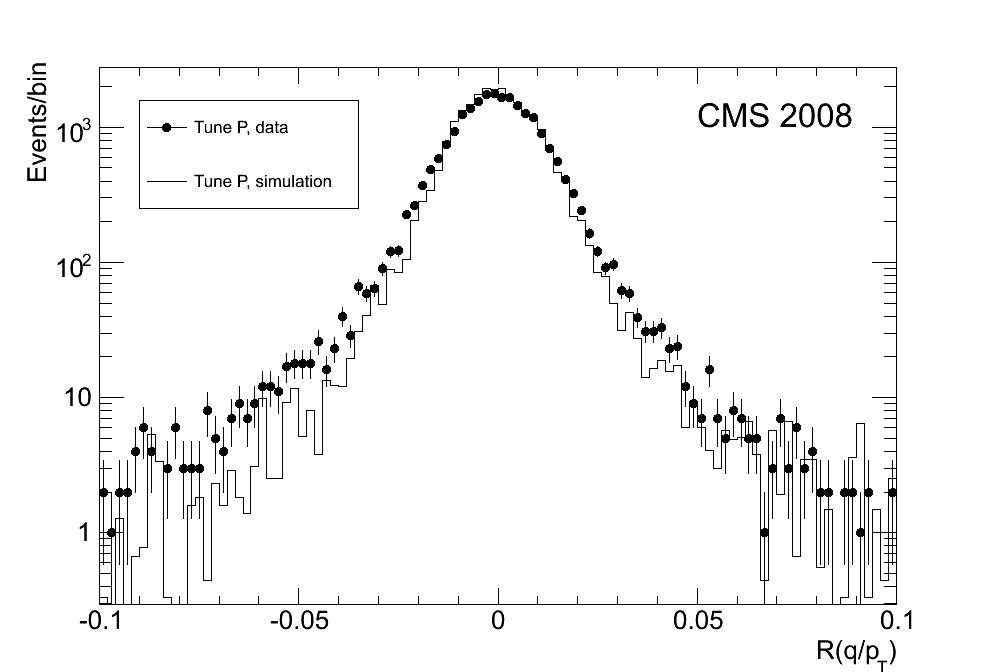}
      \caption{Distributions of the relative residual, $R(q/\pt)$, for
      muons reconstructed by the ``Tune P'' method in data (points with
      error bars) and in the Monte Carlo simulation with the
      CRAFT-based alignment (histogram).  The MC distribution is normalized
      to the number of events in the data.}
      \label{fig:resol_tuneP_dataVmc}
  \end{center}
\end{figure}

Momentum resolutions obtained for various reconstruction algorithms
were compared with those predicted by the Monte Carlo simulation
of cosmic muons.  Figure~\ref{fig:resol_vs_pt_dataVmc} shows two
examples of such data-simulation comparisons, for standalone muons (without the
beam-spot constraint) and for muons
reconstructed by the ``Tune P'' method.  Each measured distribution is
compared to two types of simulated ones: that obtained using the current
best estimates of the precision to which the tracker and the muon system
have been aligned in CRAFT, and a scenario in which all components of the
tracker and the muon system are perfectly aligned.  While the MC
simulation using the CRAFT-based alignment describes the resolution
for standalone muons rather well, its prediction for muons reconstructed
by the ``Tune P'' method (as well as other combined tracker-muon fits)
is better than the measured resolution at all \pt values.  The difference
is about 10\% at low \pt, mostly due to a too optimistic
description of the tracker alignment, and is about a factor of two
in the highest-\pt bin, where both the tracker and the muon alignment
play a role.  Comparisons between the data and MC predictions for the
ideal alignment confirm the
results of other studies~\cite{CMS_CFT_09_003, CMS_CFT_09_016}
demonstrating that the alignment precision achieved in CRAFT for the
barrel tracker and muon system is already quite good, although there
is some room for improving the resolution further, notably at high \pt.
The resolution tails are rather well reproduced, as can be
seen from the data-MC comparison for the $R(q/\pt)$ distribution for
``Tune P'' muons shown in Fig.~\ref{fig:resol_tuneP_dataVmc}.

\begin{figure}[htbp]
  \centering
  \vspace*{-0.2cm}
  \includegraphics[width=0.5\textwidth]{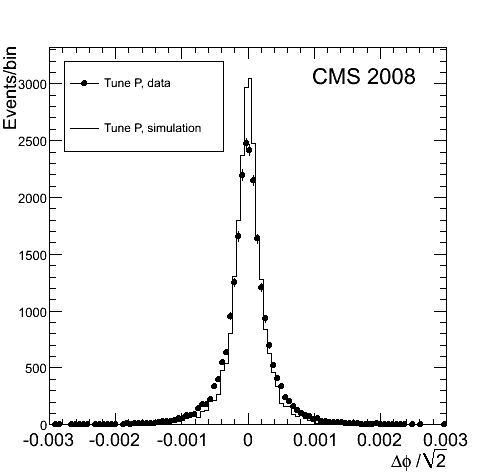}
  \put(-60,50){\bf\large a)}
  \includegraphics[width=0.5\textwidth]{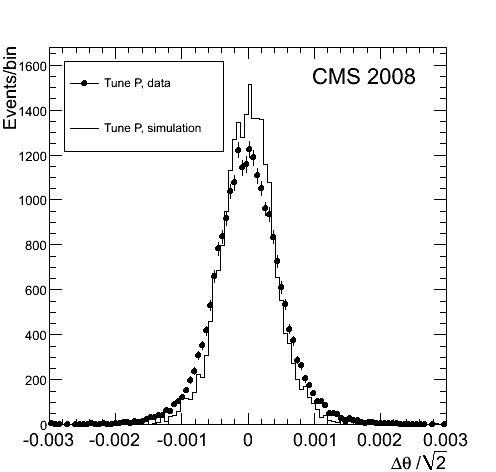}
  \put(-60,50){\bf\large b)}
  \caption{Distributions of residuals for a) the azimuthal angle
  $\phi$ and b) the polar angle $\theta$ at the point of closest
  approach to the nominal beam line, for muons reconstructed by the ``Tune P''
  method.  Residuals in data (points) are compared with the
  predictions of the Monte Carlo simulation using the CRAFT-based
  alignment (histograms).  The MC distributions are normalized to the
  number of events in the data.}
  \label{fig:resol_angles}
\end{figure}

The same sample of events was used to evaluate angular resolutions.
The absolute residuals for the azimuthal angle $\phi$ and the
polar angle $\theta$ were calculated as differences
between the corresponding angles of the upper and the lower tracks
divided by $\sqrt{2}$.  In Fig.~\ref{fig:resol_angles},
the distributions of the $\phi$ and $\theta$ residuals for ``Tune P'' muons in
data are compared to the predictions obtained from the simulation
using the CRAFT-based alignment.  As for the \pt resolution, the predictions of
the Monte Carlo simulation for angular resolutions are slightly more optimistic
than the resolutions measured in data, but the overall agreement is
reasonable.

\begin{figure}[hbt!]
  \begin{center}
      \includegraphics[width=1.0\textwidth]{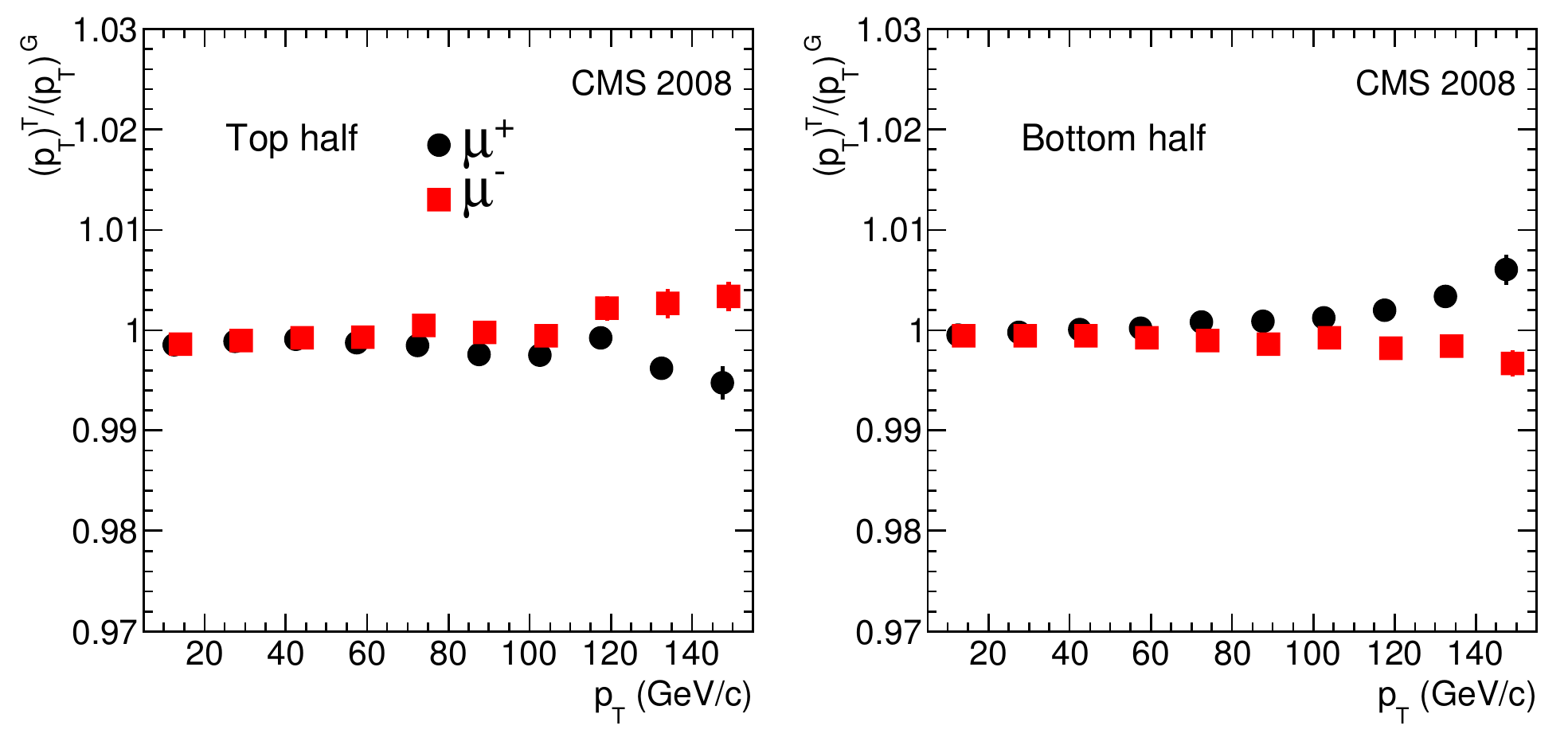}
      \put(-280,60){\bf\large a)}
      \put(-50,60){\bf\large b)}
      \caption{Mean values of Gaussian fits to $F(\pt)$ distributions
      (see text) for $\mu^+$ (circles) and $\mu^-$ (squares)
      as a function of \pt, for split global muons a) at the top half
      and b) at the bottom half of the detector.
      When plotted, points for $\mu^-$ are offset by a small distance
      in \pt for clarity.}
      \label{fig:resol_gb_vs_tt}
  \end{center}
\end{figure}

The impact of certain systematic effects on the muon reconstruction
performance can be examined by comparing results of the full muon
reconstruction with parameters of the tracks reconstructed in the
tracker alone.  Such effects include systematic
deviations of the values of the magnetic field used in the muon
reconstruction from their true values~\cite{CMS_CFT_09_015}, as well
as unaccounted shifts or rotations of the muon system with respect
to the tracker~\cite{CMS_CFT_09_016}.  For this purpose, a
track-by-track comparison was performed of
the \pt value of the global muon, $(\pt)^\text{G}$, with the \pt
of the corresponding tracker track, $(\pt)^\text{T}$.  Split global muons
in the upper and lower halves of the detector were used.
The distributions of
\begin{equation}
 F(\pt) = (\pt)^\text{T}/(\pt)^\text{G}
\end{equation}
in various \pt bins
were plotted separately for positively charged and negatively charged
mu\-ons; the mean values of the Gaussian fits to these distributions are
shown in Fig.~\ref{fig:resol_gb_vs_tt}.  Systematic biases in the
magnetic field map would affect $\mu^+$ and $\mu^-$ distributions
similarly, resulting in deviations of $\langle F(\pt)
\rangle$ from unity of the same sign and magnitude, whereas
global misalignments that are not accounted for would lead to biases of the same
magnitude but of different sign.  As one can see in
Fig.~\ref{fig:resol_gb_vs_tt}, deviations from unity do not exceed 1\%
in the transverse momentum range up to 150~GeV/$c$ and are of different
sign for $\mu^+$ and $\mu^-$.  Such deviations are consistent with
the current understanding of global alignment in CMS~\cite{CMS_CFT_09_016};
if they are attributed wholly to the effect of the global rotation
around the $z$ axis of the muon system with respect to the tracker,
they would correspond to an angle of about 0.25~mrad.
Overall, these results provide important constraints
on the impact of any remaining unknown systematic effects on the muon
reconstruction performance.

\section{Charge Assignment} \label{sec:chargeid}
The tracker and the muon system of CMS are designed to provide a reliable
determination of the muon charge sign up to very high muon
momenta.  However, some low rate of charge misidentification is
unavoidable.  Charge identification is particularly challenging at
very high energies since the muon trajectory is nearly straight, and
radiative effects can lead to high-multiplicity showers in the muon
system further complicating muon reconstruction.

The rate of charge misassignment is studied by measuring the number of
times the two measurements of the charge of the same muon, in the top
and bottom hemispheres, disagree.  In order to obtain two independent
measurements of the muon charge, 2-leg muons in the
tracker-pointing dataset are used.  Stringent selection criteria suppressing the
wrong charge assignment to a negligible level are applied to the muon
track in the upper hemisphere (used as the tag), and the charge of
the other leg in the bottom hemisphere (the probe) is compared
with the charge of the tag.

Since charge misassignment is a low-rate effect, the purity of the test sample
is crucial. The dominant background in this study comes from muon
showers, where multiple muons traverse the detector at the same time.
In that case, the top and bottom muon measurements may not correspond
to the same muon, which would affect the charge misidentification measurement.
In order to remove this background, exactly two CosmicSTA
standalone muons are required to be
present in the event,
one in the top hemisphere, the other in the bottom.
Since the standalone muon reconstruction is very efficient at
detecting muons, the resulting sample has a high purity of single muons.

The study only includes the performance of the barrel muon system and central tracker,
as the alignment and the magnetic field maps for these parts of the
CMS detector were best
understood at the time of this paper. To obtain a pure sample of
barrel muons, cosmic rays in the muon and tracker
endcaps are vetoed explicitly by requiring that there are no CSC or tracker endcap hits in the muon
fit. In order to guarantee good fits of transverse momentum and
charge, each leg of the muon must have at least 5 hits in the tracker.
To select a sample of muons resembling those expected from beam
collisions, the PCA of each track to the nominal beam line is required to lie
within $r < 50$~cm and $|z| < 30$~cm of the nominal position of
pp interactions.  Finally, the charges of the tag track assigned by
the tracker-only, global, and TPFMS algorithms must all be the same.
While the ``core'' resolution is driven by the tracker for each of these
algorithms, events in the far resolution tails
are typically different for different algorithms.  Requiring consistent
charges for the three fits reduces the charge misassignment to about a
factor of 10 lower than that for the best performing of the three algorithms,
over the full momentum range.  Since this level of charge misassignment is
below what can be probed accurately with the available number of
events, we chose to apply stringent cuts only to the top leg and
report charge confusion for the individual algorithms in the lower
leg, for which the muon propagation direction is LHC-like.

\begin{figure}[hbt!]
  \begin{center}
      \vspace*{-0.4cm}
      \includegraphics[width=1.0\textwidth]{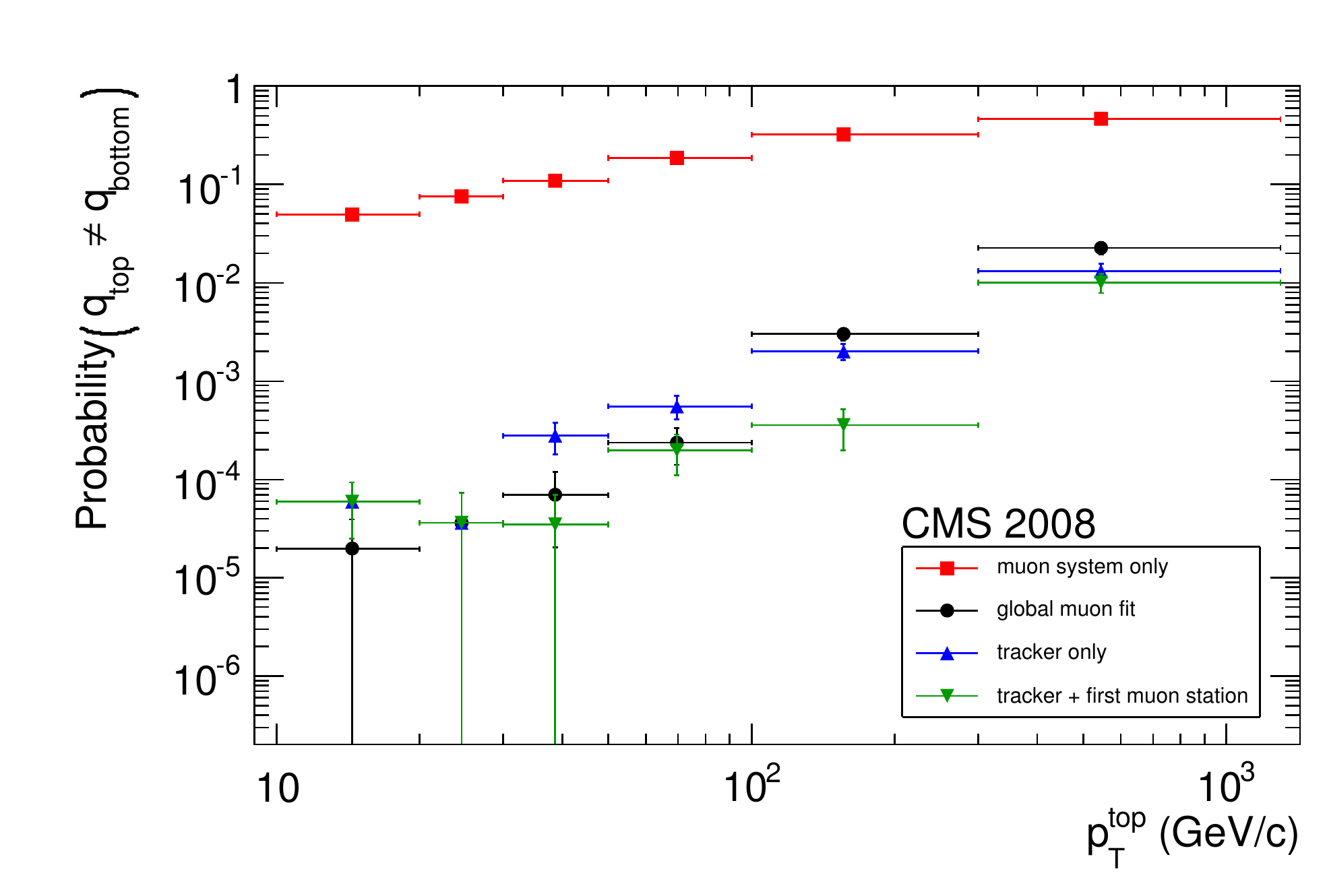}
      \vspace*{-0.6cm}
      \caption{Rate of charge misassignment as a function of \pt
       of the tracker track reconstructed in the top hemisphere, for
       standalone muons (squares), tracker tracks (triangles), global
       muons (circles), and the TPFMS refit (upside-down triangles).}
      \label{fig:chargemisid_vs_pt}
  \end{center}
\end{figure}

The results for different muon reconstruction algorithms are shown in
Fig.~\ref{fig:chargemisid_vs_pt}.  The charge misassignment fraction is
reported in bins of transverse momentum of the tracker track reconstructed
in the top hemisphere; \pt is measured at the point of closest approach
to the nominal beam line. As expected, the measurement of the
charge provided by the standalone muon reconstruction is less accurate
than that in the tracker for the entire \pt range.  Both the tracker-only
fit and the combined tracker-muon fits
provide a reliable charge measurement for the low momentum
region. At high-\pt values, the most accurate charge assignment is
given by the dedicated high-\pt muon reconstruction algorithms.
While different algorithms lead in performance over different momentum
regions, the charge misassignment remains well below 0.1\% up to
\pt = 100~GeV/$c$, becoming about 1\% at \pt $\sim$ 500 GeV/$c$.

\section{Performance of the Muon High-Level Trigger} \label{sec:hlt}
This section describes studies of the performance of the muon
reconstruction algorithms used in the online event selection.  The online
muon reconstruction carried out by the High-Level Trigger~\cite{DAQTDR}
is performed in three stages: local reconstruction
(reconstruction of hits and segments in individual chambers), \Ltwo
reconstruction, and \Lthree reconstruction.  
The \Ltwo
muon reconstruction uses only information from the muon system and is
very similar to the offline standalone muon reconstruction.  The
\Lthree muon reconstruction adds information from the silicon tracker
and shares many features with the offline global muon reconstruction.
Since the \Lone trigger used during CRAFT was configured
to maximize the acceptance for cosmic muons~\cite{CMS_CFT_09_013}
and neither \Ltwo nor \Lthree selections were applied, the performance
of the HLT algorithms can be studied offline in an unbiased way.

\subsection{\Ltwo muon reconstruction}
The main difference between the \Ltwo muon reconstruction and the
offline standalone muon reconstruction lies in the seeding algorithms:
\Ltwo muon seeds are made using the information available in
the \Lone trigger, while offline muon seeds are constructed from the
muon segments reconstructed offline.  At the time of the CRAFT data
taking in 2008, the \Lone DT trigger firmware had not yet been fully
commissioned; as a consequence, the \Ltwo seeds could be
reconstructed reliably only in the bottom hemisphere of the detector,
in the $\phi$ slice between $-$2.2 and $-$0.9 radians.

The efficiency of the \Ltwo muon reconstruction algorithm was studied by
selecting events with a good-quality tracker track in the top hemisphere
and checking whether a \Ltwo muon track was reconstructed in the bottom
hemisphere.  In order to ensure that the parameters of the reference
track are well measured, the tracker track was required to be in the
barrel region ($|\eta| <$ 0.8) and to have at
least 10 hits in the silicon-strip tracker and at least 1 hit in the pixel
detector.  Muons with a topology similar to that expected in beam collisions
at the LHC were selected by requiring that the distance between the point
of closest approach to the beam line and the nominal position of
pp interactions did not exceed 10~cm in $r$ and 30~cm in $z$.
The only modification to the standard \Ltwo reconstruction
algorithm made for this study was the removal of the beam-spot constraint
in the final track fit.  The final sample contained about 100\,000 events.

\begin{figure}[thb!]
  \centering
  \includegraphics[width=0.5\textwidth]{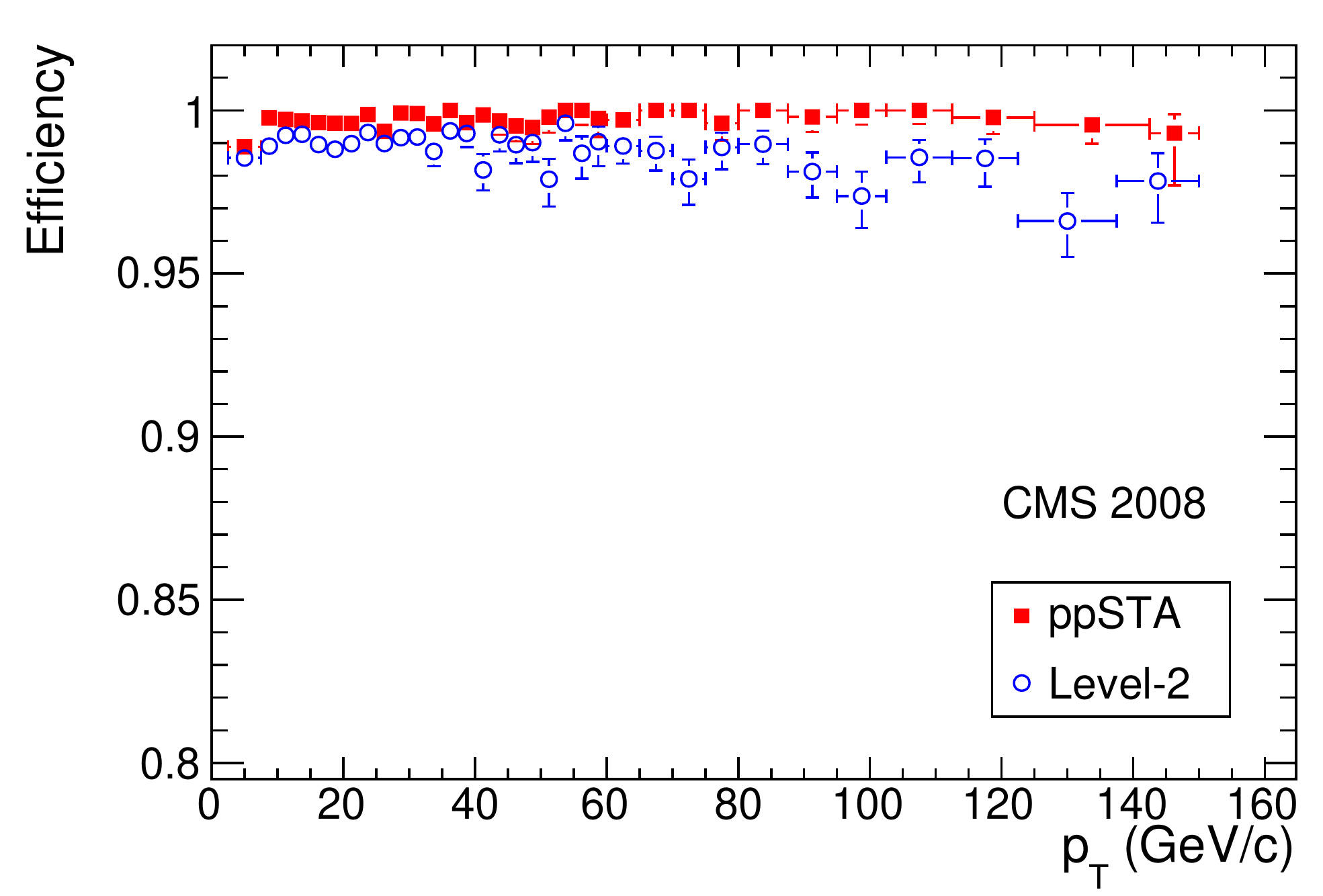}
  \put(-160,40){\bf\large a)}
  \includegraphics[width=0.5\textwidth]{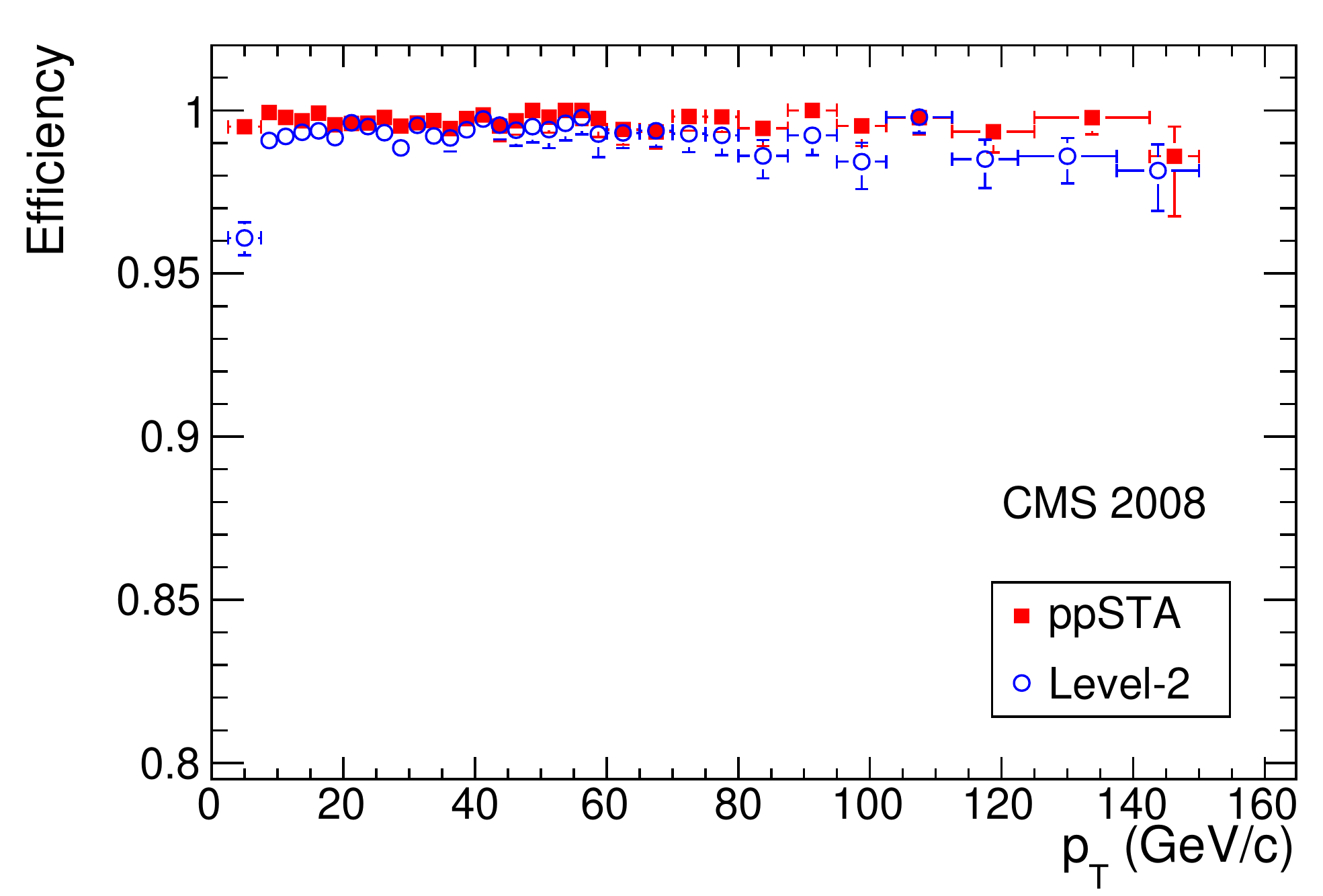}
  \put(-160,40){\bf\large b)}\\
  \includegraphics[width=0.5\textwidth]{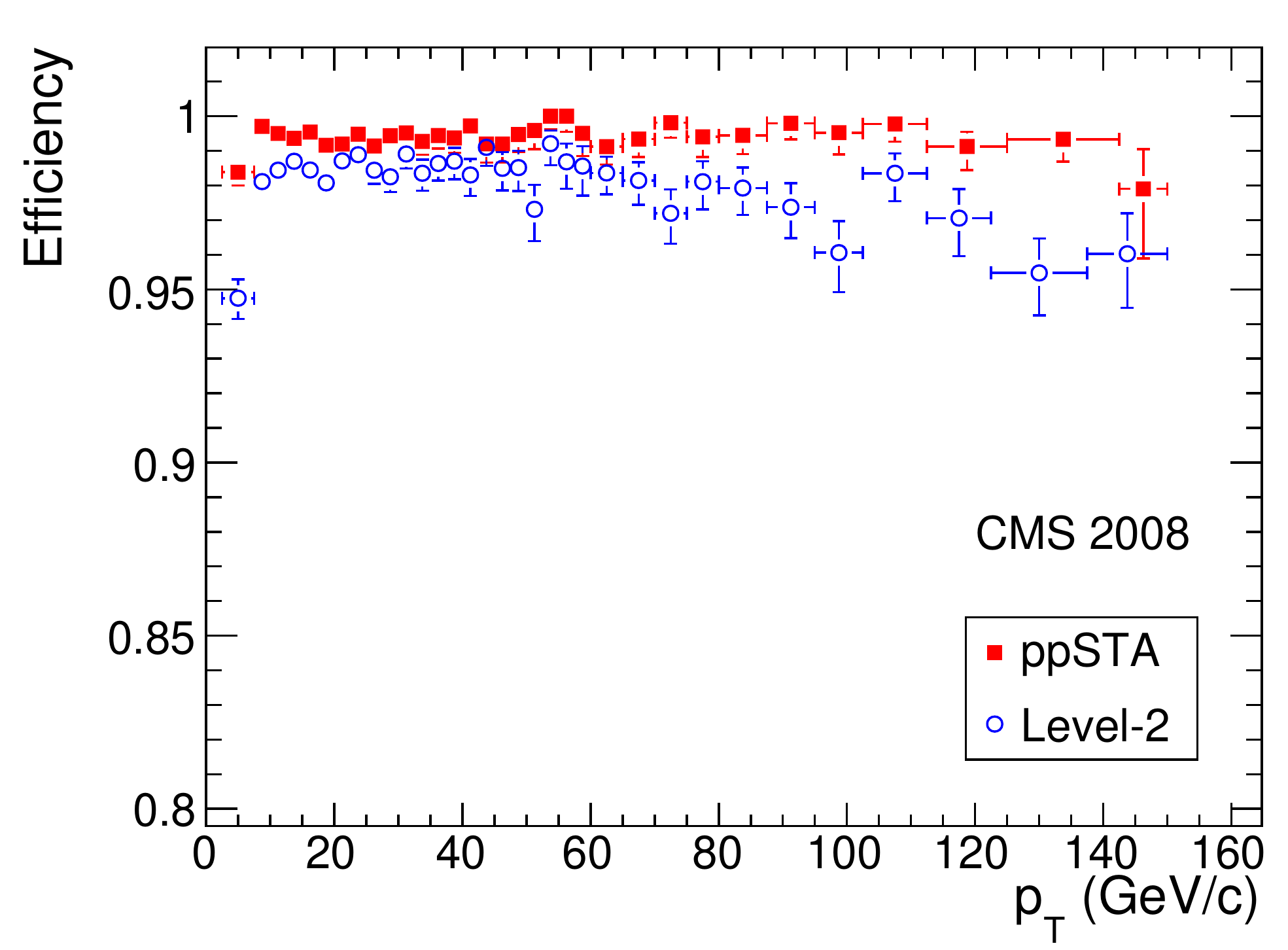}
  \put(-160,40){\bf\large c)}
  \includegraphics[width=0.5\textwidth]{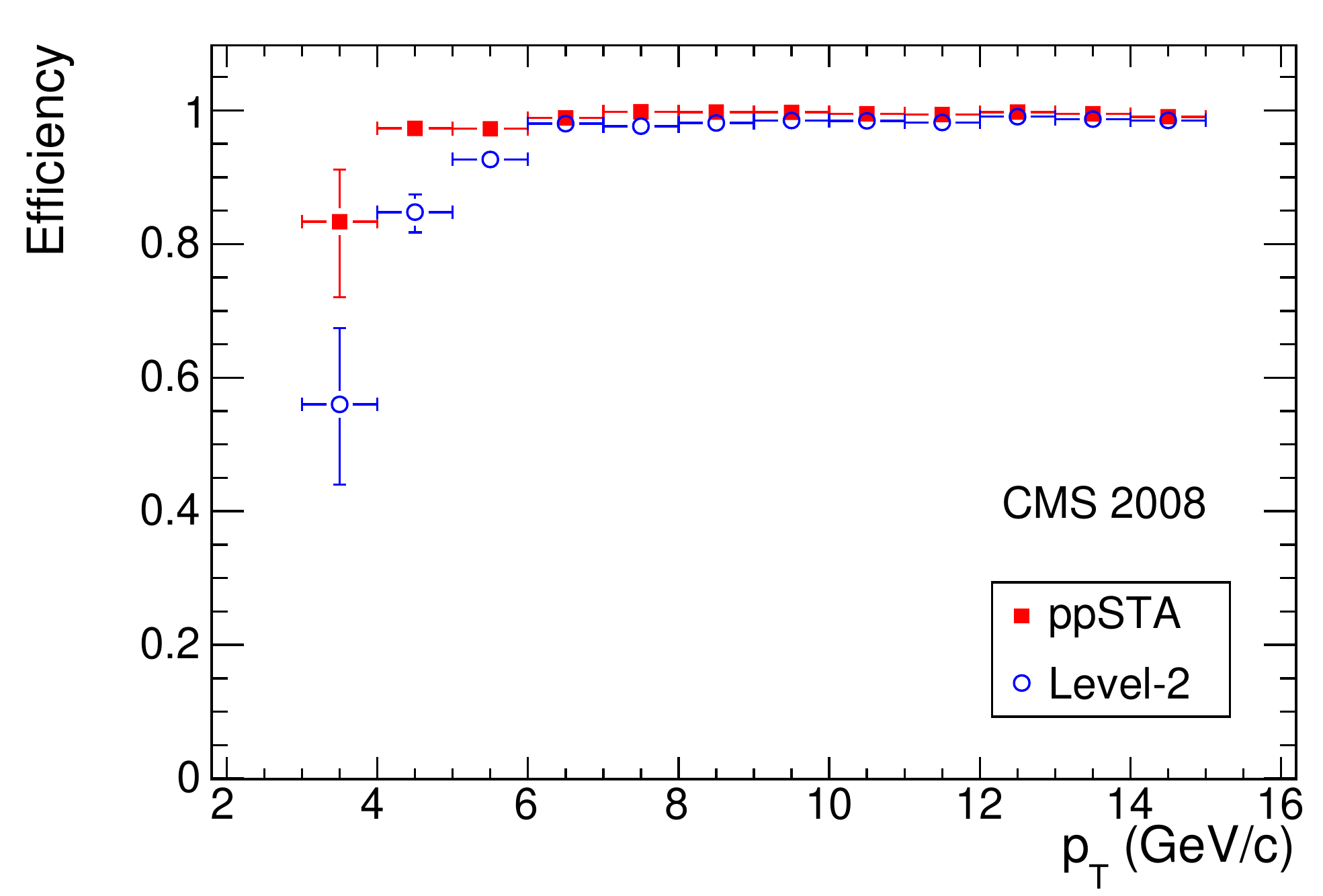}
  \put(-160,40){\bf\large d)}
  \caption{a) Seeding, b) trajectory building, and c) overall
  efficiencies for Level-2 (open circles) and ppSTA (filled
  squares) reconstruction algorithms as a function of tracker-track
  \pt, for muons in the region defined by $|\eta|$ $<$ 0.8 and $-2.2 < \phi <
  -0.9$.  Panel d) is a close-up view of the low-\pt region of the plot
  in c).}
  \label{fig:eff_l2}
\end{figure}

Figure~\ref{fig:eff_l2} shows the efficiencies of
the seeding and trajectory-building steps of the \Ltwo reconstruction,
as well as of the full \Ltwo reconstruction algorithm, as a function of
\pt of the reference track at the PCA.  For comparison, the corresponding
efficiencies of the default offline standalone muon reconstruction (ppSTA) in
the same $\phi-\eta$ detector region, also calculated relative to the
tracker tracks and with the beam-spot constraint removed, are shown
superimposed.  The overall \Ltwo efficiency reaches a plateau close
to 100\% for muons with \pt above 5~GeV/$c$, as expected, and remains very
high up to \pt values on the order of 60--70~GeV/$c$.  The \Ltwo efficiency
in this momentum region is $\sim$1\% lower than the offline
efficiency, because of the slightly less accurate seeding.  At higher \pt
values, the \Ltwo efficiency drops by a few per cent; this efficiency
reduction occurs mostly at the seeding step.

\subsection{\Lthree muon reconstruction}
The \Lthree muon reconstruction consists of three steps: seeding,
trajectory building, and global matching and refitting.  Three \Lthree
algorithms differing in how the seeding step is performed are currently
available: Inside-Out Hit-based (\textit{IOHit}), Outside-In Hit-based
(\textit{OIHit}),
and Outside-In State-based (\textit{OIState}).  In the IOHit algorithm, the
initial trajectory seed is constructed from the innermost tracker (pixel) hits
by proceeding from the interaction point towards the outer edge of the silicon
tracker.  Unlike the IOHit strategy, the OIHit algorithm forms the
trajectory seed from the outermost tracker hits, and proceeds to collect
hits for the fit from
the outer edge of the tracker towards the center of the detector.
The OIState algorithm does not use tracker hits for seeding: it
builds the trajectory seed from the
parameterization of the \Ltwo muon trajectory extrapolated to the
outermost layer of the tracker.
The CRAFT data sample is used to study the efficiencies of these
algorithms for muons resembling those produced in collisions at the LHC.


Since reconstruction of hits and tracks in the tracker is an integral
part of the \Lthree reconstruction, only the runs from the ``period B''
of CRAFT (see Section~\ref{sec:eff_Ivan}) were used.
As the \mbox{\Lthree} muon algorithms are designed to reconstruct muons
originating from the nominal beam-interaction point, a sample of
collision-like muons is needed to evaluate their performance.
Event selection started by requiring that there be exactly one CosmicTF track
reconstructed in the tracker.  The point of closest approach of
this track was required to be inside a cylinder with boundaries
$r < 4$~cm and $|z| < 26$~cm,
corresponding to the innermost layer of the pixel detector.  This
requirement ensures that the muons studied resemble muons produced
by colliding beams; furthermore, the PCA position serves as a
stand-in for the beam spot position needed by the
\Lthree algorithms.
Collision-like \Ltwo muons were selected by demanding that the PCA of
the \Ltwo muon lies within $r < 20$~cm and $|z| < 30$~cm.
Finally, the presence of at least two pixel hits in different pixel
layers in the detector hemisphere opposite to that containing the
\Ltwo muon was required.  Table~\ref{tab:selection} shows the 
summary of the event selection.

\begin{table}[ht]
\centering
\caption{Summary of event selection used in studies of the \Lthree
 reconstruction.}
\begin{tabular}{|c|r|} \hline
Selection criteria & Number of events \\ \hline
Initial dataset (``period B'') & 88\,015 \\
One CosmicTF track per event & 68\,318 \\
PCA position requirement & 6\,224 \\
\Ltwo pointing requirement & 3\,363 \\
Pixel hit requirement & 2\,527 \\ \hline
\end{tabular}
\label{tab:selection}
\end{table}

\begin{figure}[htb]
\begin{center}
  \hspace*{-0.45cm}\rotatebox{90}{\includegraphics[width=0.359\linewidth]{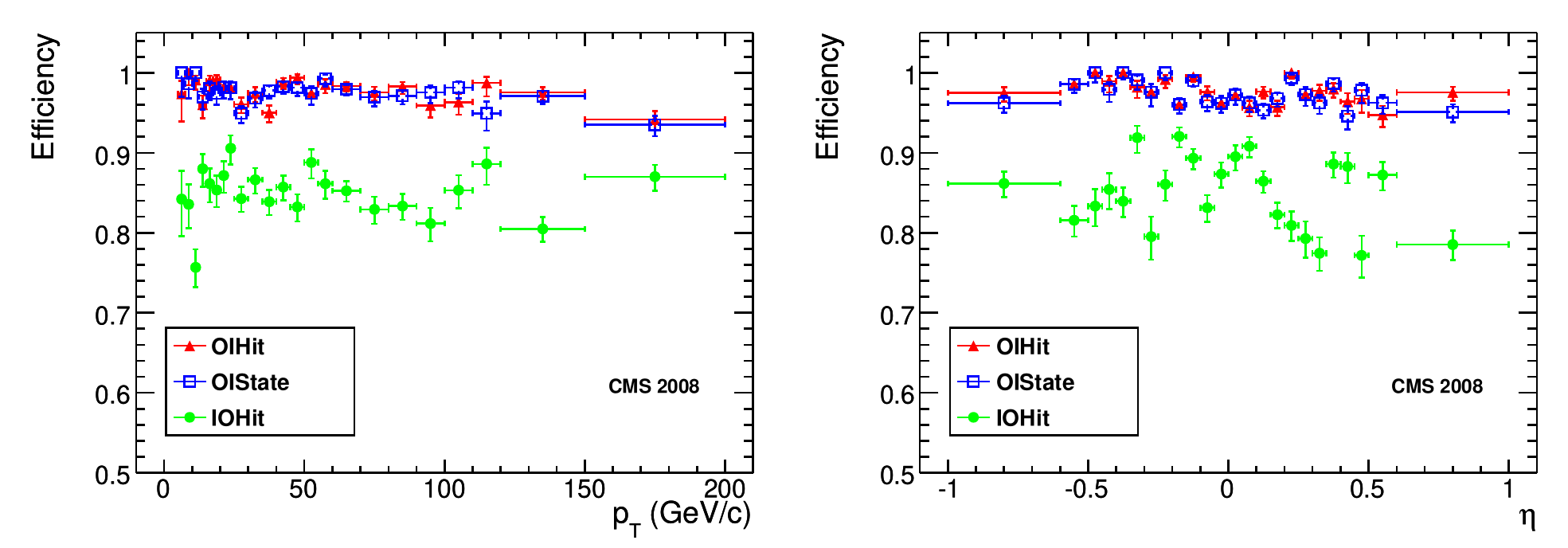}}
  \put(-345,45){\bf\large a)}
  \put(-110,45){\bf\large b)}
  \caption{Reconstruction efficiency for \Lthree muons as a function of
  a) \pt and b) $\eta$ of the \Ltwo muon for three algorithms: IOHit
  (circles), OIHit (triangles), OIState (squares).  Error bars represent
  statistical uncertainties only.}
  \label{fig:eff_l3}
\end{center}
\end{figure}

A total of 2527 events were selected. They contain 3445 \Ltwo muons
satisfying the pointing requirements used in this study.  For each of these
\Ltwo muons, the reconstruction of a \mbox{\Lthree} muon was attempted.
The \Lthree
reconstruction was performed individually with each algorithm (IOHit,
OIHit, and OIState), and the \Lthree efficiency given a pointing \Ltwo muon
was calculated as a function of the \Ltwo muon track parameters.
Figure~\ref{fig:eff_l3} shows the \Lthree
reconstruction efficiency as a function of \Ltwo muon \pt and
$\eta$ for each algorithm.  Both the OIHit and the OIState algorithms
show similar performance
with a high efficiency, and only a small dependence on track
parameters is observed.  The IOHit algorithm has a lower efficiency than
the other two algorithms because its performance strongly depends
on the hit-detection efficiency in the pixel tracker, which in CRAFT
was affected by random arrival times of cosmic rays and some other
factors discussed in Ref.~\cite{CMS_CFT_09_001}.

\section{Conclusion} \label{sec:concl}
The performance of the CMS muon reconstruction has been evaluated
using the large sample of cosmic muons collected during CRAFT.
Measured distributions of basic muon-track quantities are fairly well
reproduced by the Monte Carlo simulation.  Efficiencies of all
available high-level trigger, identification, and reconstruction
algorithms have been measured in a broad range of muon momenta, and
were found to be in good agreement with expectations from Monte Carlo
simulation.  The momentum resolution and the charge assignment in the
barrel part of the CMS detector have been studied up to the TeV momentum
region.  A relative momentum resolution better than 1\% at low \pt values
and of about 8\% at \pt $\sim$ 0.5 TeV/$c$ has been obtained with the
initial CRAFT-based alignment of the tracker and the muon chambers.
Charge misassignment has been measured to be less than 0.01\% at
10~GeV/$c$ and about 1\% at 0.5~TeV/$c$.

The analysis of cosmic-ray muons from CRAFT has provided detailed
insight into the performance of the CMS muon reconstruction
algorithms.  The experience gained is valuable in the preparation for
data from LHC collisions, where reconstruction and identification of
muons will be crucial to achieve the physics goals of the CMS
collaboration.

\section*{Acknowledgements}
We thank the technical and administrative staff at CERN and other CMS
Institutes, and acknowledge support from: FMSR (Austria); FNRS and FWO
(Belgium); CNPq, CAPES, FAPERJ, and FAPESP (Brazil); MES (Bulgaria);
CERN; CAS, MoST, and NSFC (China); COLCIENCIAS (Colombia); MSES
(Croatia); RPF (Cyprus); Academy of Sciences and NICPB (Estonia);
Academy of Finland, ME, and HIP (Finland); CEA and CNRS/IN2P3
(France); BMBF, DFG, and HGF (Germany); GSRT (Greece); OTKA and NKTH
(Hungary); DAE and DST (India); IPM (Iran); SFI (Ireland); INFN
(Italy); NRF (Korea); LAS (Lithuania); CINVESTAV, CONACYT, SEP, and
UASLP-FAI (Mexico); PAEC (Pakistan); SCSR (Poland); FCT (Portugal);
JINR (Armenia, Belarus, Georgia, Ukraine, Uzbekistan); MST and MAE
(Russia); MSTDS (Serbia); MICINN and CPAN (Spain); Swiss Funding
Agencies (Switzerland); NSC (Taipei); TUBITAK and TAEK (Turkey); STFC
(United Kingdom); DOE and NSF (USA). Individuals have received support
from the Marie-Curie IEF program (European Union); the Leventis
Foundation; the A. P. Sloan Foundation; and the Alexander von Humboldt
Foundation.

\bibliography{auto_generated}   
\cleardoublepage\appendix\section{The CMS Collaboration \label{app:collab}}\begin{sloppypar}\hyphenpenalty=500\textbf{Yerevan Physics Institute,  Yerevan,  Armenia}\\*[0pt]
S.~Chatrchyan, V.~Khachatryan, A.M.~Sirunyan
\vskip\cmsinstskip
\textbf{Institut f\"{u}r Hochenergiephysik der OeAW,  Wien,  Austria}\\*[0pt]
W.~Adam, B.~Arnold, H.~Bergauer, T.~Bergauer, M.~Dragicevic, M.~Eichberger, J.~Er\"{o}
% ...
\vskip\cmsinstskip
% ...

\end{sloppypar}
\end{document}